\documentclass[a4paper,11pt]{article}
\pdfoutput=1
\usepackage{jcappub}
\usepackage{slashed}
\usepackage[T1]{fontenc}
\usepackage{adjustbox}
\usepackage{multirow}
\usepackage{hhline}
\usepackage{tabularx,booktabs,caption}
\usepackage{float}
\usepackage{placeins}
\usepackage{caption}
\usepackage{verbatim}
\usepackage{soul}
\usepackage{color}
\usepackage[utf8]{inputenc}
\usepackage{xcolor}
\newcommand{\expt}[1]{\left\langle #1 \right\rangle}
\newcommand{\ie}{$i.e.,$}
\newcommand{\eg}{$e.g.,$}
\newcommand{\kmax}{k_{\rm max}}
\newcommand{\klim}{k_{\rm lim}}
\newcommand{\mwdm}{m_{\rm WDM}}
\newcommand{\mwdmr}{m_{\rm WDM}^{\rm ref}}
\newcommand{\nn}{\nonumber{\nonumber}}
\newcommand{\abs}[1]{\left| #1 \right|}
\newcommand{\beq}{\begin{equation}}
\newcommand{\eeq}{\end{equation}}
\newcommand{\beqn}{\begin{eqnarray}}
\newcommand{\eeqn}{\end{eqnarray}}

\mathchardef\mhyphen="2D 

\title{\boldmath Distinguishing thermal histories of dark matter from structure formation}

\author[b]{Fei Huang,}
\author[a,c]{Yuan-Zhen Li}
\author[a,c,d,e,f]{and Jiang-Hao Yu} 

\affiliation[a]{CAS Key Laboratory of Theoretical Physics, Institute of Theoretical Physics,\\
Chinese Academy of Sciences, Beijing 100190, China}
\affiliation[b]{Department of Particle Physics and Astrophysics, Weizmann Institute of Science, Rehovot 7610001, Israel}
\affiliation[c]{School of Physical Sciences, University of Chinese Academy of Sciences, Beijing 100049, P.\ R.\ China}
\affiliation[d]{Center for High Energy Physics, Peking University, Beijing 100871, China}
\affiliation[e]{School of Fundamental Physics and Mathematical Sciences, Hangzhou Institute for Advanced
Study, UCAS, Hangzhou 310024, China}
\affiliation[f]{International Centre for Theoretical Physics Asia-Pacific, Beijing/Hangzhou, China}

\emailAdd{fei.huang@weizmann.ac.il}
\emailAdd{liyuanzhen@itp.ac.cn}
\emailAdd{jhyu@itp.ac.cn}

\abstract{It is important to understand the implications of current observational constraints and potential signatures on the thermal history of dark matter.
In this paper, we build the connection between the present-day velocities and the production mechanism of dark matter and find that the current observation on structure formation can be imposed to constrain the decoupling temperatures and the phase-space distribution of dark matter.
We further explore the potential of distinguishing different possible thermal histories of dark matter with hypothetical future observational data. 
Using the freeze-in/-out scenarios as templates, we find that future precision data may uniquely identify the allowed parameter spaces for freeze-in and freeze-out, or even completely rule out one of the scenarios.
This method can be more generally applied to other scenarios.}

\begin{document} 
\captionsetup[figure]{labelfont={bf},labelformat={default},labelsep=period,name={FIG.}}
\maketitle
\flushbottom

\section{Introduction}\label{sec:intro}

Dark matter is a long-standing mystery in modern physics whose full picture involves delicate interplay between the physics on the smallest and the largest scales, and the dynamical evolution from the earliest times to the present day.
With ample evidence from across many magnitudes in space and time,
we now know that it does not interact appreciably with the standard-model (SM) fields, nor with itself, it is non-relativistic today and is non-baryonic,
and it has a relic density that is about a quarter of the total energy of the universe.

Our knowledge about dark matter is however still limited, despite the fact that it comprises the majority of the matter component in the universe.
We do not know its fundamental nature such as its mass and spin, whether it is elementary or composite, and whether it is single-component or multi-component.
Likewise, although dark matter plays an important role in the dynamical evolution of the universe especially in shaping the CMB power spectrum and in seeding the structure formtion, the thermal history of dark matter remains unknown.
We do not know how dark matter is produced in the early universe, 
nor do we know whether or not the dark-matter particles have ever been in thermal equilibrium.

Numerous models have been proposed to provide viable dark-matter candidates with some potential detection signatures,
and a large suite of detection techniques, such as direct detection and indirect detection, have been developed to discover or constrain these candidates.
While this top-down approach is important and might shed light on the properties of dark matter, it is of equal importance to consider the bottom-up approach, to figure out 1) what information about the early-universe dynamics of dark matter might be encoded in the observational data, and 2) to what extent one can extract this information from the current or future data to learn about the thermal history of dark matter and discriminate different models or scenarios.

Currently, most of the experimental or observational efforts are based on the non-gravitational interactions of dark matter with the SM particles.
For example, direct-detection experiments assume that dark matter can scatter off the target material (which is made of electrons and nuclei) of a detector with a appreciable rate.
Indirect-detection experiments require dark matter to annihilate sufficiently rapidly in astrophysical environment where the dark-matter density is large.
In addition, collider searches demand that dark matter can be produced via collisions of SM particles and manifest as missing transverse energy.
On the other hand, all pieces of the known evidence for dark matter, such as 
the velocity dispersion of galaxies in the Coma Cluster \cite{Zwicky:1933gu,Zwicky:1937zza},
the galactic rotation curve \cite{Rubin:1970zza}, the cosmic microwave background (CMB) \cite{Planck:2018vyg}, the gravitational lensing \cite{Tyson:1990yt} by galaxy clusters and the Bullet Cluster \cite{Clowe:2006eq},
originate from the gravitational interaction between dark matter and the SM.
In addition, the measurements on CMB and the Bullet cluster also indicate that dark matter has negligible non-gravitational interaction strength.
It is therefore important to explore other possible means of probing dark matter which do not rely on interactions with the SM particles other than gravity.

One of such methods is via observations on the structure of our universe since dark matter seeds structure formation by contributing predominantly to the gravitational potential of density perturbations.
Indeed, it is through studies on structure formation that hot-dark-matter models like the SM neutrinos are ruled out as a candidate that constitutes the majority of dark matter \cite{Peebles:1982ib,Bond:1983hb,White:1983fcs},
while numerical simulations with cold dark matter (CDM) which has negligible free-streaming distance are shown to be consistent with the large scale structure \cite{Gao:2005ca,Springel:2005nw,Springel:2006vs}.
Despite great success on large scales, in recent years, inconsistencies between observations and simulations seem to appear on small scales \cite{Weinberg:2013aya,DelPopolo:2016emo,Tulin:2017ara,Bullock:2017xww,Perivolaropoulos:2021jda}.
The potential deviation from the CDM predictions on these scales may reveal properties of dark matter
which include but are not limited to its self interaction in self-interacting dark-matter models (for example, see Refs.~\cite{Carlson:1992fn,Spergel:1999mh,Dave:2000ar,Rocha:2012jg,Peter:2012jh,Kaplinghat:2015aga,Tulin:2017ara,Garani:2022yzj} and references therein), its de Broglie wavelength or mass in ultralight dark-matter models (see, \eg~Refs.~\cite{Hu:2000ke,Arvanitaki:2009fg,Hui:2016ltb,Irsic:2017yje,Ferreira:2020fam,Hui:2021tkt}), 
its late-decay lifetime \cite{Cen:2000xv,Kaplinghat:2005sy,Cembranos:2005us,Abdelqader:2008wa,Peter:2010au,Peter:2010jy,Peter:2010sz,Bell:2010fk,Wang:2012eka,Wang:2013rha,Wang:2014ina,Audren:2014bca,Blackadder:2015uta,Poulin:2016nat,Boddy:2016bbu,FrancoAbellan:2020xnr,DES:2022doi},
and the free-streaming scale from non-negligible velocities \cite{Colombi:1995ze,Colin:2000dn,Bode:2000gq,Avila-Reese:2000nqd,Viel:2005qj,Abazajian:2005xn,Boyarsky:2008xj,Lovell:2013ola,Schneider:2013ria,Kennedy:2013uta,Schneider:2014rda,Irsic:2017ixq,Baur:2017stq}
or even the detailed shape of its distribution function \cite{Konig:2016dzg,Murgia:2017lwo,Huang:2019eyi,Dienes:2020bmn,Dienes:2021itb,Dienes:2021cxp,Decant:2021mhj}.
Such properties can then be further exploited to understand the thermal history of dark matter.

We are primarily interested in the connection between structure formation and the thermal history of dark matter through the primordial phase-space distribution of dark matter.\footnote{By ``primordial'', we specifically mean the distribution that is redshifted from the time when dark matter is produced to the present day without considering any perturbation due to structure formation such as the virialization inside dark-matter halos.}
On the one hand, simulations of structure formation and observations that depend on the cosmic structure such as measurements on the Lyman-$\alpha$ forest, the 21-cm signal, the halo mass function and the Milky-Way (MW) satellites can provide useful constraints on the dark-matter phase-space distribution.
On the other hand, since the phase-space distribution is ultimately related to the production mechanism of dark matter, constraints from observations may have very different implications for different production scenarios.
If an actual deviation from the CDM predictions is detected in future observations,
the observational data can even be exploited to uniquely identify different possible thermal histories if predictions from different scenarios are sufficiently distinct.

Current studies of this sort have been focused on the warm-dark-matter (WDM) scenario in which dark matter freezes out while it is still relativistic, but becomes non-relativistic before the matter-radiation equality (MRE) to allow a consistent picture with the observed large scale structure.
Since the phase-space distribution of WDM can be fully determined by the WDM mass,
constraints in the literature are often quoted as bounds on the WDM mass.
The most stringent constraints are derived from the observed Lyman-$\alpha$ forest 
with recent studies giving two lower bounds $\mwdm\geq 3.5~\rm keV$ or $5.3~\rm keV$ based on different assumptions on the temperature evolution of the intergalactic medium \cite{Irsic:2017ixq}.

For more general scenarios in which the phase-space distribution of dark matter is different from that of WDM, in principle numerical simulations are needed to obtain robust constraints.
Nevertheless, several effective methods have been developed to recast the WDM bounds to the constraints in the corresponding scenarios \cite{Schneider:2016uqi,Murgia:2017lwo,Kamada:2019kpe,Hooper:2022byl}.
In addition, numerical simulations based parametrized transfer functions have also shown important progress \cite{Murgia:2017lwo,Murgia:2018now,Hooper:2022byl}.

In this work, we shall focus on two most common dark-matter production mechanism, namely, thermal freeze-out and freeze-in.
These two scenarios can be realized by the same particle-physics process but with different interaction strength.
In the former scenario, the interaction between dark-matter particles and the thermal bath where these particles reside is rapid enough such that thermal equilibrium can be established between them,
whereas in the latter scenario such interaction is too feeble to allow equilibration.
Notice that, while WDM is one subset of the former scenario,
our primary interest lies in a different subset, that is, the case in which freeze-out occurs non-relativistically.
It turns out that 
in both the freeze-in and the non-relativistic freeze-out scenarios, dark-matter particles may acquire non-negligible velocities to impact structure formation if the dark-matter mass is sufficiently small,
and thus can be potentially constrained by structure formation. 
Nevertheless, we shall see that the bounds on these two scenarios can be very different since the functional form of the dark-matter phase-space distributions and its dependence on physical parameters such as the dark-matter mass and the decoupling temperature differ significantly between the two scenarios.
We shall then further explore this difference and discuss the potential in distinguishing these two distinct types of thermal histories with future observational data.

This paper is organized as the following.
In Sec.~\ref{sec:PSD_PROD}, we study the relation between the present-day velocities and physical quantities at the decoupling time via the phase-space distribution of dark matter resulting from different production mechanism.
In Sec.~\ref{sec:constraints},
we take a vanilla dark-matter model and numerically calculate the phase-space distribution of dark matter in different regimes while focusing on the freeze-in and freeze-out cases.
We also place constraints on these scenarios using observables from structure formation and discuss the different implications for the freeze-in and freeze-out cases.
Then, in Sec.~\ref{sec:distinguish}, we hypothesize several forms of future observational data on small scale structure,
and evaluate the possibility of differentiating different scenarios from the future data.
Finally, we conclude in Sec.~\ref{sec:conclusion}.

\section{Dark-matter velocities from different production mechanisms}\label{sec:PSD_PROD}

The phase-space distribution $f_\chi(\vec{x},\vec{p},t)$ of the dark-matter particle $\chi$ is a function of spatial coordinates, momentum, and time.
Since the universe is spatially homogeneous and isotropic at the zeroth order, the distribution function can be reduced to a function that only depends on time and the magnitude of momentum $f_\chi(p,t)$.
With $f_\chi(p,t)$, quantities that are relevant for cosmology such as the number density $n_\chi(t)$, the energy density $\rho_\chi(t)$, and the pressure $P_\chi(t)$ can be calculated via
\beqn
n_\chi(t)&=&g_\chi\int \frac{d^3p}{(2\pi)^3} f_\chi(p,t)\,,\\
\rho_\chi(t)&=&g_\chi\int \frac{d^3p}{(2\pi)^3} f_\chi(p,t)E\,,\label{eq:rho_integral0}\\
P_\chi(t)&=&g_\chi\int \frac{d^3p}{(2\pi)^3} f_\chi(p,t)\frac{p^2}{3E}\,,
\eeqn
in which $E=\sqrt{p^2+m_\chi^2}$ is the energy of the dark-matter particle with mass $m_\chi$, and $g_\chi$ is the internal degrees of freedom.

The dynamical evolution of the phase-space distribution is governed by the Boltzmann equation
\beq
\frac{\partial f_\chi}{\partial t}=Hp\frac{\partial f_\chi}{\partial p}+\mathcal{C}[f]\,,\label{eq:Boltzmann_f}
\eeq
in which $H\equiv \dot{a}/a$ is the Hubble parameter with $a$ being the scale factor, and $\mathcal{C}[f]$ is the collision term which contains integrals of the phase-space distribution of all relevant particle species.
In general, the collision term describes how particles are added to, subtract from, or reshuffled within the phase space due to all possible interactions, 
whereas the Hubble term describes how particles redshift from higher momentum to lower momentum due to the expansion of the universe.
Therefore, apparently, the phase-space distribution of dark matter depends on how dark matter is produced.
In what follows, we discuss the phase space distribution resulting from two types of scenarios --- the freeze-out scenario and the freeze-in scenario.

\subsection{Production mechanism: freeze-out}

In the thermal freeze-out scenarios, dark matter is assumed to be in thermal equilibrium with a thermal bath at earlier times 
during which the phase-space distribution is expected to be Fermi-Dirac or Bose-Einstein, \ie
\beq
f_\chi(p,t)=\frac{1}{\exp\left(\frac{E-\mu}{T_\chi}\right)\pm 1}\,,\label{eq:g_thermal}
\eeq
where $T_\chi$ is the temperature of the thermal bath that dark matter is in equilibrium with, and $\mu$ is the chemical potential.
Note that, $T_\chi$ is not necessarily identical to the temperature $T$ of the SM thermal bath.
At later times, dark matter decouples both chemically and kinematically.
Let us assume that the decoupling occurs instantaneously at $t_{\rm dec}$ when the temperature of dark matter is $T_\chi(t_{\rm dec})$.
Since the momentum $p\sim a^{-1}$, the distribution at a later time $t$ can be mapped to the distribution at the decoupling time $t_{\rm dec}$ via
\beqn
f_\chi(p,t)=f_\chi(p_{\rm dec}, t_{\rm dec})= 
\frac{1}{\exp\left(\frac{E_{\rm dec}-\mu_{\rm dec}}{T_\chi(t_{\rm dec})}\right)\pm 1}\,,\label{eq:f_dec}
\eeqn
where $p_{\rm dec}\equiv pa(t)/a(t_{\rm dec})$.

To recast this distribution into a thermal form as Eq.~\eqref{eq:g_thermal} with physical quantities taken at time $t$, one would require the exponent in the brackets to match:
\beq
\frac{\sqrt{p^2+m_\chi^2}-\mu}{T_\chi}=\frac{\sqrt{p^2 a^2_{\rm dec}/a^2(t)+m_\chi^2}-\mu_{\rm dec}}{T_\chi(t_{\rm dec})}.\label{eq:matching}
\eeq
Obviously, if $\mu$ is independent of $p$, the above equation can only be satisfied by a single value of the momentum.
In other words, although the distribution $f_\chi$ is simply related to its what it is at the decoupling time by redshift, it is in principle not evolving towards a thermal distribution with a different temperature and a single-valued chemical potential.

However, for a distribution of the type in Eq.~\eqref{eq:f_dec}, most of the dark-matter particles in the phase space are concentrated at the average momentum $\expt{p}_{\rm dec}$ at the decoupling time if $\mu_{\rm dec}$ is not much larger than $T_\chi(t_{\rm dec})$.
In other words, after integrating the angular dependence, the integrated distribution function $p_{\rm dec}^2f_\chi({p_{\rm dec}, t_{\rm dec}})$ only has non-negligible support at momenta that are not too far away from the average momentum.
Thus, for the part of $p_{\rm dec}^2f_\chi({p_{\rm dec}, t_{\rm dec}})$ that is non-vanishing, it is possible to define an effective temperature and chemical potential in the cases of relativistic and non-relativistic decoupling.

\subsubsection{Relativistic decoupling vs.\ non-relativistic decoupling}

If dark matter decouples relativistically, \ie~$\expt{p}_{\rm dec}\sim T_\chi(t_{\rm dec})\gg m$, the energy of the dark-matter particle is dominated by the kinetic energy
\beq
E_{\rm dec}\approx p_{\rm dec}=p\frac{a(t)}{a_{\rm dec}}\,.
\eeq
By defining
\beqn
\mu^{\rm eff}(t)= \mu_{\rm dec}\frac{a_{\rm dec}}{a(t)}\,,~~T_\chi^{\rm eff}(t)= T_\chi(t_{\rm dec}) \frac{a_{\rm dec}}{a(t)}\,,\label{eq:mu_T_rel}
\eeqn
it is easy to find that the factor
\beq
\frac{E_{\rm dec}-\mu_{\rm dec}}{T_\chi(t_{\rm dec})}\approx
\frac{p_{\rm dec}-\mu_{\rm dec}}{T_\chi(t_{\rm dec})}=\frac{p-\mu^{\rm eff}}{T_\chi^{\rm eff}}\,,
\eeq
namely, the distribution after relativistic decoupling still takes the approximate form of a relativistic thermal distribution characterized by effective temperature $T^{\rm eff}$ and chemical potential $\mu^{\rm eff}$ which both scale like $a^{-1}$. 
Notice that, the shape of the distribution does not depend on the dark-matter mass even after dark-matter particles become non-relativistic at later times.

On the other hand, if dark matter decouples when it is still non-relativistic, \ie~$m\gg T_\chi(t_{\rm dec})$, we have 
\beq
E_{\rm dec}\approx m_\chi+\frac{p_{\rm dec}^2}{2m_\chi}=m_\chi+\frac{p^2}{2m_\chi}\frac{a(t)^2}{a^2_{\rm dec}}\,.
\eeq
In this case, we can define 
\beqn
\mu^{\rm eff}=m_\chi-(m_\chi-\mu_{\rm dec})\frac{a_{\rm dec}^2}{a^2(t)}\,,~~T_\chi^{\rm eff}=T_\chi(t_{\rm dec})\frac{a_{\rm dec}^2}{a^2(t)}\,, \label{eq:mu_T_nrel}
\eeqn
and thus Eq.~\eqref{eq:matching} can also be satisfied approximately by all momenta that are close to the average momentum 
\beq
\frac{E_{\rm dec}-\mu_{\rm dec}}{T_\chi(t_{\rm dec})}\simeq
\frac{m_\chi+\frac{p_{\rm dec}^2}{2m_\chi}-\mu_{\rm dec}}{T_\chi(t_{\rm dec})}=\frac{m_\chi+\frac{p^2}{2m_\chi}-\mu^{\rm eff}}{T_\chi^{\rm eff}}\,.
\eeq
Therefore, for non-relativistic decoupling, the resulting distribution is also effectively thermal with $T_\chi^{\rm eff}\sim a^{-2}$.
If $m_\chi\gg\mu_{\rm dec}$, the quantum effect from Bose enhancement or Pauli blocking is negligible, and then the distribution is effectively Maxwell-Boltzmann as $f_\chi(p,t)\sim \exp(-p^2/(2m_\chi T_\chi^{\rm eff}))$.

Knowing that the temperature and chemical potential can only be defined effectively, we shall from now on drop the superscript ``eff''.

\subsection{Production mechanism: freeze-in}

If the interaction between dark matter and the SM particles is too feeble to establish thermal equilibrium, dark-matter particles may still be produced through the freeze-in mechanism (see Refs.~\cite{McDonald:2001vt,Hall:2009bx,Elahi:2014fsa,Bernal:2017kxu} and references therein).
In principle, it is possible for these particles to establish thermal equilibrium within a dark sector sourced by the SM sector \cite{Cheung:2010gj,Cheung:2010gk,Chu:2011be,Bernal:2015ova,Bernal:2017kxu,Krnjaic:2017tio,Berger:2018xyd,Evans:2019vxr,Du:2020avz,Hryczuk:2021qtz,Ghosh:2021wrk}.
However, in this type of scenarios, as dark matter equilibrates with the dark thermal bath, the distribution function of dark matter is simply thermal,
which is no different from what is studied in the previous subsection.
In other words, the equilibration process erases the ``memory'' before the dark freeze-out.
Given these considerations, we shall only take into account scenarios in which dark-matter particles, once produced, never equilibrates.

The freeze-in production can proceed through various channels, and the analytical approximation of the final distribution function in many different cases has been studied in literature \cite{Heeck:2017xbu,Boulebnane:2017fxw,Bae:2017dpt,Kamada:2019kpe,DEramo:2020gpr}.
Usually, such analytical expressions are written as a function of the comoving quantity $p/(Ma_{\rm dec}/a)$, where $M$ is the mass of the heaviest particle in the production channel and thus marks the temperature scale at which the dark-matter production is complete.
{At lower temperatures, the production of dark matter is suppressed due to an insufficiency in either the kinetic energy or the number density of the source particles depending on whether $M$ is one of the masses of the source particles or the products.}
Therefore, one can define a quantity $T_\chi(t)\equiv Ma_{\rm dec}/a(t)$ which plays the role of an effective ``temperature'' after dark matter is produced at $t_{\rm dec}$.
The decoupling temperature in the dark sector is then simply $T_\chi(t_{\rm dec})=M$.
For concreteness, we shall use the following analytical formula which is suitable for freeze-in through two-body decay or $2\to 2$ scattering:
\beq
f_\chi(p,t)\approx {C} \frac{e^{-p/T_\chi}}{\sqrt{p/T_\chi}}\,, \label{eq:f_fi}
\eeq
where ${C}$ is a constant that depends on the internal degrees of freedom, the decay width/cross section, and the masses of the particles involved in the production channel.
For simplicity, we shall treat it as a normalization factor to match the relic abundance of dark matter.

\subsection{Relation between mass and velocity} \label{sc:mass and velocity}

Since the phase-space distribution of dark matter after production takes different functional forms in different scenarios,
the impact on structure formation and its dependence on other physical parameters, especially the dark-matter mass, is expected to be different among these scenarios.
It is therefore of great interest to find the relation between the dark-matter mass and velocity in each case.

We first notice that
the distribution function after decoupling allows us to calculate the number density and average momentum of dark matter straightforwardly.
Assuming that $\mu_{\rm dec}$ is negligible for simplicity, we have
\beq
n_{\chi}=
\begin{cases}
A_n g_\chi T_\chi^3 & (\textbf{RFO})\\
\displaystyle g_\chi\left(\frac{m_\chi T_\chi}{2\pi}\right)^{3/2}e^{(\mu-m_\chi)/T_\chi} & (\textbf{NRFO})\\
\displaystyle\frac{3CT_\chi^{3}}{8\pi^{3/2}} & (\textbf{FI})
\end{cases},\label{eq:n_cases}
\eeq
and
\beq
\expt{p}=
\begin{cases}
A_p T_\chi & (\textbf{RFO})\\
\displaystyle\sqrt{\frac{8m_\chi T_\chi}{\pi}} & (\textbf{NRFO})\\
\displaystyle\frac{5T_\chi}{2}  & (\textbf{FI})
\end{cases}\label{eq:p_cases}\,,
\eeq
where we use ``RFO'', ``NRFO'' and ``FI'' for relativistic freeze-out, non-relativistic freeze-out and freeze-in, respectively,
the values of $A_n$ and $A_p$ in each case are listed in Table~\ref{tab:A}, and we have approximated both the Bose-Einstein and the Fermi-Dirac distributions by the Boltzmann distribution in cases of non-relativistic decoupling.
Notice that we have kept $\mu$ in the second line of Eq.~\eqref{eq:n_cases} since the chemical potential increases and asymptotes $m_\chi$ for non-relativistically decoupled dark matter.

\begin{table}[t]
\centering
\begin{tabular}{|c||c|c|c|}
\hline
     & FD & BE & MB \\
\hline
$A_n$ & $3\zeta(3)/(4\pi^2)$ & $\zeta(3)/\pi^2$ & $1/\pi^2$\\     
\hline
$A_p$ & $7\pi^4/(180\zeta(3))$ & $\pi^4/(30\zeta(3))$ & 3\\
\hline
\end{tabular}
\caption{Values of $A_n$ and $A_p$ in each case, where FD, BE and MB stand for Fermi-Dirac, Bose-Einstein and Maxwell-Boltzmann distributions, respectively.}
\label{tab:A}
\end{table}

The above equations allow us to estimate the present-day dark matter velocity\footnote{Again, we emphasize that we only consider the redshift effect, and thus this is not the velocity of dark matter in a virialized halo.} 
$\expt{v}_0\approx \expt{p}_0/m_\chi$ from two different perspectives.
On the one hand, since the energy density of dark matter today $\rho_\chi(t_0)\approx n(t_0) m_\chi$, 
one can use Eq.~\eqref{eq:n_cases} to relate the effective temperature of dark matter today to the well-measured relic density of dark matter: 
\beq
T_\chi(t_0)=
\begin{cases}
\displaystyle\left[\frac{\rho_\chi(t_0)}{A_n g_\chi m_\chi}\right]^{1/3} & (\textbf{RFO})\\
\displaystyle\frac{2\pi}{m_\chi^{5/3}}\left[\frac{\rho_\chi(t_0)e^{x_{\rm dec}}}{g_\chi}\right]^{2/3} & (\textbf{NRFO})\\
\displaystyle2\sqrt{\pi}\left(\frac{\rho_\chi(t_0)}{3C g_\chi m_\chi}\right)^{\frac 1 3} & (\textbf{FI})
\end{cases},
\label{eq:Tchi0}
\eeq
where $x_{\rm dec}\equiv (m_\chi-\mu_{\rm dec})/T_\chi(t_{\rm dec})=(m_\chi-\mu_0)/T_\chi(t_0)$.
Then, using Eq.~\eqref{eq:p_cases}, the present-day average velocity can be expressed as:
\beqn
\expt{v}_0 &\approx& \left\{
\begin{array}{l}
\displaystyle\frac{A_p}{m_\chi}\left(\frac{\Omega_\chi \rho_{\rm crit}(t_0)}{A_n g_\chi m_\chi}\right)^{\frac 1 3} \\
\displaystyle \frac{4}{m_\chi}\left(\frac{\Omega_\chi\rho_{\rm crit}(t_0)e^{x_{\rm dec}}}{g_\chi m_\chi}\right)^{\frac{1}{3}} \\
\displaystyle\frac{5\sqrt{\pi}}{m_\chi}\left(\frac{\Omega_\chi \rho_{\rm crit}(t_0)}{3C g_\chi m_\chi}\right)^{\frac 1 3}
\end{array}\right.\nn\\
&\approx& \left\{
\begin{array}{ll}
\displaystyle 1.1\times 10^{-7}\times\left(\frac{2}{g_\chi}\right)^{\frac 1 3}\left(\frac{\Omega_\chi}{0.25}\right)^{\frac 1 3}\left(\frac{1~\rm keV}{m_\chi}\right)^{\frac 4 3} & \hspace{0.5cm}(\textbf{RFO})\\
\displaystyle  1.9\times 10^{-6}\times 
\left(\frac{2}{g_\chi}\right)^{\frac{1}{3}}
\left(\frac{\Omega_\chi}{0.25}\right)^{\frac{1}{3}}\left(\frac{1~\rm keV}{m_\chi}\right)^{\frac{4}{3}} \left(\frac{e^{x_{\rm dec}}}{e^{10}}\right)^{\frac{1}{3}} &
\hspace{0.5cm}(\textbf{NRFO})\\
\displaystyle 1.0\times 10^{-7}\times
\left(\frac{2}{g_\chi}\right)^{\frac 1 3}
\left( \frac{\Omega_\chi}{0.25} \right)^{\frac 1 3}\left( \frac{1~{\rm keV}}{m_\chi}\right)^{\frac 4 3}\left(\frac{1}{C}\right)^{\frac{1}{3}} &
\hspace{0.5cm}(\textbf{FI})
\end{array}\right.,
\label{eq:v0_rel_nrel_1}
\eeqn
where $\Omega_\chi\equiv \rho_\chi(t_0)/\rho_{\rm crit}(t_0)$ is the dark-matter abundance today with $\rho_{\rm crit}$ being the critical energy density.

On the other hand,
Eq.~\eqref{eq:p_cases} alone enables us to relate the average velocity of dark matter today to relevant physical quantities at the decoupling time.
Using the scaling relation between $T_\chi$ and $a$ in each case
and the conservation of entropy in the SM sector, it is easy to find that 
\beqn
\expt{v}_0 &\approx& 
\left\{
\begin{array}{l}
\displaystyle A_p\left(\frac{g_{*,s}(T_0)}{g_{*,s}(T_{\rm dec})}\right)^{\frac 1 3}\frac{T_0}{m_\chi}\eta_{\rm dec} \\
\displaystyle \sqrt{\frac{8}{\pi}}\left(\frac{g_{*,s}(T_{\rm now})}{g_{*,s}(T_{\rm dec})}\right)^{\frac 1 3}\sqrt{\frac{T_\chi(t_{\rm dec})}{m_\chi}}\frac{T_0}{T_{\rm dec}}\\
\displaystyle\frac{5}{2}
\frac{T_0}{m_\chi}\left(\frac{g_{\star,s}(T_0)}{g_{\star,s}(T_{\rm dec})}\right)^{\frac 1 3}\eta_{\rm dec}
\end{array}
\right.\nn\\
&\approx& 
\left\{
\begin{array}{ll}
\displaystyle  6.5\times 10^{-7}\times\left(\frac{5}{g_{*,s}(T_{\rm dec})}\right)^{\frac 1 3}\frac{1 \rm keV}{m_\chi}~\eta_{\rm dec} & \hspace{0.5cm}(\textbf{RFO})\\
\displaystyle  1.1\times 10^{-6}\times
\left(\frac{5}{g_{*,s}(T_{\rm dec})}\right)^{\frac 1 3}
\left(\frac{x_{\rm dec}}{10}\right)^{\frac 1 2}
\frac{1~{\rm keV}}{m_\chi}~\eta_{\rm dec} & \hspace{0.5cm}(\textbf{NRFO})\\
\displaystyle 5.4\times 10^{-7}\times 
\left(\frac{5}{g_{\star s}(T_{\rm dec})}\right)^{\frac 1 3}\frac{1~\rm keV}{m_\chi}\eta_{\rm dec} & \hspace{0.5cm}(\textbf{FI})
\end{array}
\right.,\label{eq:v0_rel_nrel_2}
\eeqn
where $g_{\star,s}$ is the relativistic degrees of freedom for entropy density in the SM thermal bath, and
we have defined
\beq
\eta(t)\equiv \frac{T_\chi(t)}{T(t)}\,~~\text{and}~~\eta_{\rm dec}\equiv \frac{T_\chi(t_{\rm dec})}{T_{\rm dec}}\,.\label{eq:eta_dec}
\eeq
Note that, the use of entropy conservation implies our assumption that the SM sector is not transferring energy with any other sector after dark matter decouples.

The relations in Eqs.~\eqref{eq:v0_rel_nrel_1} and \eqref{eq:v0_rel_nrel_2} both suggest that, given one functional form of the phase-space distribution, a constraint on dark-matter velocities can be recast as a constraint on the dark-matter mass.
For typical values of the parameters, the average velocity of dark-matter particles can be larger by roughly one order of magnitude in the case of non-relativistic decoupling than in the case of freeze-in or relativistic freeze-out.
This means, for a constraint on the average velocity $\expt{v}_0$, the corresponding constraints on the dark-matter mass $m_\chi$ is typically stronger in the case of non-relativistic decoupling.

\begin{figure}
\centering
\includegraphics[width=0.6\textwidth]{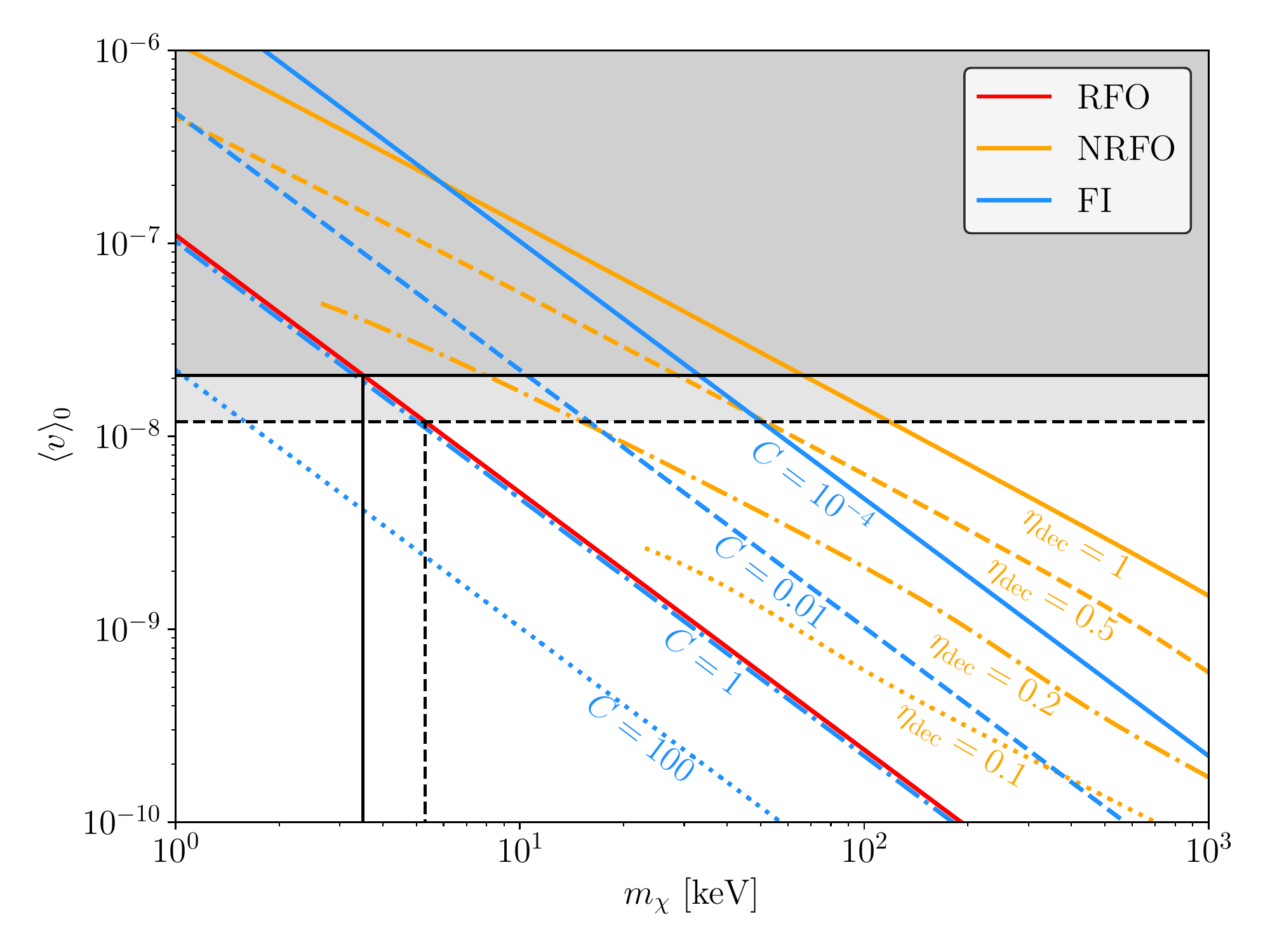}
\caption{Relation between the dark-matter mass and present-day average velocity for different decoupling scenario, where we have chosen $\Omega_\chi=0.25$. 
The black horizontal lines indicate the critical velocities associated with the two WDM constraints $\mwdm\geq 3.5~\rm keV$ and $\mwdm\geq 5.3~\rm keV$, and thus the part of the curve in the shaded region can be considered as excluded. Notice that, in the case of non-relativistic freeze-out, the curves cannot extend to arbitrarily small mass without violating the constraint on the dark-matter relic abundance.}
\label{fg:mx_v0}
\end{figure}

To allow a better comparison, we plot the relation between $\expt{v}_0$ and $m_\chi$ with the condition from the relic abundance imposed in FIG.~\ref{fg:mx_v0}.
To be specific, 
we directly use the relations in Eq.~\eqref{eq:v0_rel_nrel_1} for freeze-in and relativistic freeze-out since the relation can be uniquely determined with at most one additional parameter $C$.
On the other hand, we use both Eq.~\eqref{eq:v0_rel_nrel_1} and Eq.~\eqref{eq:v0_rel_nrel_2} for the non-relativistic freeze-out scenario
in order to eliminate $x_{\rm dec}$.
The current Lyman-$\alpha$ constraints on WDM mass, $\mwdm\geq 3.5~\rm keV$ and $\mwdm\geq 5.3~\rm keV$ \cite{Irsic:2017ixq}, can then be converted to preliminary constraints on the present-day average velocity via the relation in the relativistic freeze-out scenario,
which gives $\expt{v}_0\lesssim 2.1\times 10^{-8}$ or $\expt{v}_0\lesssim 1.2\times 10^{-8}$, as indicated by the black horizontal lines.\footnote{In fact, these constraints are fundamentally constraints on the WDM velocity distribution, but are converted to constraints on the WDM mass. 
In addition, such constraints are preliminary since the average velocity of dark matter is not a direct observable.} 
We can then apply these constraints on the average velocity to the cases of non-relativistic freeze-out and freeze-in.

We first notice that the constraints on $m_\chi$ in the freeze-in scenario is very similar to that in the WDM scenario (which is exactly the relativistic freeze-out scenario) if $C=1$.
However, these lower bounds can drastically increase or decrease if $C$ becomes smaller or bigger which is consistent with Eq.~\eqref{eq:v0_rel_nrel_1}. 
In the non-relativistic freeze-out scenario,
the constraints on $m_\chi$ are overall much stronger than that in the WDM scenario,
where the bounds can be $m_\chi\gtrsim 5~\rm keV$ or $m_\chi\gtrsim 120~\rm keV$ depending on different choices of $\eta_{\rm dec}$. 

It is worth pointing out that the orange curves in the case of non-relativistic freeze-out do not extend arbitrarily to the left.
This is because
while the mass-velocity relations in Eq.~\eqref{eq:v0_rel_nrel_1} are constrained by the abundance $\Omega_\chi$,
the relations in Eq.~\eqref{eq:v0_rel_nrel_2} are derived directly from the relation between momentum and velocity and thus do not necessarily respect the constraint on the dark-matter relic abundance.
Therefore, parameters in Eq.~\eqref{eq:v0_rel_nrel_2}, $\{m_\chi,~T_{\rm dec},~\eta_{\rm dec},~x_{\rm dec}\}$, in principle cannot vary independently from each other without violating the observational bound on the dark-matter relic abundance.
Indeed, while holding $\eta_{\rm dec}$ fixed and vary $m_\chi$ in Eq.~\eqref{eq:v0_rel_nrel_2}, one would need to adjust $x_{\rm dec}$ (and thus $T_{\rm dec}$) in order to satisfy the constraint in Eq.~\eqref{eq:v0_rel_nrel_1} at the same time.
This is reasonable since $x_{\rm dec}$ is essentially determined by comparing the interaction rate and the Hubble expansion rate when actually solving the Boltzmann equation, and thus depends on $m_\chi$ through the cross-section and number density.
The dependence on $m_\chi$ in $x_{\rm dec}$ can also be seen by noticing that the slopes of the orange curves differ distinctively from that of the red and blue curves, despite that all relations in Eq.~\eqref{eq:v0_rel_nrel_1} contains a factor of $m_\chi^{-4/3}$.

It is also worth mentioning that, for the same decoupling temperatures, one in general does not expect $\expt{v}_0$ from freeze-in to be larger than that from non-relativistic freeze-out.
This can be seen from Eq.~\eqref{eq:v0_rel_nrel_2}.
To make the present-day average velocity from freeze-out smaller than that from freeze-in, one need $x_{\rm dec}\lesssim 3$ --- a value close to the regime of relativistic freeze-out which makes our estimate unreliable.
Therefore, the only possibility for any of the blue curves to surpass the $\eta_{\rm dec}=1$ orange curve is to have $\eta_{\rm dec}>1$ in the freeze-in case.
While the information about $\eta_{\rm dec}$ is encoded in the parameter $C$, and having $\eta_{\rm dec}>1$ does not explicitly violate any assumption in the calculation that we have presented so far, the existence of a dark thermal bath hotter than the SM thermal bath would be severely constrained by current observations.
We shall therefore neglect any part of the blue curve above the solid orange curve.

Other than constraining the dark-matter mass, 
the constraint on the average velocity also have non-trivial implication on physical quantities at the decoupling time which are extremely important for understanding the thermal history of dark matter.
Equating Eq.~\eqref{eq:v0_rel_nrel_1} with Eq.~\eqref{eq:v0_rel_nrel_2},
we find the relation between the dark-matter mass and the temperatures at decoupling
\beq
m_\chi \approx\left\{
\begin{array}{ll}
1.9\times 10^{-3} {~\rm keV}\times \displaystyle\frac{g_{\star,s}(T_{\rm dec})}{g_\chi} 
\frac{\Omega_\chi}{0.25}
~\eta_{\rm dec}^{-3} & \hspace{1cm}(\textbf{RFO})\\
2.1{~\rm keV}\times 
\displaystyle\frac{g_{\star,s}(T_{\rm dec})}{g_\chi}  
\frac{\Omega_\chi}{0.25}~
\eta_{\rm dec}^{-3}
\left(\frac{10}{x_{\rm dec}}\right)^{\frac 3 2} \frac{e^{x_{\rm dec}}}{e^{10}} & \hspace{1cm}(\textbf{NRFO})\\
\displaystyle 2.7\times 10^{-3} {~\rm keV}\times \displaystyle\frac{g_{\star,s}(T_{\rm dec})}{g_\chi}
\frac{\Omega_\chi}{0.25}~C^{-1}
\eta_{\rm dec}^{-3} & \hspace{1cm}(\textbf{FI})
\end{array}\right..
\label{eq:mx_Tdec_eta}
\eeq

\begin{figure}
\centering
\includegraphics[width=0.5\textwidth]{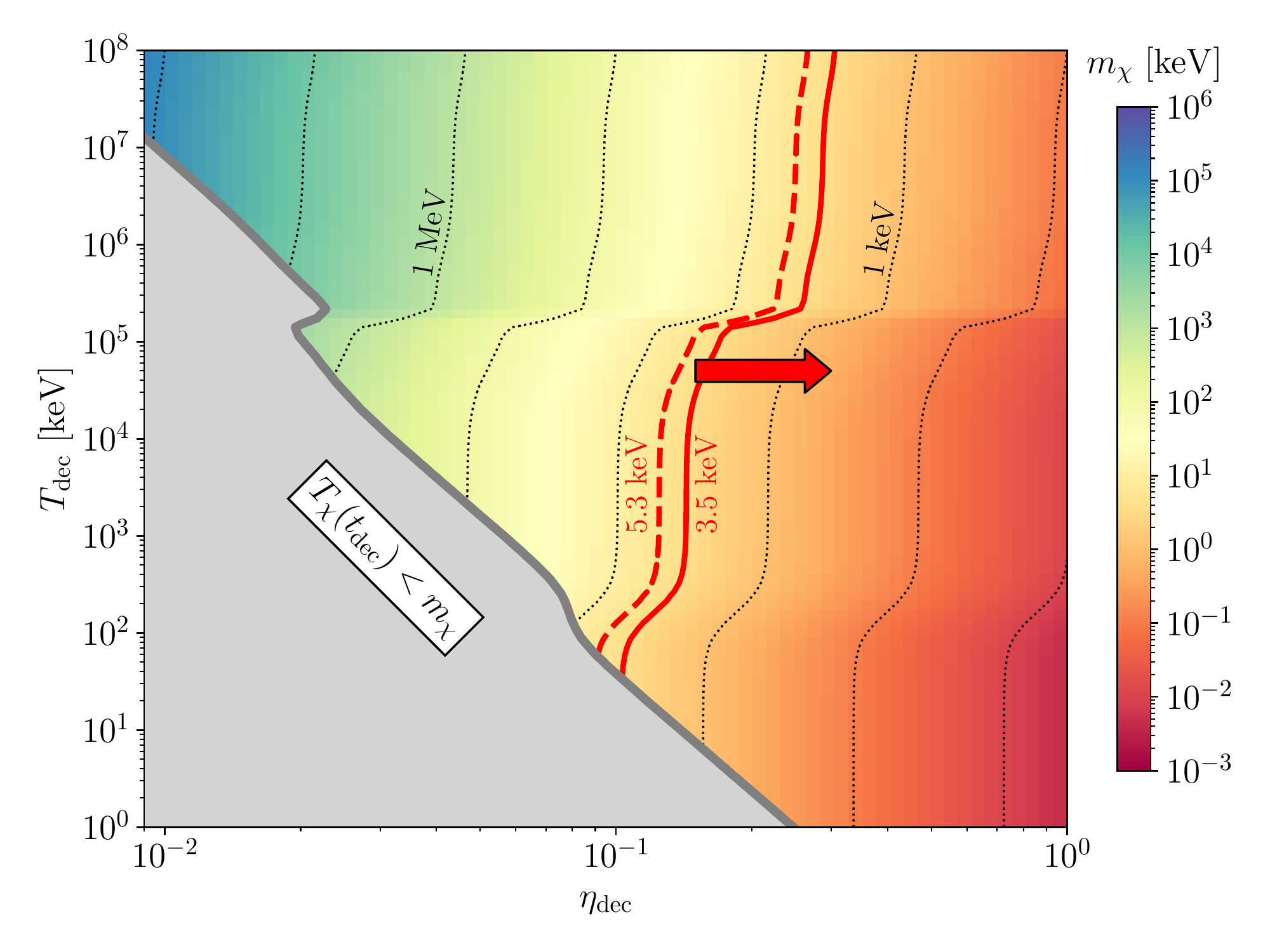}\includegraphics[width=0.5\textwidth]{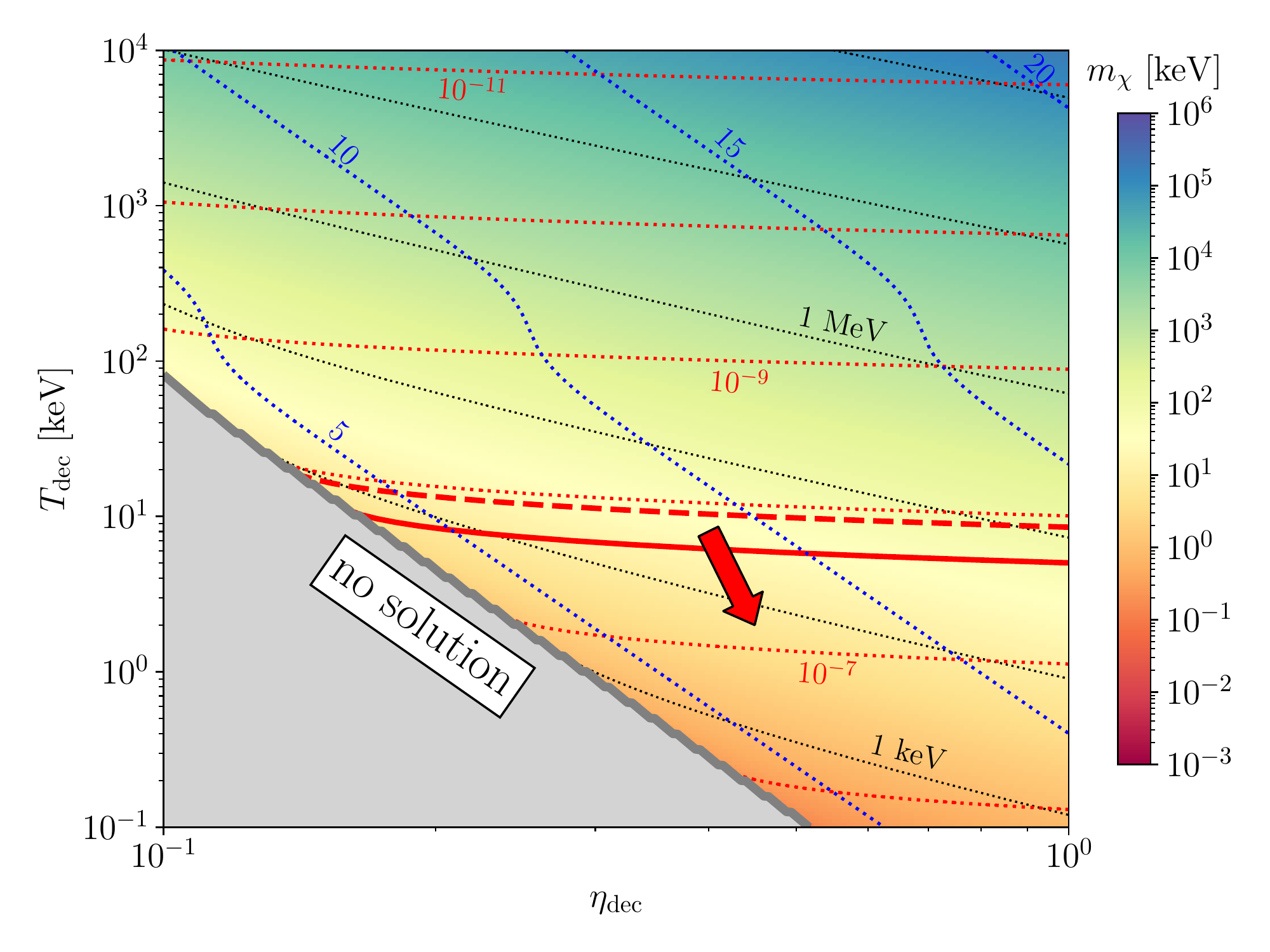}
\includegraphics[width=0.5\textwidth]{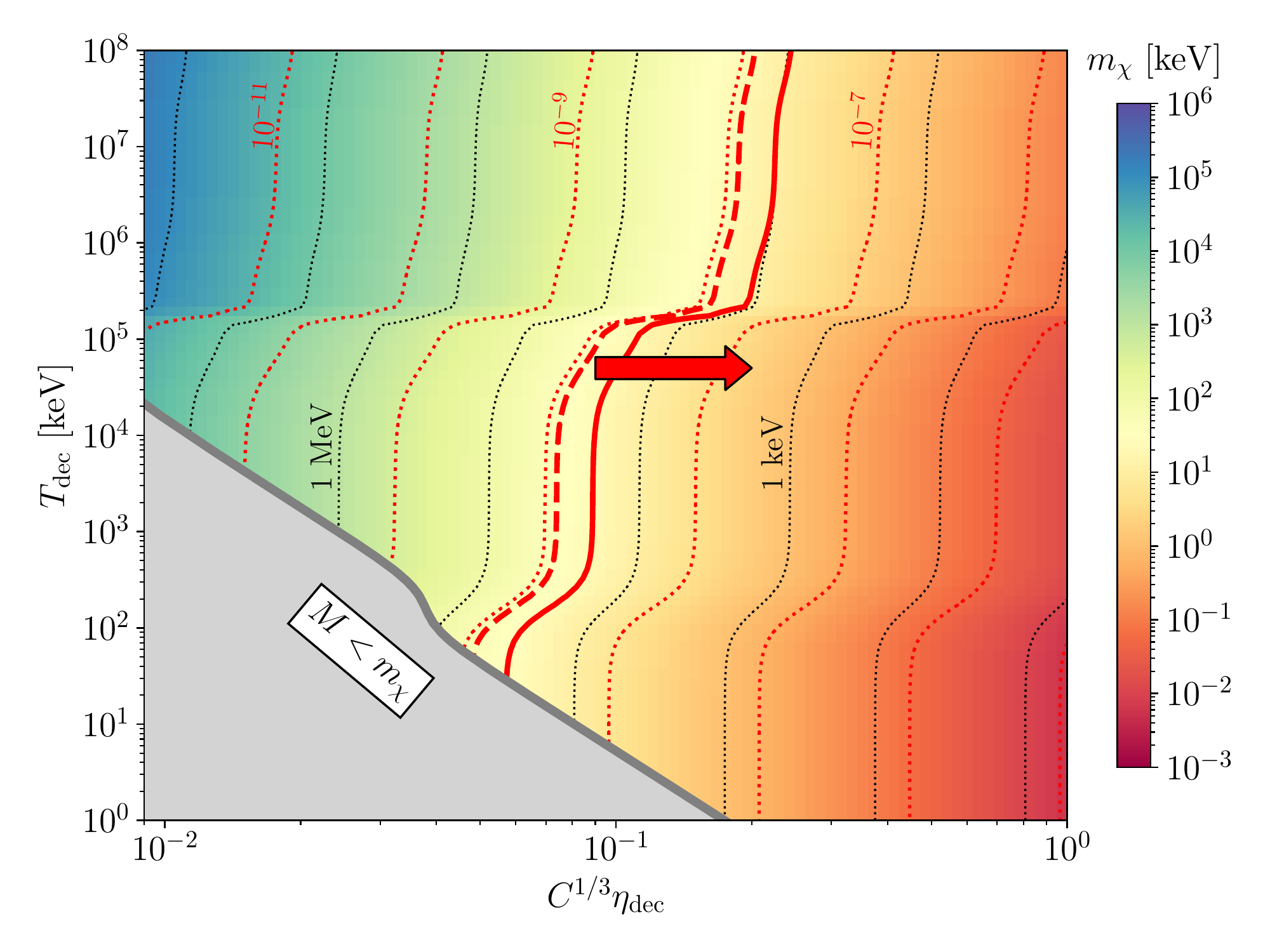}
\caption{Constraints on decoupling temperatures in the cases of relativistic freeze-out ({\bf top left}), non-relativistic freeze-out ({\bf top right}), and freeze-in ({\bf bottom}). 
The color map and the black dotted contours indicate the masses of dark matter.
They are also contours of the present-day average velocity $\expt{v}_0$ in the top left panel.
On the other hand, $\expt{v}_0$ is shown by the red dotted contours in the top right and bottom panels, where we have assumed that $\eta_{\rm dec}=1$ in the freeze-in case.
The red dashed and solid contours thus represent the current constraints on $\expt{v}_0$, with the arrow pointing to the constrained region of the parameter space.
In addition, the blue contours in the top right panel shows the value of $x_{\rm dec}$.
As described by the text, the gray areas indicate that, when using Eq.~\eqref{eq:mx_Tdec_eta}, either the solution of $m_\chi$ is inconsistent with the assumption of the calculation (top left and bottom panels), 
or no solution can be found in this region (right panel).
}
\label{fg:T_dec_eta}
\end{figure}

The above equations suggest that, for a given dark-matter mass $m_\chi$, one cannot uniquely determine the temperatures $T_{\rm dec}$ and $T_{\chi}(t_{\rm dec})$ at the moment of decoupling.
Nevertheless, 
if a bound from observations can be place on the dark-matter mass,
regardless of how this bound is obtained,
such information can then be used to constrain the parameter space spanned by $T_{\rm dec}$ and $T_{\chi}(t_{\rm dec})$ in a given production scenario which reveals information about the thermal history of dark matter.
This can be seen in FIG.~\ref{fg:T_dec_eta},
where we show such constraints in each case in different panels.

For the relativistic freeze-out (or WDM) scenario in the upper left panel, 
the contours of $m_\chi$ are also contours of the present-day velocity of dark matter $\expt{v}_0$.
Therefore, the current constraint on $m_\chi$ from structure formation, as indicated by the dashed and solid red contours, follows the same trend as the dotted black contours.
One immediate conclusion that can be drawn from such constraint is that WDM cannot decouple directly from the SM thermal bath.
Instead, dark matter can only decouple relativistically from a dark thermal bath with a lower temperature.
Depending on $T_{\rm dec}$, the temperature of the SM sector at the decoupling time, $T_\chi(t_{\rm dec})$ has to be smaller by a factor of $\sim 0.3$ or $0.1$ at least.

In the case of non-relativistic freeze-out, 
we follow the procedure in FIG.~\ref{fg:mx_v0} to turn the constraint on $m_\chi$ in the case of relativistic freeze-out into the constraint on the present-day average velocity, \ie~$\expt{v}_0\lesssim 2.1\times 10^{-8}$ or $\expt{v}_0\lesssim 1.2\times 10^{-8}$, which is again indicated by the dashed and solid red curves.
Similar to examples shown in FIG.~\ref{fg:mx_v0}, we also observe that the constraints on dark-matter mass are now depending on the decoupling temperatures.
The constraint for relativistic freeze-out, $\mwdm\geq 3.5~\rm keV$ (or $\mwdm\geq 5.3~\rm keV$), now covers a broad range from roughly $5$ to $70~\rm keV$ (or from $8$ to $120~\rm keV$) in the case of non-relativistic freeze-out.
However, the exact constraints on mass cannot be unambiguously determined until the decoupling temperatures are specified.
Moreover,
since $\eta_{\rm dec}$ can approach unity above the solid or dashed red curves, the current WDM bound from structure formation cannot completely rule out the possibility that dark matter might decouple from the SM thermal bath if such decoupling is non-relativistic.
In addition, we also notice that the second line of Eq.~\eqref{eq:mx_Tdec_eta} does not always have a solution on the entire $\eta_{\rm dec}$-$T_{\rm dec}$ plane.
This simply means that, for certain combination of decoupling temperatures, it is not possible to obtain the correct relic abundance without introducing additional assumptions.
The corresponding region of parameter space is therefore shaded in gray.

For the freeze-in scenario, we first notice that there is an additional degree of freedom from the normalization factor $C$.
Indeed, comparing the freeze-in case in Eq.~\eqref{eq:v0_rel_nrel_1} with that in Eq.~\eqref{eq:v0_rel_nrel_2}, we see that a change in $C$ in the former equation amounts to a change in the product $\eta_{\rm dec}/g_{\star,s}^{1/3}(T_{\rm dec})$ in the latter equation.
Physically, a variation in $C$ (while holding the masses of the particles in the corresponding process fixed) would result in a change in the number density of dark matter at the decoupling temperature $T_\chi(t_{\rm dec})=M$,
and thus one needs to adjust the mapping between $T_\chi(t_{\rm dec})$ and $T_{\rm dec}$ in order to obtain the correct relic abundance.
Therefore, instead of plotting against $\eta_{\rm dec}$, we choose $C^{1/3}\eta_{\rm dec}$ as the horizontal axis such that we can unambiguously determine contours of $m_\chi$ on the plane.
For the average velocity, the red contours correspond to the choice of fixing $\eta_{\rm dec}=1$, which can be seen as having dark matter produced directly from the SM thermal bath.
In this situation, the contours of $\expt{v}_0$ differs slightly from that of $m_\chi$ due to the dependence on $T_{\rm dec}$ in Eq.~\eqref{eq:v0_rel_nrel_2},
and the current bound on average velocity covers a range from $10$ to $30~\rm keV$ (or from $17$ to $50~\rm keV$).
Equivalently, one can also fix $C=1$, and then the contours of $m_\chi$ can be directly turned into contours of $\expt{v}_0$ using Eq.~\eqref{eq:v0_rel_nrel_1}.
In this situation, the present constraint is very similar to the case of relativistic freeze-out, which is $m_\chi\gtrsim 3.3~\rm keV$ (or $m_\chi\gtrsim 5.0~\rm keV$). 

\subsubsection{\texorpdfstring{Comments on $\Delta N_{\rm eff}$}{}}
In the previous discussion, we have only re-interpreted and applied the WDM constraint from studies of the large scale structure.
When exploring the implications of such constraint on the production of dark matter, other factors might also play a role and can be even more restrictive. 

The major constraint relevant for the production of dark matter comes from the big bang nucleosynthesis (BBN) which occurs when the temperature of the SM sector drops below $\sim 2~\rm MeV$.
First, dark matter (and other dark-sector species, if there is any) provides energy density in addition to the SM thermal bath which can affect the expansion rate of the universe during BBN.
Second, if there is any energy transfer between the SM thermal bath and dark matter (or the dark sector) after neutrino decoupling, the temperature ratio between photons and neutrinos would be changed.
Both effects are going to modify the prediction for the abundance of light elements.
Moreover, if there is any dark species that stay relativistic after the matter radiation equality, the anisotropy power spectrum of the cosmic microwave background (CMB) will also be affected.
Nevertheless, a systematic study on these effects is model-dependent and beyond the scope of this work.
Here, we shall only briefly comment on the possible effects on $N_{\rm eff}$, the effective number of relativistic neutrino species.

In the case of relativistic freeze-out, we have seen from FIG.~\ref{fg:T_dec_eta} that dark matter can only decouple from a dark thermal bath with a smaller temperature.
However, this implies the existence of extra relativistic species which can potentially be constrained by $N_{\rm eff}$, the effective number of relativistic neutrino species.
We can estimate this effect by noticing that the ratio between the energy densities of the dark sector and the SM neutrinos after neutrino decoupling is $\sim \left(11/4\right)^{4/3}\eta^4$, up to corrections from the number of relativistic freedoms in the two sectors and the quantum statistics.
Suppose there is no appreciable change in $\eta$ from the beginning of BBN to the CMB time, the allowed range of $\eta_{\rm dec}$ in FIG.~\ref{fg:T_dec_eta} ($\eta_{\rm dec}\lesssim 0.3$ or $0.1$ depending on $T_{\rm dec}$) indicates that the energy density of the dark-sector radiation is at most $\sim 3\%$ that of the SM neutrinos,
which can contribute to $N_{\rm eff}$ at most comparable to the uncertainty of current measurements, and thus can easily evade the current bound by slightly decreasing $\eta_{\rm dec}$.
Such a rough estimate can also be applied to the case of non-relativistic freeze-out or freeze-in.

For the case in which dark matter decouples directly from the SM thermal bath, which is not excluded by structure formation in the case of non-relativistic freeze-out and freeze-in, $N_{\rm eff}$ will be modified only if dark matter decouples after neutrinos decouple.
In the non-relativistic freeze-out scenario, if dark matter is always in thermal equilibrium with the SM thermal bath during BBN but before decoupling, for $m_\chi$ below a few $\rm MeV$, dark matter can contribute appreciably to the total energy density of the universe when it is relativistic or close to being relativistic.
As temperature drops, this energy density will be transferred to the SM sector.
This will increase $N_{\rm eff}$ if dark matter is primarily converted into neutrinos, or decrease $N_{\rm eff}$ if it is preferentially converted into photons.
In this situation, we would expect the $\eta_{\rm dec}=1$ edge in the top right panel in FIG.~\ref{fg:T_dec_eta} to be severely constrained by $N_{\rm eff}$ below $\mathcal{O}(\rm MeV)$.
Nevertheless, if, during the BBN epoch, dark matter is decoupled at first, but then equilibrate with the SM while being non-relativistic before its eventual freeze-out, the impact on $N_{\rm eff}$ can then be significantly relaxed.
This type of scenario has been explored in \cite{Berlin:2018ztp},
and its impact on BBN has been discussed in \cite{Berlin:2019pbq}.
We therefore do not consider the $\eta_{\rm dec}=1$ region of sub-$\rm MeV$ dark matter in the non-relativistic freeze-out scenario as ruled out.

In the freeze-in scenario, the constraint from BBN is much easier to evade as the ratio $\rho_\chi/\rho_{\rm R}\sim n_\chi \expt{E}/(g_{\star}(T)T^4)$ do not increase as we go back in time, if we neglect the possibility of equilibration within the dark sector which might exponentially suppress $n_\chi$ as dark matter becomes non-relativistic.
Therefore, during the entire BBN epoch, the fraction of the total energy density of the universe occupied by dark matter is maximum by the end of BBN, \ie~when $T\sim \mathcal{O}(10)~\rm keV$.
Suppose dark matter decouples before the end of BBN (which is consistent with the allowed region in FIG.~\ref{fg:T_dec_eta}),
a rough estimate shows that the ratio $nm_\chi/\rho_{R}\sim a/a_{\rm eq}\sim T_{\rm eq}/T$, is at most $\mathcal{O}(10^{-4})$ during the BBN epoch,
where $T_{\rm eq}\sim \mathcal{O}(1)~\rm eV$ is the temperature at the matter-radiation equality.
Thus, unless the kinetic energy of dark matter is orders of magnitude larger than its mass energy by the end of BBN, which might again run into trouble with constraints from structure formation,
we do not expect any observable effect on BBN in the freeze-in scenario.

\section{Constraints on phase-space distribution}\label{sec:constraints}

The analysis in the previous section provides a useful estimate in constraining the mass and decoupling-temperatures of dark matter in several different scenarios.
However, such constraints are preliminary as they are obtained by simply converting the current Lyman-$\alpha$ constraints on WDM mass into constraints on the average velocity of dark matter,
and the latter provides very limited information 
other than a characteristic length scale below which the formation of structure is suppressed.
On the other hand, the actual astronomical observables such as the Lyman-$\alpha$ forest often concern the structure of the universe over a wide range of length scales.
Therefore,
to more rigorously constrain the scenarios considered in this work, 
it is important to go beyond the estimate from a single average velocity and study structure formation in more detail.

For this purpose,
in what follows, we shall first demonstrate how to obtain 
the phase space distribution of dark matter by solving the Boltzmann equation at the level of the phase-space distribution $f_\chi(p,t)$.
The phase-space distribution is then used as an input to compute the density perturbations and the matter power spectrum via the public code \texttt{CLASS} \cite{Lesgourgues:2011re,Lesgourgues:2011rh}.
Eventually, we shall use the matter power spectrum to derive more rigorous constraints in our scenarios.

\subsection{Phase-space distributions from freeze-in and freeze-out}

As we have shown in Sec.~\ref{sec:PSD_PROD}, the general form of the Boltzmann equation for the phase-space distribution $f_\chi(p,t)$ reads
\beq
\left(\frac{\partial }{\partial t} - Hp \frac{\partial }{\partial p } \right)f_\chi(p,t) =\mathcal{C}[f]\,,\label{eq:Boltzmann_fx}
\eeq
where $\mathcal{C}[f]$ is the collision term, and for a general process $a+b+\dots\leftrightarrow i+j+\dots$ it takes the form of
\beqn
\mathcal{C}_a[f]&=&-\frac{1}{2E_{a}}\int d\pi_b\dots d\pi_i d\pi_j\dots (2\pi)^4 \delta^{(4)}(p_a+p_b+\dots-p_i-p_j-\dots)\nn\\
&~&\times\bigg[\abs{\mathcal{M}_{a+b\dots\rightarrow i+j+\dots}}^2 f_a f_b\dots(1\pm f_i)(1\pm f_j)\dots\nn\\
&~&~~~~ - \abs{\mathcal{M}_{i+j+\dots\rightarrow a+b+\dots}}^2f_i f_j\dots (1\pm f_a)(1\pm f_b)\dots \bigg]\,,\label{eq:collision_general}
\eeqn
where {$d\pi_i=\frac{g_i}{(2\pi)^3}\frac{d^3p_{i}}{2E_{i}}$}, ``$\pm$'' describes the Bose-enhancement/Pauli-blocking effects from the quantum statistics, and the two amplitudes correspond to the forward and inverse processes respectively. 
In principle, the collision term in Eq.~\eqref{eq:Boltzmann_fx} contains all the relevant processes such as the annihilations, scatterings or decays that dark-matter particles participate.

To proceed further, we consider a concrete example in which the dark-matter particle $\chi$ is predominantly produced through a $2\to2$ annihilation $\psi +\psi \leftrightarrow \chi +\bar\chi$, 
where we assume that $\chi$ and $\bar\chi$ are distinct particles and set $g_\chi =2$.
For the particle $\psi$, we assume that it is either a massless species in the SM sector or a massless species that thermalizes in the dark sector.
In the latter case, we assume that the total energy density of the dark sector is negligible compared with that of the SM thermal bath,
so that the Hubble parameter is dominated by the contribution from SM-sector energy density. 
And this can be easily satisfied as long as $\eta$ is not too close to unity.
Note that, by choosing $\psi$ to be massless, dark matter will decouple \emph{non-relativistically} in either the freeze-in scenario or the freeze-out scenario since the decoupling condition is $T_\chi\lesssim m_\chi$ for the former, and $n_\chi$ being Boltzmann suppressed, \ie~$e^{-m_\chi/T_\chi}\ll 1$, for the latter.
Nevertheless, we shall simply call the latter scenario freeze-out for brevity when there is no confusion.
For simplicity, we shall also neglect elastic scattering processes like $\psi +\chi \longleftrightarrow \psi +\chi$ in this work.
The effects of elastic scatterings on the phase-space distribution have been discussed in Ref.~\cite{Du:2021jcj}. 
Therefore, in our example, the collision term for the $2\to2$ annihilation process can be simplified as 
\beqn
 C_{\rm ann}(T_\chi,p_\chi)&=&\frac{{g_{\bar\chi}g_\psi^2}}{512\pi^3 E_\chi p_\chi}\int_{\max \{4m_\chi^2, 4m_\psi^2\} }^{\infty} ds \int_{E_{\bar\chi}^{\rm min}(s)}^{E_{\bar\chi}^{\rm max}(s)} \frac{dE_{\bar\chi}}{\sqrt{s}p_{\chi}^*(s)}\nn\\
&~&
\times \int^{t^{\rm max}(s)}_{t^{\rm min}(s)} dt~\Big[f^{\rm eq}_\chi(p_\chi) f^{\rm eq}_\chi(p_{\bar\chi})-f_\chi(p_\chi) f_{\chi}(p_{\bar\chi})\Big]\overline{\abs{\mathcal{M}}^2} \, , \label{eq:col_ann_balance}
\eeqn
where $s$ and $t$ are the Mandelstam variables, $p_\chi^*(s)$ is the momentum of $\chi$ in the center-of-mass frame, $\overline{\abs{\mathcal{M}}^2}$ is the squared amplitude for the annihilation $\chi\bar\chi\to\psi\psi$,
the superscript ``eq'' denotes the value of $f_\chi$ in thermal equilibrium, and we have approximated $f_\chi^{\rm eq}$ by the Maxwell-Boltzmann distribution, \ie~$f_\chi^{\rm eq}(p_\chi)=e^{-E_\chi/T_\chi}$, and ignored the quantum-statistics factors.
The explicit form of $E_{\bar\chi}^{\rm max/min}(s)$ and $t^{\rm max/min}(s)$ can be found in Ref.~\cite{Du:2021jcj} and are not repeated here for brevity.

Note that, the above equation is general in the sense that the freeze-out and freeze-in scenarios are unified under one single framework --- it allows one to switch from the freeze-out scenario to the freeze-in scenarios by simply tuning the size of the squared amplitude $\overline{\abs{\mathcal{M}_{\chi\bar\chi\to\psi\psi }}^2}$.
In particular, the collision term always tends to bring $f_\chi$ to its equilibrium value $f_\chi^{\rm eq}$ as can be seen from the part in the squared brackets in Eq.~\eqref{eq:col_ann_balance}.
In the freeze-out scenario, the squared amplitude is large enough such that the annihilation rate of $\chi$ can be sufficiently rapid to set $f_\chi=f_\chi^{\rm eq}$ until $\chi$ decouples.
On the contrary, in the freeze-in scenario, the squared amplitude is too small such that $f_\chi\ll f_\chi^{\rm eq}$ 
and the annihilation of $\chi$ has negligible effect on the final dark-matter density.
As is pointed out in Ref.~\cite{Du:2021jcj}, the transition between the two regimes can be conveniently quantified by introducing a quantity 
\beq
\gamma_\chi \equiv\frac{n_\chi^{\rm eq}\expt{\sigma_{\chi\chi\to\psi\psi} v}_{T_\chi}}{H(T)} \Bigg|_{T_\chi = m_\chi}\,
\eeq
to specify the interaction rate of the annihilation process, where $\expt{\sigma_{\chi\chi\to\psi\psi} v}$ is the thermally averaged cross (see Appendix~\ref{sec:formula} for definition).
Thus, the freeze-out scenario correspond to $\gamma_\chi \gg 1$, whereas the freeze-in scenario corresponds to $\gamma_\chi \ll 1$.
In addition, there exists a transition regime when $\gamma_\chi \sim 1$,
where the inverse process $\chi\bar\chi\to\psi\psi$  has significant impact on the production of $\chi$, even though the chemical equilibrium between $\chi$ and $\psi$ is never established.

\begin{figure}
\centering
\includegraphics[width=0.6\textwidth]{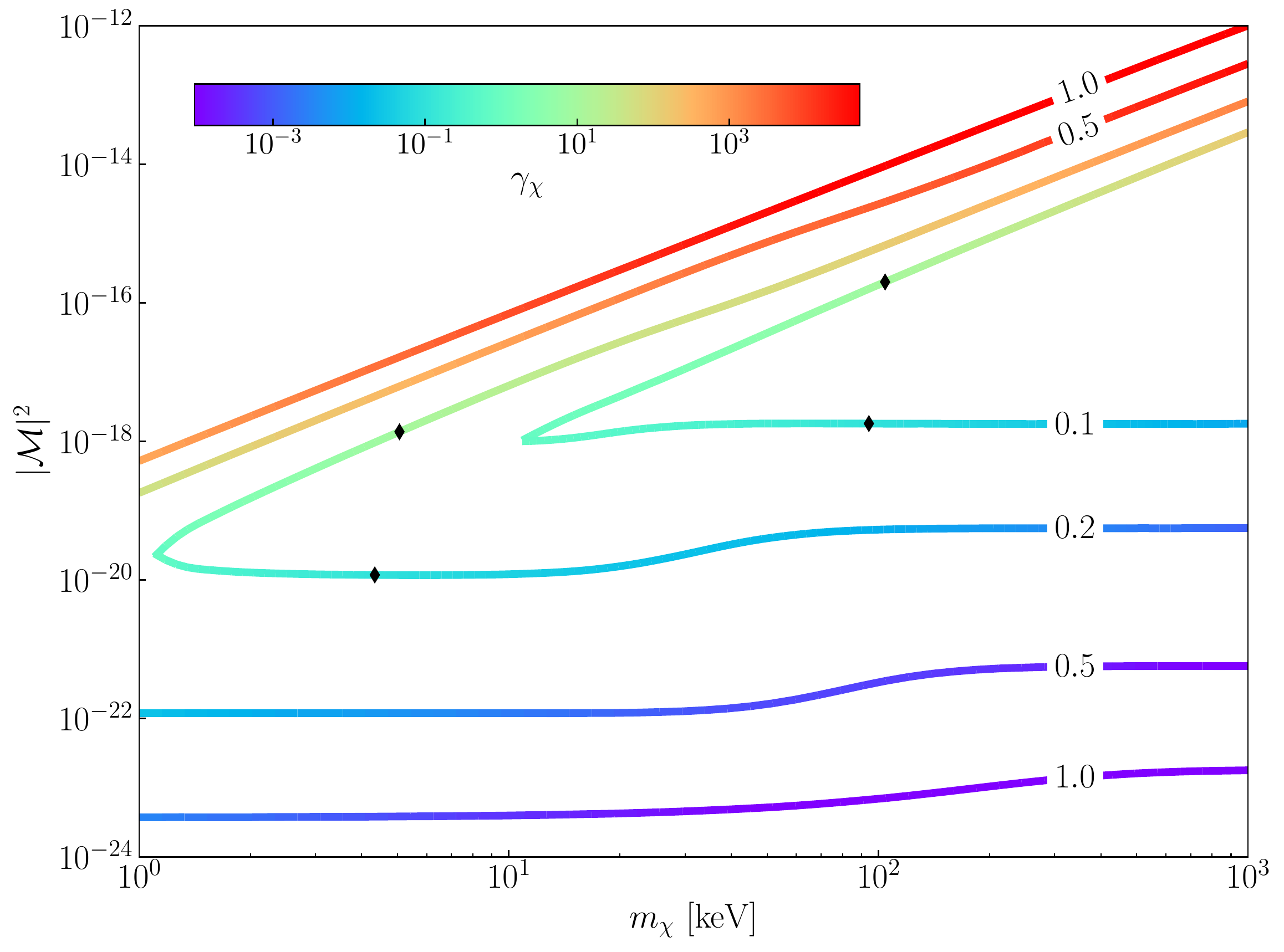}
\caption{
Contours of $\Omega\chi=0.25$ in the $m_\chi$-$\abs{\mathcal{M}}^2$ plane for different values of $\eta_{\rm dec} $. 
For given a $m_\chi$, the (constant) squared amplitude $\abs{\mathcal{M}}^2$ is determined by solving the Boltzmann equation to obtain $\Omega\chi=0.25$. 
As indicated by the color bar, the interaction rate $\gamma_\chi$ varies from $\gamma_\chi \ll 1$ to $\gamma_\chi \gg 1$ along each contour. 
Thus, each contour corresponds to a path that goes through the freeze-in, transition, and freeze-out regimes.
The diamond markers are placed at $\gamma_\chi=0.1$ and $10$ to roughly identify these three regimes.}\label{fg:contour2to2}
\end{figure}

In Fig.~\ref{fg:contour2to2}, we present contours of fixed relic abundance, $\Omega_\chi =0.25$, in the $m_\chi$-$\abs{\mathcal{M}}^2$ plane for cases with $\eta_{\rm dec} =0.1 ,\, 0.2,\, 0.5 $ and $1$. 
In all the cases, we assume that $\abs{\mathcal{M}}^2$ is a constant for simplicity, and its value is adjusted to obtain $\Omega_\chi =0.25$ after solving the Boltzmann equation.
As indicated by the color, the interaction rate characterized by $\gamma_\chi$ increases from $\gamma_\chi \ll 1$ to $\gamma_\chi \sim 1$, and eventually to $\gamma_\chi \gg 1$ along each contour as one goes from the lower right part to the upper right part in a clockwise direction.
For each contour, there is a lower limit $m_\chi^{\rm min}$ on the dark-matter mass below which the correct relic abundance cannot be obtained, and this smallest mass increases when $\eta_{\rm dec}$ decreases.
Above this lower limit, there are always two values of the squared amplitude that are able to generate the correct relic abundance at a fixed mass, which give rise to two possible thermal histories.

Motivated by the results in Ref.~\cite{Du:2021jcj}, we can roughly identify the freeze-in, the transition and the freeze-out regimes with $\gamma_\chi < 0.1$, $0.1 \leq\gamma_\chi \leq 10$, and $\gamma_\chi > 10$.
It is easy to notice that, for each fixed $\eta_{\rm dec}$, the transition regime only takes a small portion in the entire mass range, whereas most of the available parameter space corresponds to the freeze-in and the freeze-out regimes as there is no upper limit for the $m_\chi$ other than the potential unitarity bound \cite{Griest:1989wd}.
Based on this observation, we shall therefore focus on the freeze-in and the freeze-out regimes in what follows.
We shall assume $\eta_{\rm dec} = 1$ in these two regimes,
and show in Sec.~\ref{sec:improve} how to extend our results in the parameter space with $\eta_{\rm dec} < 1$.

\begin{figure}\centering
\includegraphics[width=0.49\textwidth]{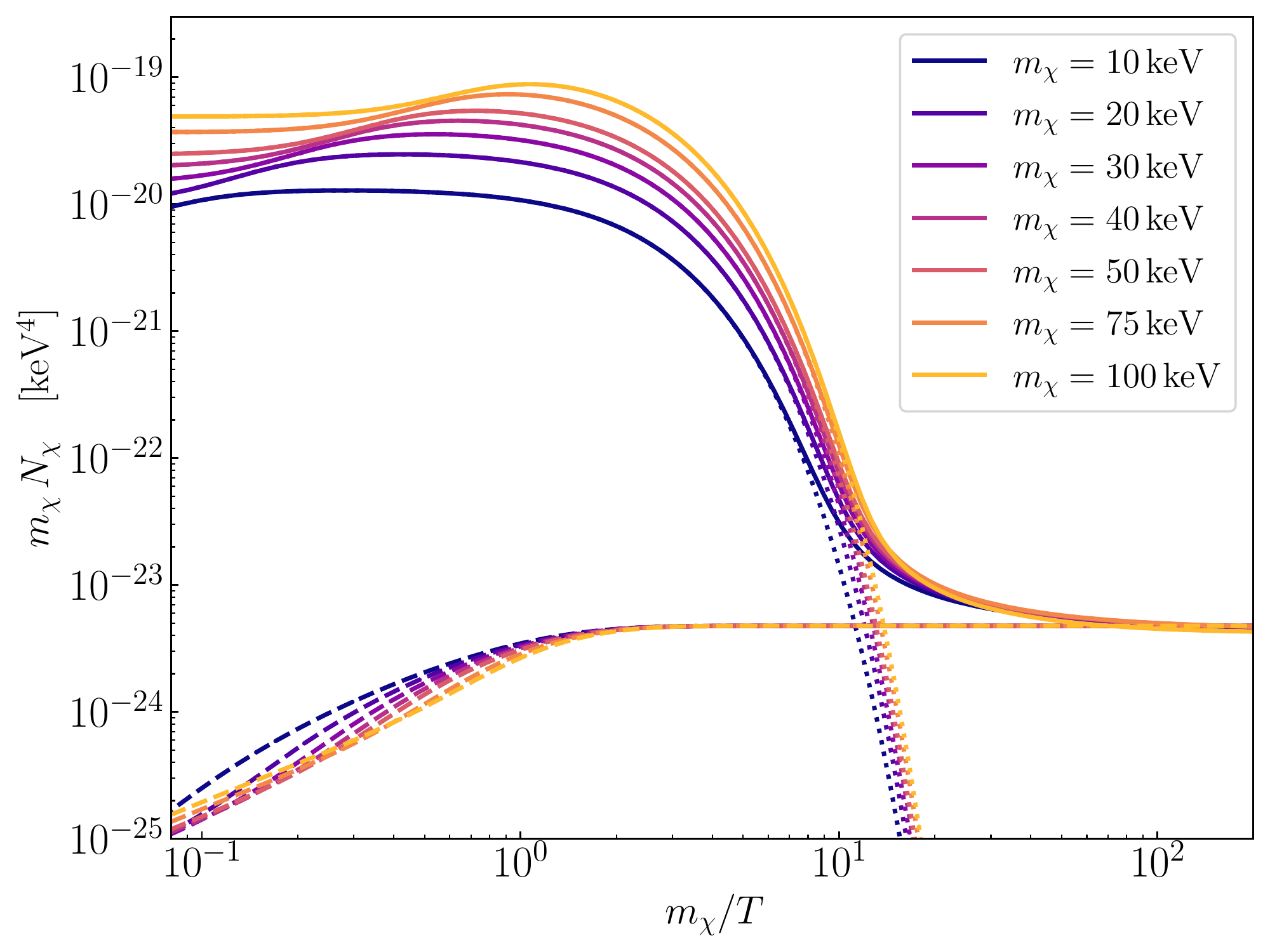} \includegraphics[width=0.49\textwidth]{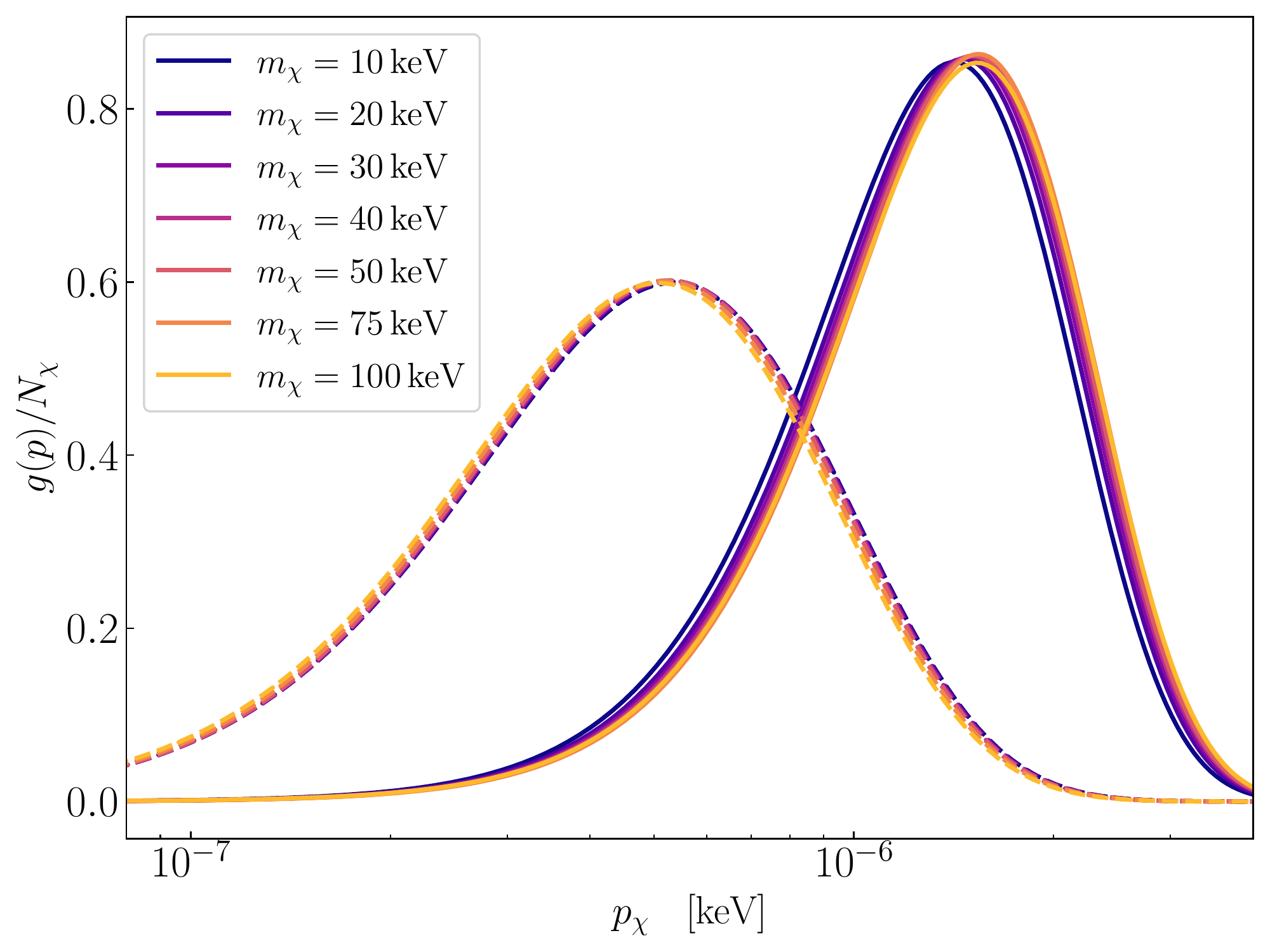} 
\caption{
\textbf{Left panel}: The evolution of the comoving mass density in the freeze-out (solid) and the freeze-in (dashed) scenarios, and when dark matter is in thermal equilibrium with the SM sector (dotted). 
Note that the mass $m_\chi$ is different for each curve, and so is the corresponding temperature along the horizontal axis. 
\textbf{Right panel}:
Normalized momentum distributions of dark matter today associated with the examples shown on the left.
The freeze-out and the freeze-in scenarios are represented by the solid and the dashed curves, respectively.
}
\label{fg:Nx_fx}
\end{figure}

In the left panel of Fig.~\ref{fg:Nx_fx}, we show the 
evolution of dark-matter comoving mass density $N_\chi m_\chi$, where $N\equiv n_\chi a^3$ is the comoving number density.
We have normalized the present-day value of the scale factor $a(t_{0})=1$ such that $N_\chi(t_{0})m_\chi\simeq \rho_\chi(t_{0})$.
The examples shown here are selected from the data points on the $\eta_{\rm dec}=1$ contour in Fig.~\ref{fg:contour2to2}.
The solid and dashed bunches of curves correspond to the freeze-out and the freeze-in scenarios, respectively,
whereas the dotted bunch of curves shows the value of $N_\chi^{\rm eq}m_\chi$ --- the comoving mass density if $\chi$ is in thermal equilibrium.
Notice that, the dotted curves, and thus the solid curves which overlap with the dotted curve before freezing out, are lowered as $m_\chi$ decreases.
On the other hand, the dashed curves appear to have the opposite behavior before freezing in.
The freeze-out and the freeze-in bunches become closer when $m_\chi$ is dialed smaller.
Nevertheless, the distinct separation of them suggests that for each $m_\chi$ within this mass range there are always two different thermal histories that give rise to the same relic-abundance of dark matter, even though the dark-matter particles are created via the same $2\to 2$ process.

In the right panel of Fig.~\ref{fg:Nx_fx}, we plot the resulting phase-space distribution
$g_\chi(p)\equiv p^3 f_\chi(p,t_0)$ (normalized by the comoving number density $N_\chi$)
which is the appropriate distribution with respect to the momentum on a logarithmic scale \cite{Dienes:2020bmn,Dienes:2021itb,Dienes:2021cxp}.
We find that, despite having the same relic abundance, the two production mechanisms result in two distinct sets of momentum distributions, among which the momentum distributions from the freeze-out production always possess an overall higher momentum and a smaller width.
Another noticeable feature is that the momentum distributions from the same production mechanism tend to be extremely similar which is consistent with our discussion in Sec.~\ref{sec:PSD_PROD}.
The variation in the dark-matter mass $m_\chi$ only causes a very slight horizontal shift.
In the freeze-in scenario, this slight shift to the left as $m_\chi$ increases is only caused by the change of $g_\star(T)$ after dark matter is produced since the effective temperature of dark matter today $T_\chi(t_0)=Ma_{\rm dec}/a_0=T_0[g_{\star,s}(T_0)/g_{\star,s}(M)]^{1/3}$.
In the freeze-out scenario, the same effect from $g_\star(T)$ also exists and is even stronger since the effective temperature $T_\chi\sim a^{-2}$.
However, as is shown in the left panel, the requirement on producing the correct relic abundance forces dark matter to decouple ``later'' (at a smaller $m_\chi/T$) when $m_\chi$ increases.
Since this additional effect allows dark matter to thermalize with the radiation bath for a longer time, it tends to work in the opposite direction by shifting the distribution to higher momentum.
It turns out that the second effect is always stronger than the first one within the mass range that we are showing since $g_\star(T)$ does not vary appreciably below $\mathcal{O}(100)~\rm keV$. 
Thus, the distribution for freeze-out always shifts to the right when $m_\chi$ increases.
Analytically, such behavior can be understood from the second line of Eq.~\eqref{eq:v0_rel_nrel_2} where the competition between the dependence on $m_\chi$ and the dependence on $x_{\rm dec}$ is explicit.

Although both the thermal history and the final phase-space distribution in the freeze-in and freeze-out regimes 
are extremely different,
in principle, one can always dial down the mass to enter the transition regime and find a continuous and smooth interpolation between the two types of distribution functions.
This is similar to the results shown
in Ref.~\cite{Du:2021jcj}, where, for fixed $m_\chi$, the production of $\chi$ goes through the three regimes continuously by simply increasing $\gamma_\chi$, and the resulting phase-space distributions after dark-matter production also interpolate smoothly between the two limits set by the freeze-in and freeze-out mechanism.
However, for the $\eta_{\rm dec}=1$ case here, the transition regime correspond to $m_\chi \lesssim \mathcal{O}(1)~\rm eV$, which is well excluded by current constraints.
We thus do not plot the results in the transition regime for $\eta_{\rm dec}=1$.

\begin{figure}\centering
\includegraphics[width=0.49\textwidth]{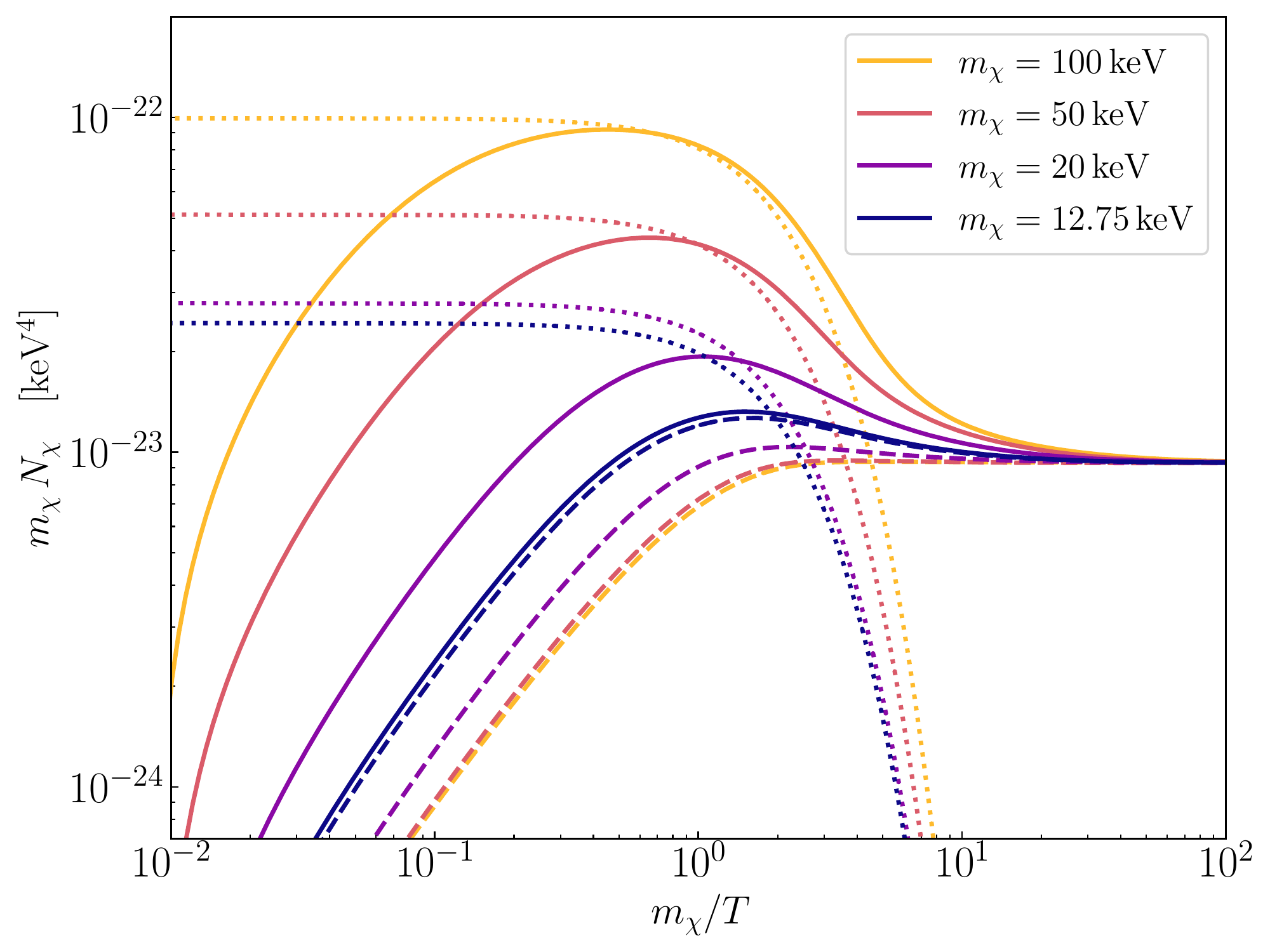}
\includegraphics[width=0.49\textwidth]{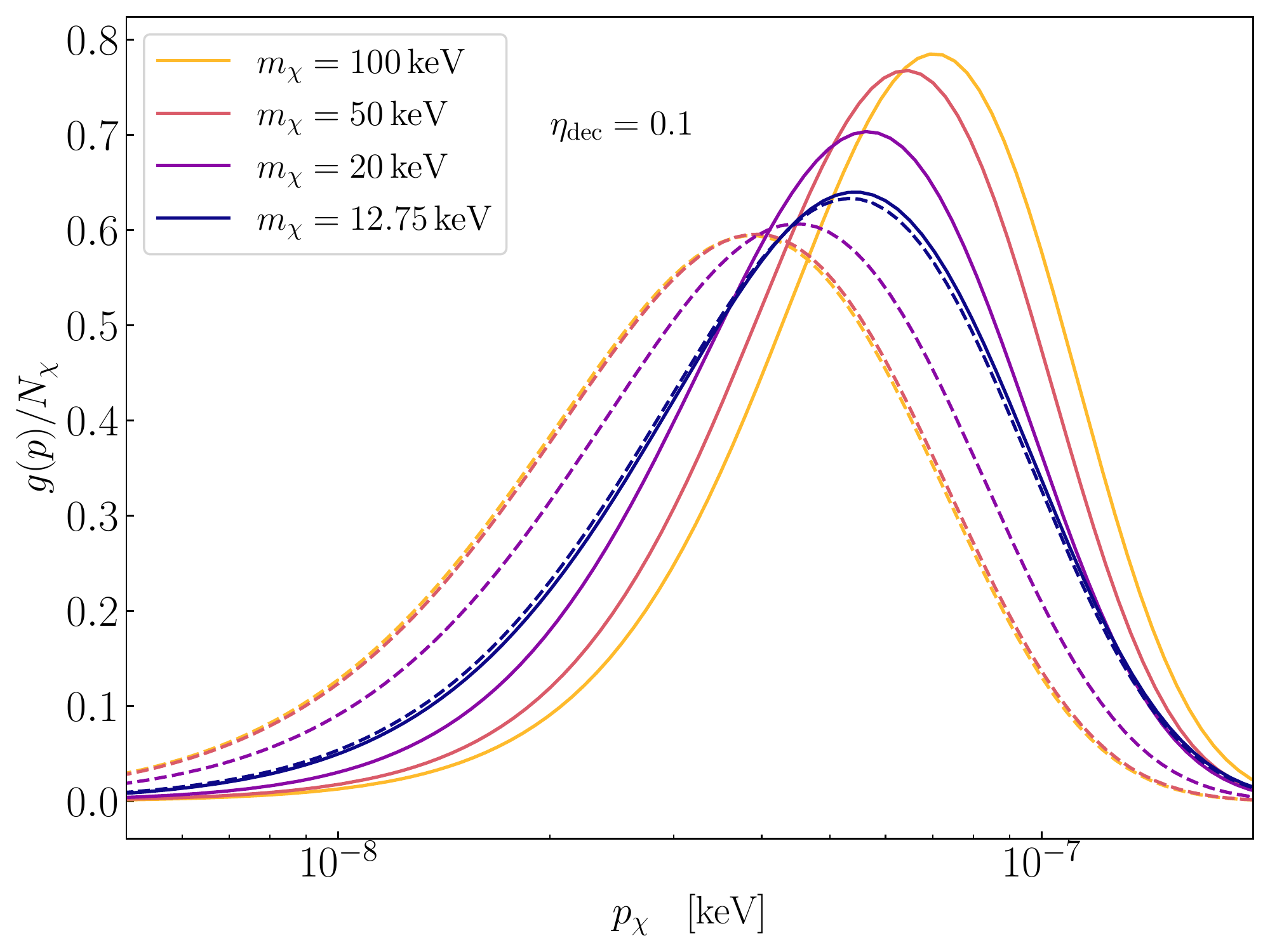}
\caption{
\textbf{Left panel}:
The evolution of the comoving mass density in the transition regime where, at the same dark-matter mass, the solution with a larger/smaller amplitude is represented by the solid/dashed curve.
The evolution of $m_\chi N_\chi$ when $\chi$ is kept in equilibrium with the thermal bath is shown by the dotted curves.
\textbf{Right panel}:
The resulting phase-space distributions of dark matter associated with the examples in the left panel.
All the distributions are normalized
by the their comoving number density $N_\chi$.}\label{fg:Nx_fx_transition}
\end{figure}

Instead, for completeness, we present in Fig.~\ref{fg:Nx_fx_transition} the results in the transition regime for $\eta_{\rm dec}=0.1$ which is more likely to be allowed.
Just like the previous figure, for each $m_\chi>m_\chi^{\rm min}$, there are always two solutions with different $\gamma_\chi$ that give rise to the same relic abundance --- one with a larger $\gamma_\chi$ and thus is closer to establishing thermal equilibrium, and the other with a smaller $\gamma_\chi$ and is thus further away from equilibrium, even though the annihilation of dark-matter particles are not negligible in both solutions.
Such behavior is also reflected in the final phase-space distributions shown in the right panel, where the solutions from having a larger $\gamma_\chi$ always results in a momentum distribution with overall higher momenta but smaller width.
As $m_\chi$ decreases towards $m_\chi^{\rm min}$, the two solutions approach each other, which is consistent with our observation in Fig.~\ref{fg:contour2to2}.
Eventually, when $m_\chi\rightarrow m_\chi^{\rm min}\simeq 12.75~\rm keV$, the two solutions would merge. 
This is evident in both panels as the curves from the two solutions for $m_\chi= 12.75~\rm keV$ almost overlap with each other.

The reason that there is no solution for $m_\chi<m_\chi^{\rm min}$ can also be glimpsed from the left panel of Fig.~\ref{fg:Nx_fx_transition}.
As we have mentioned before, the particle interactions always tend to bring the $\chi$ particles into thermal equilibrium.
Thus, we see in all cases that the $m_\chi N_\chi$ curve goes up and evolves towards $m_\chi N_\chi^{\rm eq}$ before reaching its maximum.
However, since the maximum is bounded by the value of $m_\chi N_\chi^{\rm eq}$ in the relativistic regime, this also means that it is impossible to obtain the correct relic abundance if $m_\chi N_\chi^{\rm eq}<\rho_\chi(t_0)$.
In fact, as can be inferred by the behavior when $m_\chi\rightarrow m_\chi^{\rm min}$, 
further lowering $m_\chi$ would bring down $m_\chi N_\chi^{\rm eq}$ and render it impossible to generate enough dark-matter particles 
even if the plateau of $m_\chi N_\chi^{\rm eq}$ is still higher than $\rho_\chi(t_0)$.

\subsection{Phase space distribution and structure formation}

\subsubsection{Density perturbations}
Beyond the relic abundance of dark matter, the thermal history of dark matter also manifests itself in the process of structure formation via the phase-space distribution. 
To better understand how this works, let us briefly review how the phase-space distribution affects structure formation.
Following Ref.~\cite{Ma:1995ey}, the perturbed distribution function can be written as
\begin{equation}\label{eq:f_1st}
f_\chi \left(\vec x, \vec p, \tau\right)=f_{\chi,0}(q)\left[1+\Psi\left(\vec x, q, \hat n, \tau\right)\right]\,,
\end{equation}
where $\tau\equiv \int dt/a(t)$ is the conformal time, $q \equiv a p$ is the comoving momentum, $\hat{n} \equiv \vec p / p$ is the direction of the momentum, and $f_{\chi,0}(q)$ is the unperturbed zeroth-order distribution, \ie the distribution obtained in last section. 
In the Synchronous gauge, the line element is given by
\beq
ds^2 = a^2(\tau) \{ - d\tau^2 + (\delta_{ij} + h_{ij} )d x^i dx^j \}\,,
\eeq
where $\tau$ is the conformal time, and the perturbation of the metric $h_{ij} $ can be decomposed as $h_{ij} = h \delta_{ij}+ h_{ij}^{\parallel}+h_{ij}^{\perp} +h_{ij}^{T} $, where $h \equiv h_{ii}$ is the trace part of $h_{ij}$ (not to be confused with the little $h$ in the Hubble constant), and $h_{ij}^{\parallel}$, $h_{ij}^{\perp}$ and $ h_{ij}^{T}$ are the traceless parts, satisfying
\beq
\epsilon_{ijk} \partial_j \partial_l h_{lk}^{\parallel} = 0\,, \quad \partial_i \partial_j h_{ij}^{\perp} = 0\,,\quad \partial_i h_{ij}^{T} = 0\,.
\eeq
By definition, $h_{ij}^{\parallel}$ and $h_{ij}^{\perp}$ can be written in terms of a scalar filed $\mu$ and a divergenceless vector $\vec A$, respectively:
\beq
\begin{aligned}
    h_{ij}^{\parallel} &= \left( \partial_i \partial_j - \frac{1}{3} \delta_{ij} \nabla^2 \right) \mu\,, \\
    h_{ij}^{\perp} &=\partial_i A_j + \partial_j A_i\,, \quad \partial_i A_i = 0\,.
\end{aligned}
\eeq
Therefore, the scalar mode of the metric perturbations are characterised by the two scalar fields $h $ and $\mu$, while the vector and the tensor modes are characterised by $\vec A$ and $h_{ij}^T $, respectively.

In this work we will only consider the scalar modes of $h_{ij}$, which can be Fourier transformed as
\beq
 h_{ij} (\vec{x},\tau) = \int d^3 k~e^{i \Vec{k} \cdot \Vec{x}} \left \{ \frac{{k}_i {k}_j}{k^2} h(\vec{k},\tau) + 6 \eta (\vec{k},\tau)\left(\frac{{k}_i {k}_j}{k^2}-\frac{1}{3} \delta_{ij}\right)  \right \}\,,
\eeq
where $h(\vec{k},\tau)$ and $\eta(\vec{k},\tau)$ are the trace part and the traceless part of the scalar mode in the Fourier space which are related to the scalar fields $h$ and $\mu$ in the real space. 
Since we have assumed that $\chi$ is collisionless after production, the evolution of $\Psi$ obeys the first-order collsionless Boltzmann equation in the Fourier-space
\begin{equation}\label{eq:Boltzmann_1st}
\frac{\partial \Psi}{\partial \tau}+i \frac{q}{\epsilon}(\vec{k} \cdot \hat{n}) \Psi+\frac{d \ln f_{\chi,0}}{d \ln q}\left[\dot{\eta}-\frac{\dot{h}+6 \dot{\eta}}{2}(\hat{k} \cdot \hat{n})^2\right]=0\,,
\end{equation}
where $\epsilon \equiv a E = \sqrt{q^2 + a^2 m^2}$ is the comoving energy.
Since Eq.~\eqref{eq:Boltzmann_1st} is independent of the azimuth angle, we can expand $\Psi$ in a Legendre series
\begin{equation}\label{eq:Psi_l}
\Psi(\vec{k}, \hat{n}, q, \tau)=\sum_{l=0}^{\infty}(-i)^l(2 l+1) \Psi_l(k, q, \tau) P_l(\hat{k} \cdot \hat{n})\,.
\end{equation}
Inserting Eq.~\eqref{eq:Psi_l} into Eq.~\eqref{eq:Boltzmann_1st} and using the orthonormality of the Legendre polynomials, we can obtain the Boltzmann hierarchies for $\Psi_l$:
\begin{equation}
\begin{aligned}
\dot{\Psi}_0 & =-\frac{q k}{\epsilon} \Psi_1+\frac{\dot{h}}{6} \frac{{d} \ln f_{\chi,0}}{{d} \ln q}\,, \\
\dot{\Psi}_1 & =\frac{q k}{3 \epsilon}\left(\Psi_0-2 \Psi_2\right)\,, \\
\dot{\Psi}_2 & =\frac{q k}{5 \epsilon}\left(2 \Psi_1-3 \Psi_3\right)-\left(\frac{\dot{h}}{15}+\frac{2 \dot{\eta}}{5}\right) \frac{{d} \ln f_{\chi,0}}{{d} \ln q}\,, \\
\dot{\Psi}_{l \geq 3} & =\frac{q k}{(2 l+1) \epsilon}\left[l \Psi_{l-1}-(l+1) \Psi_{l+1}\right]\,.
\end{aligned}
\end{equation}
Notice that, even though we assume that dark-matter particles are collisionless,
they are still coupled with other species such as photons and baryons through metric perturbations encoded in $h$ and $\eta$.

\begin{figure}\centering
\includegraphics[width=0.32\textwidth]{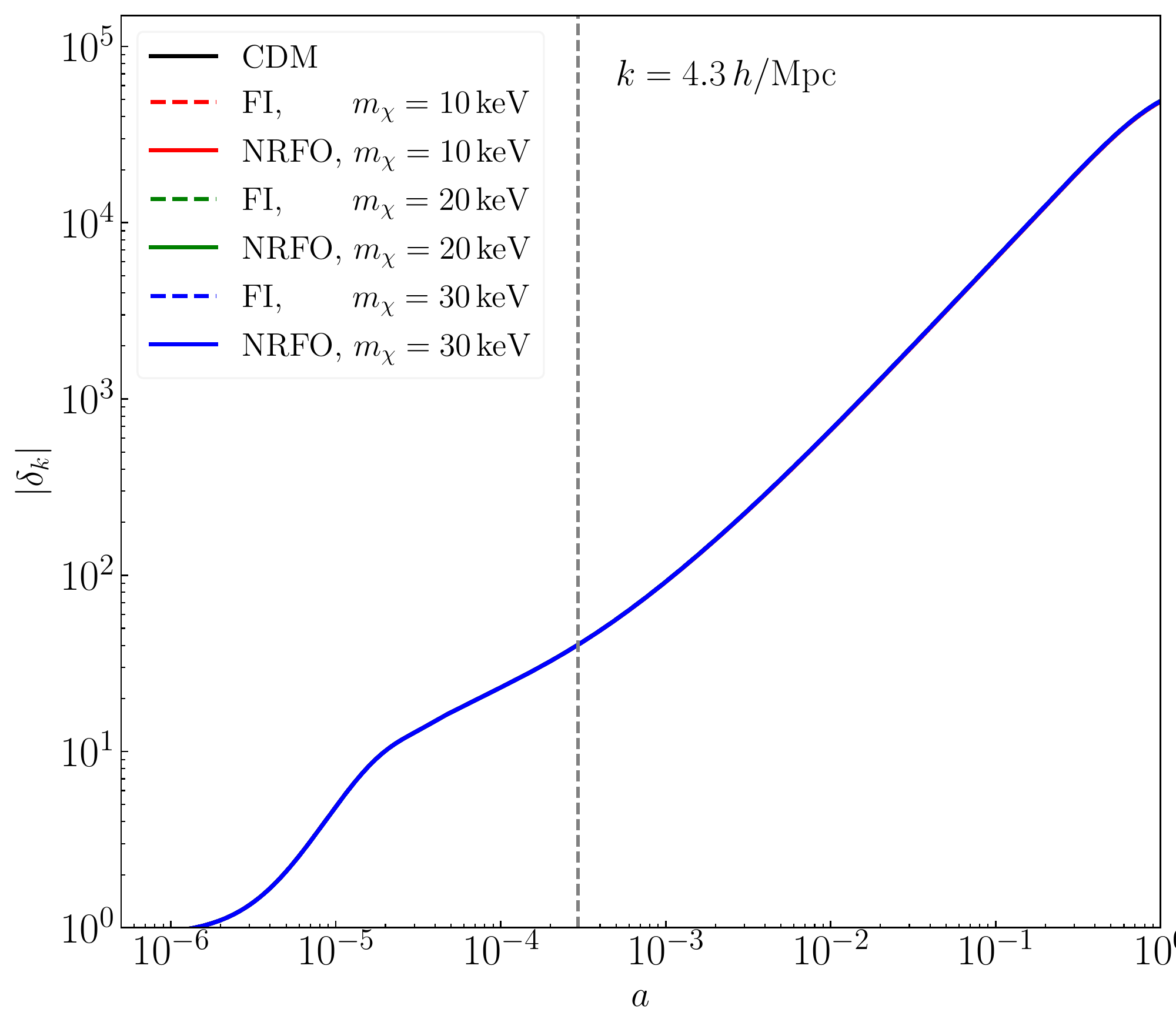}
\includegraphics[width=0.32\textwidth]{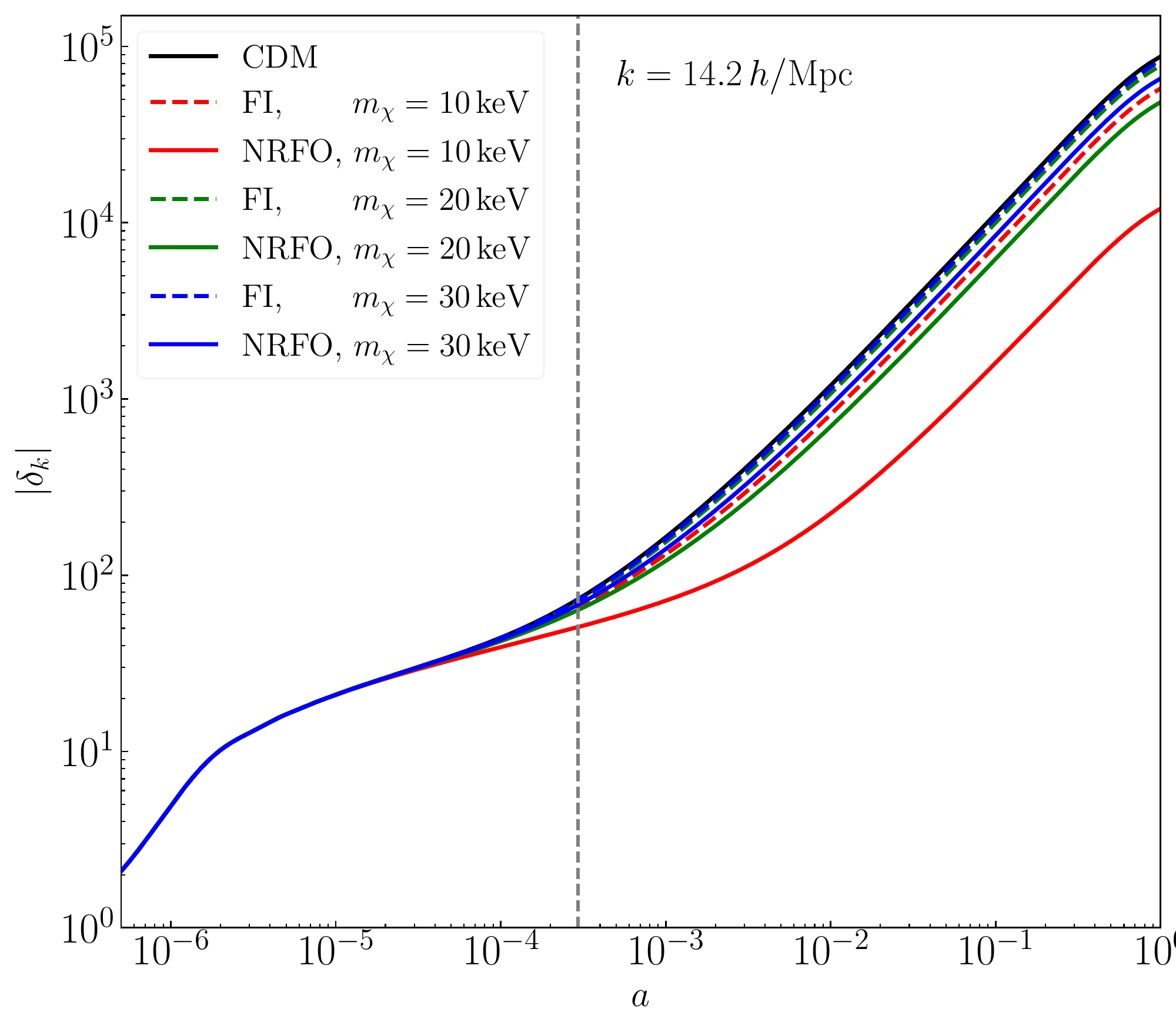} 
\includegraphics[width=0.32\textwidth]{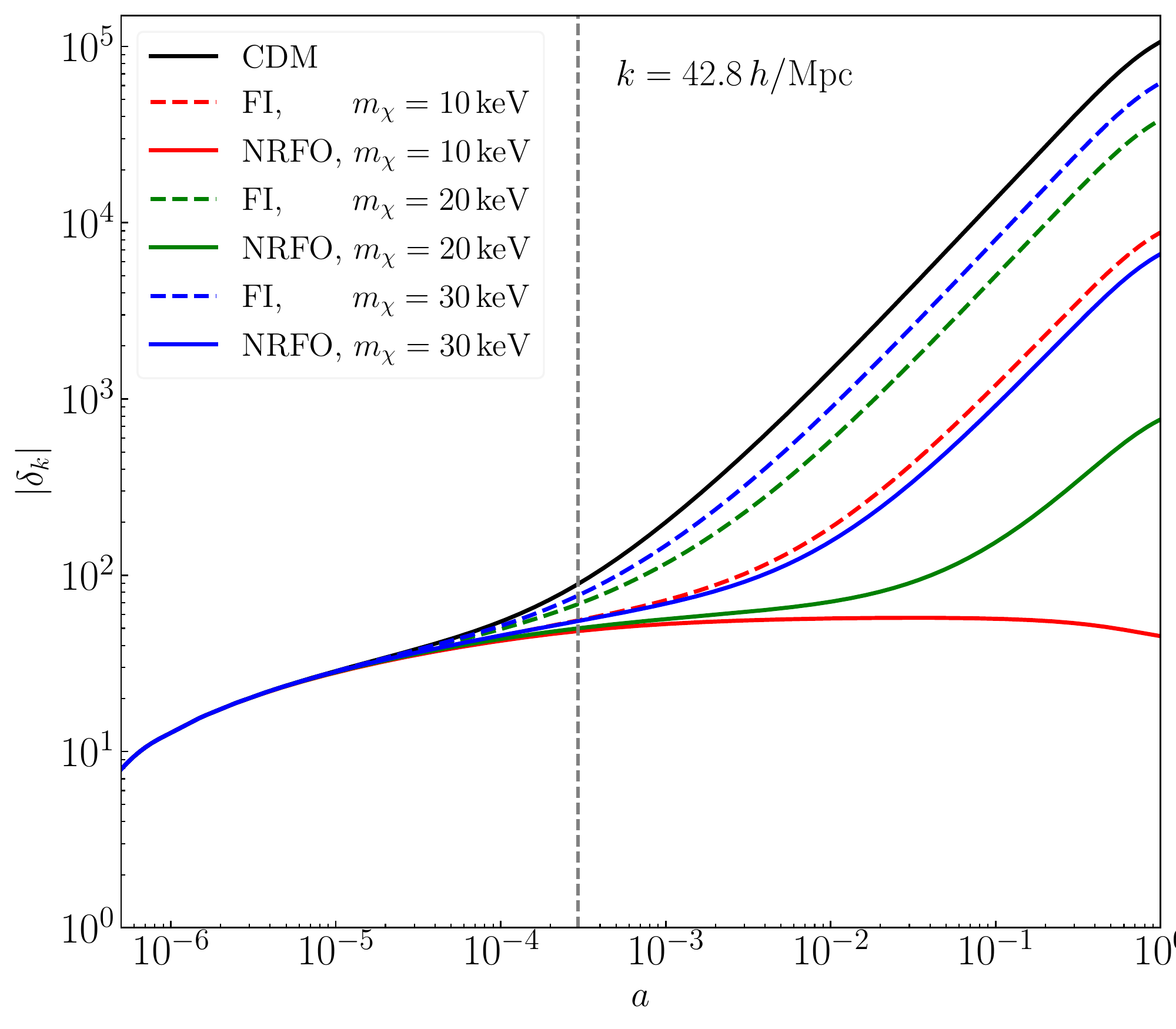}
\caption{Evolution of the density perturbation modes $|\delta_k |$ with $ k =4.3,\,14.2$ and 42.8$~h/{\rm Mpc}$ for the CDM, freeze-out and freeze-in scenarios.
For the latter two scenarios, three dark-matter mass, $m_\chi = 10, \, 20 $ and $30~{\rm keV}$ are chosen.
The vertical dashed line marks the matter-radiation equality.}
\label{fg:delta_k}
\end{figure}

With the moments $\Psi_l$, the perturbation in physical quantities, such as the energy density, pressure, energy flux and shear stress, can be computed straightforwardly by integrating different moments.
For our purpose, the fractional density perturbation can be expressed as
\beq
\delta(\vec x, t)\equiv \frac{\rho_\chi(\vec x, t)-\bar\rho_\chi(t)}{\bar\rho_\chi(t)} =\frac{4 \pi a^{-4} \int d q~q^2 f_{\chi,0}(q)\Psi_0 \epsilon}{\bar\rho_\chi(t)}\,,
\eeq
where $\bar{\rho}_\chi$ is the zeroth-order homogeneous background energy density defined in Eq.~\eqref{eq:rho_integral0}.
The effects from the phase-space distribution at different length scales can then be reflected by the evolution of the Fourier modes $\delta_k$ which we obtain via the \texttt{CLASS} code.
In particular, 
we have truncated the Boltzmann hierarchy at $\ell=50$, and have found that our general results are not affected significantly as long as $\ell>8$.
We present in Fig.~\ref{fg:delta_k} the evolution of $\delta_k$ for the CDM scenario and the freeze-in/freeze-out scenarios with three dark-matter masses $m_\chi=10,~20$ and $30~\rm keV$ 
at three wavenumbers $k=7.1,~14.2$ and $42.8~h/\rm Mpc$.

Overall, the suppression effect from having non-negligible velocities become weaker at larger length scales which correspond to smaller wavenumbers $k$.
For small enough $k$ whose corresponding length scale is larger than the maximum free-streaming distance of the dark-matter particles, no suppression in $\delta_k$ is expected.
This is evident in the left panel, where $k=4.3~h/\rm Mpc$, as all the curves overlap with the black curve which shows the evolution in the CDM scenario.
As $k$ increases to larger values, the deviation from CDM occurs.
Thus, in the middle panel with $k=14.2~h/\rm Mpc$, we see the red, green and blue curves all peel off from the black curve.
For the same production scenario (freeze-in or freeze-out), a smaller mass corresponds to a stronger suppression.
On the other hand, for the same $m_\chi$, the suppression in the freeze-out scenario is stronger than that in the freeze-in scenario.
Both behavior can be understood from the right panel of Fig.~\ref{fg:Nx_fx} --- 1) although the momentum distribution is nearly the same for the same production mechanism, the velocity is larger for smaller dark-matter mass; 2) for the same $m_\chi$, phase space distribution from freeze-out scenario possesses overall larger momenta than that from the freeze-in scenario.
As $k$ further increases, the deviation from the CDM scenario become more distinct in all the cases 
while the differences among all the cases also become very significant.
Thus, in the right panel, we see that at $k=42.8~h/\rm Mpc$ all the curves are well separated, and the suppression in the freeze-out scenario is always stronger than that in the freeze-in scenario.

The examples in Fig.~\ref{fg:delta_k} demonstrate three important points.
First, for a particular type of production mechanism, the suppression on the perturbation modes $\delta_k$ increases with the wavenumber $k$.
Second, such suppression is stronger for smaller dark-matter mass $m_\chi$.
Third, dependence of the suppression on the wavenumber $k$ and the dark-matter mass $m_\chi$ is significantly different
in the freeze-in and the freeze-out scenarios.
The differences in these perturbation modes $\delta_k$ therefore offer us an avenue to examine the predictions of different thermal histories, which, in turn, can be used to constrain or even distinguish these thermal histories.

\subsubsection{Matter power spectrum and transfer function}\label{sec:T2}

The observations from the previous section shows that it is reasonable to examine the whole spectrum of $\delta_k$ when studying the impact of dark-matter velocities on structure formation.
Such physical quantity is called the matter power spectrum:
\beq
P(k)\equiv \abs{\delta_k^2}\,.
\eeq
Since we are always interested in studying the deviation from the CDM scenario, it is convenient to focus on the squared transfer function defined as
\beq
T^2(k) \equiv \frac{P(k)}{P_{\rm CDM} (k)}\,,
\eeq
where $P_{\rm CDM} (k) $ is the matter power spectrum obtained for cold dark matter. 

\begin{figure}
	\centering
	\includegraphics[width=0.49\textwidth]{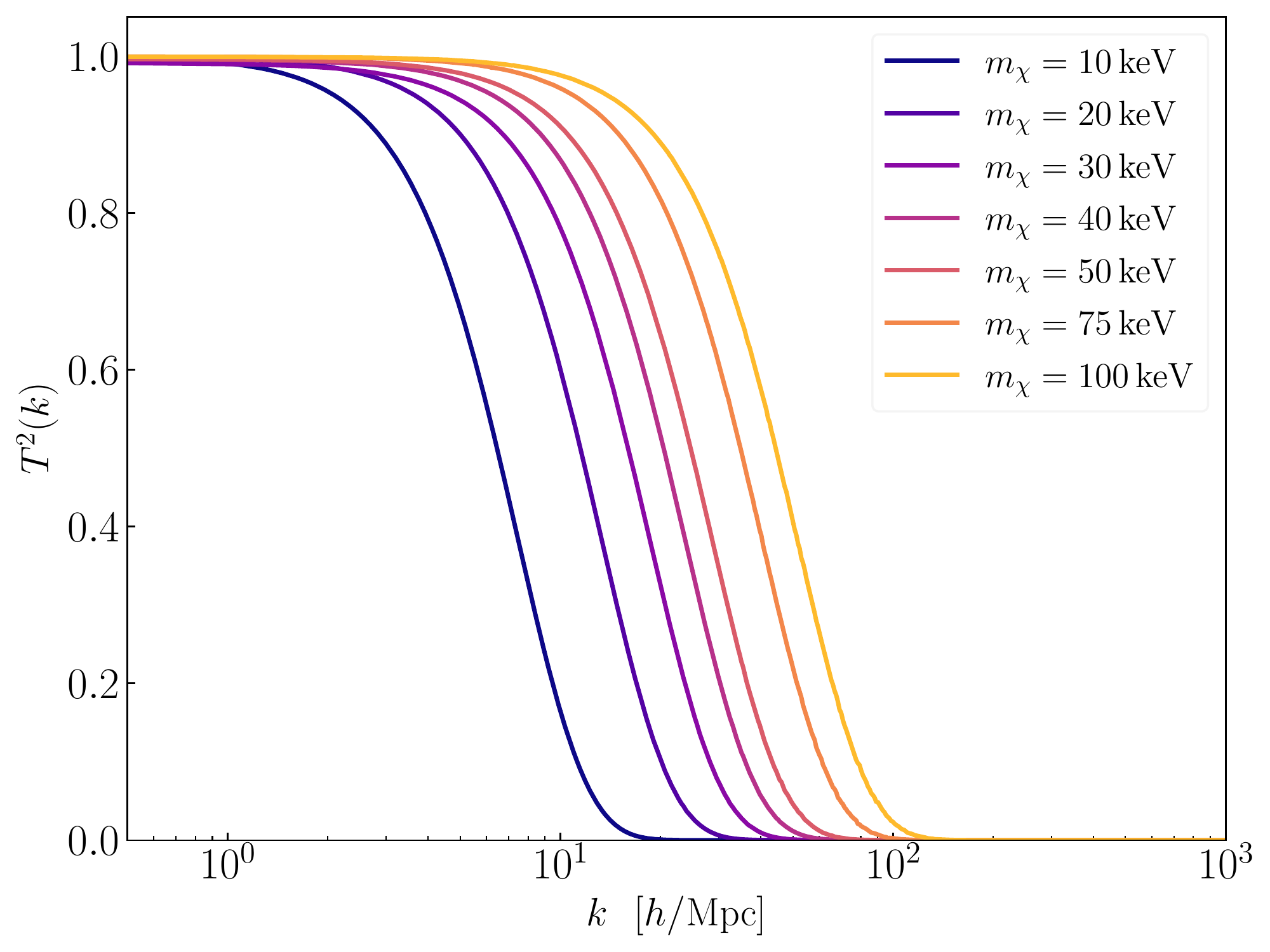}
	\includegraphics[width=0.49\textwidth]{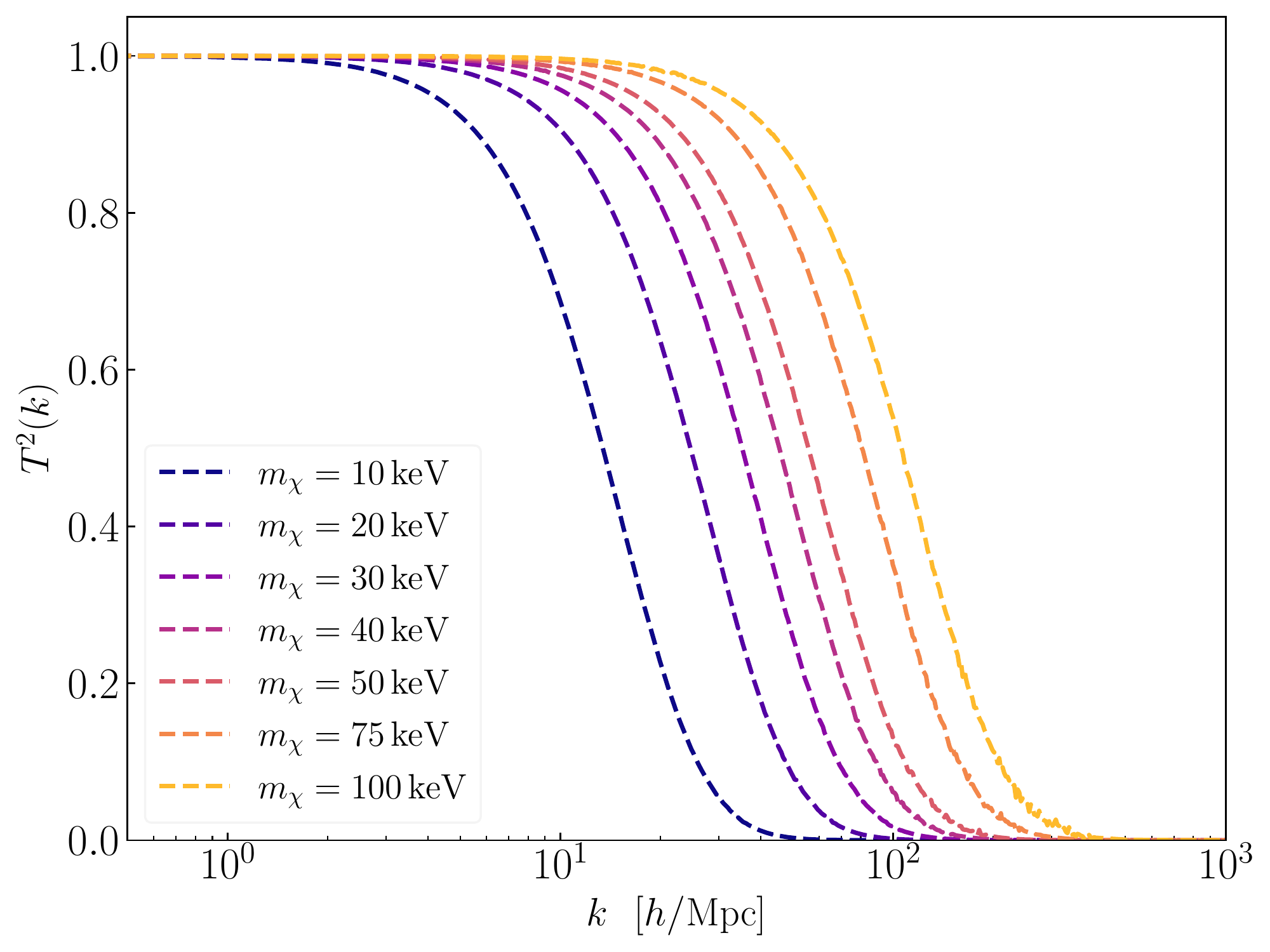}
	\caption{
The squared transfer functions resulting from the freeze-out (left panel) and freeze-in (right panel) scenarios for a series of dark-matter masses from $m_\chi = 10 \,{\rm keV} $ to $100 \, {\rm keV} $.}\label{fg:T2_FIFO}
\end{figure}

In Fig.~\ref{fg:T2_FIFO}, we show a few examples of the squared transfer function associated with the cases studied in Fig.~\ref{fg:Nx_fx}.
The results from the freeze-out and the freeze-in scenarios are presented in left and right panels, respectively. 
As expected, we observe that in each panel the suppression on $T^2(k)$ is stronger as the dark-matter mass decreases.
Comparing across the panels, we also find that
for fixed $m_\chi$, the small-scale suppression on $T^2(k)$ is always greater in the freeze-out scenario than in the freeze-in scenario.

Similar to the WDM scenario, where the transfer function can be fitted by \cite{Bode:2000gq,Viel:2005qj}
\begin{equation}\label{eq:T2_wdm}
    T^2_{\rm WDM} (k) \simeq \left[ 1+ (\alpha_{\rm WDM} k)^{2 \nu} \right]^{-10/\nu} \, ,
\end{equation}
with $\nu = 1.12$ and
\beq
\alpha_{\rm WDM} = 0.049 \times\left(\frac{\mwdm}{\rm keV}\right)^{-1.11}\left(\frac{\Omega_\chi}{0.25}\right)^{0.11}\left(\frac{h}{0.7}\right)^{1.22}~\frac{{\rm Mpc}}{h}\,,\label{eq:alpha_wdm}
\eeq
it proves convenient to analytically fit the transfer function resulting from the freeze-in and the non-relativistic freeze-out scenarios.
We find that, although the WDM fitting formula can fit well the transfer functions in the non-relativistic freeze-out scenario with a different $\alpha$, 
the fit for the freeze-in scenario is not as good.
A better fit for the freeze-in scenario can be done using the $\{\alpha,\beta,\gamma\}$ formula proposed in Refs.~\cite{Murgia:2017lwo,Murgia:2018now}:
\begin{equation}\label{eq:T2_abc}
	T^2(k) = [1+(\alpha \, k)^{\beta}]^{2\gamma}\,.
\end{equation}
We shall therefore adopt Eq.~\eqref{eq:T2_abc} for both the freeze-in and freeze-out cases.
In analogous to Eq.~\eqref{eq:alpha_wdm}, we choose $\alpha$ to be a function the dark-matter mass.
In order to allow the parameter $\alpha$ to have a different mass dependence in different scenarios, 
we further decompose $\alpha$ as\footnote{We fix $\Omega_\chi=0.25$ and $h=0.67$ since we are primarily interested in the effects from varying $m_\chi$ or $\expt{v}_0$ while holding the relic abundance fixed.}
\begin{equation}
	\alpha = \alpha_1 \, \left( \dfrac{m_\chi}{ {\rm keV}}\right) ^{\alpha_2} \, \left( \frac{g_{\star,s}(T_0)}{g_{\star,s}(m_\chi)}\right) ^{1/3} \frac{\rm Mpc}{h} \,,\label{eq:alpha12}
\end{equation}
where we have inserted the factor $\left( g_{*,s}(T_0)/g_{*,s}(m_\chi)\right)^{1/3} $ to include the shift effect due to the change of $g_{\star,s}$ after dark matter is produced.
With $\alpha$ specified, the parameters $\beta$ and $\gamma$ can be fixed based on the goodness of fit.
We find that the transfer functions can be fitted to high accuracy by choosing $\{\alpha_1,\alpha_2,\beta,\gamma\}=\{ 0.32,-0.80, 2.24,-4.46\}$ 
for freeze-out, and 
$\{\alpha_1,\alpha_2,\beta,\gamma\}=\{ 0.30,-0.87, 2.28,-1.51\}$ for freeze-in.
Note that the values of $\beta$ and $\gamma$ for freeze-out are chosen to match the corresponding exponents $2 \nu$ and $-10/\nu$ for the WDM fitting functions.
Therefore, the overall shape of the fitted transfer function is the same in the freeze-out and the WDM scenarios.
In addition, Eq.~\eqref{eq:T2_wdm} or Eq.~\eqref{eq:T2_abc} suggests a symmetry transformation:
\beqn
\alpha\to \alpha'\,, &~&
k\to k'=\frac{\alpha}{\alpha'}k\,,\nn\\
T^2(k)&\to& {T'}^2(k')=T^2(k)\,,\label{eq:trans_sym}
\eeqn
which holds approximately true in the actual transfer functions and is exactly true in the fitting formula.  
Therefore, when plotted against $k$ on logarithmic scale, a variation in $\alpha$ caused by a change in $m_\chi$ amounts to sliding $T^2(k)$ rigidly in the horizontal direction.
Indeed, the reason for this behavior lies in the similarity between the momentum distributions  
within either the freeze-out or the freeze-in scenario,
which we have already seen in Fig.~\ref{fg:Nx_fx}.
We shall prove this in the Appendix~\ref{sec:shift}.

The above fitting formulas allow us to conveniently compare the transfer functions in our cases with the standard WDM scenario over a large range of dark-matter mass.
To estimate how the Lyman-$\alpha$ constraints in the WDM scenario are projected to scenarios studied in this work, namely, freeze-in and non-relativistic freeze-out, we adopt the widely used ``half-mode analysis'' \cite{Konig:2016dzg} and the ``$\delta A$ analysis'' \cite{Murgia:2017lwo}.

In the half-mode analysis, the half-mode $k_{1/2}$ is defined via $T^2(k_{1/2})=1/2$.
It is then straightforward to find that
\beq
k_{1/2}\approx 
\begin{cases}
\alpha_{\rm WDM}^{-1}\left(2^{\nu/10}-1 \right)^{1/(2\nu)}, & \text{for WDM fit}\,;\\
\alpha^{-1}\left( 2^{-1/(2\gamma)}-1 \right)^{1/\beta}, & \text{for $\{\alpha,\beta,\gamma\}$ fit}\,.
\end{cases}
\eeq
The Lyman-$\alpha$ constraint for a particular model is then set by demanding that
\beq
T^2(k)\geq T^2_{\rm WDM}(k)\,, \text{~~~for all}~0\leq k\leq k_{1/2}\,,
\eeq
where $T^2_{\rm WDM}(k)$ is obtained from Eq.~\eqref{eq:T2_wdm} by assuming a certain reference WDM mass.

On the other hand,
the $\delta A$ criterion is based on comparing the quantity $A$ defined as
\beq
A\equiv \int_{k_{\rm min}}^{k_{\rm max}}dk~ \frac{P_{\rm 1D}(k)}{P_{\rm 1D}^{\rm CDM}(k)}\,,
\eeq
where $k_{\rm min/max}=0.5$ and $20~h/{\rm Mpc}$ respectively, corresponding to the MIKE/HIRES and XQ-100 combined dataset used in \cite{Irsic:2017ixq},
and $P_{\rm 1D}(k)$ is the ``one-dimensional'' matter power spectrum
\beq
P_{\rm 1D}(k)\equiv \frac{1}{2\pi}\int_k^{\infty} dk'~k'P(k')\,.
\eeq
The quantity $\delta A$ is then defined as the fractional deviation from the CDM scenario
\beq
\delta A\equiv 1-A/A_{\rm CDM}\,.
\eeq
The Lyman-$\alpha$ constraint for a particular scenario is then estimated by requiring that $\delta A\leq \delta A_{\rm ref}$,
where $\delta A_{\rm ref}$ is a reference value in a WDM scenario.
In particular, we find that $\delta A_{\rm ref}\approx 0.50$ for $\mwdm=3.5~\rm keV$, and $\delta A_{\rm ref}\approx 0.36$ for $\mwdm=5.3~\rm keV$.

With these two methods, 
the current constraints on WDM, $\mwdm\geq 3.5$ (or 5.3) keV, can be mapped easily to the scenarios of freeze-in and non-relativistic freeze-out.
We find that, in the freeze-in scenario, the constraints are given by
\beq
m_\chi \gtrsim
\begin{cases}
21.9~(\text{or}~37.0)~{\rm keV}& (\text{half-mode}) \\
21.3~(\text{or}~36.1)~{\rm keV}& (\delta A)
\end{cases}\,;\label{eq:bounds_FI}
\eeq
using the two different methods.
In the non-relativistic freeze-out scenario, 
both the half-mode analysis and the $\delta A$ criterion give the same results:
\beq
m_\chi\gtrsim 56.5~(\text{or}~94.5)~\rm keV\,,\label{eq:bounds_NRFO}
\eeq
which is expected as the transfer functions possess the same shape as the WDM ones.
In each case, the larger (smaller) value in each case corresponds to the projection of the more stringent (conservative) bound on WDM.
Overall, the results obtained from the two methods are extremely similar with the relative difference in the freeze-in case no more than 5\%.

The results above shows that the dark-matter mass in both the freeze-in and the non-relativistic freeze-out scenario are constrained at much larger values than those in the WDM scenario.
This is consistent with our discussion in Sec.~\ref{sc:mass and velocity}, which is based on the present-day average velocity $\expt{v}_0$.
The reason roots in the fact that the WDM can only be generated from a dark sector that is much colder than the SM thermal bath, whereas the cases we are comparing here all have $\eta_{\rm dec} =1$.
It is therefore also foreseeable that such constraints can be weaker if $\eta_{\rm dec} < 1$ which is indeed the case as we can see by the end of this section.
In addition, we have also find that the constraints obtained here are considerably weaker than the preliminary results in Sec.~\ref{sc:mass and velocity}, where the constraints are estimated to be $m_\chi\gtrsim 30$ (or 50) keV for freeze-in, and $m_\chi\gtrsim 70$ (or 120) keV.
This therefore suggests that while a single free-streaming velocity can provide a good order of magnitude estimate on the constraints from structure formation, the constraints it provides is in practice not very accurate.

\begin{figure}
\centering
\includegraphics[width=0.5\textwidth]{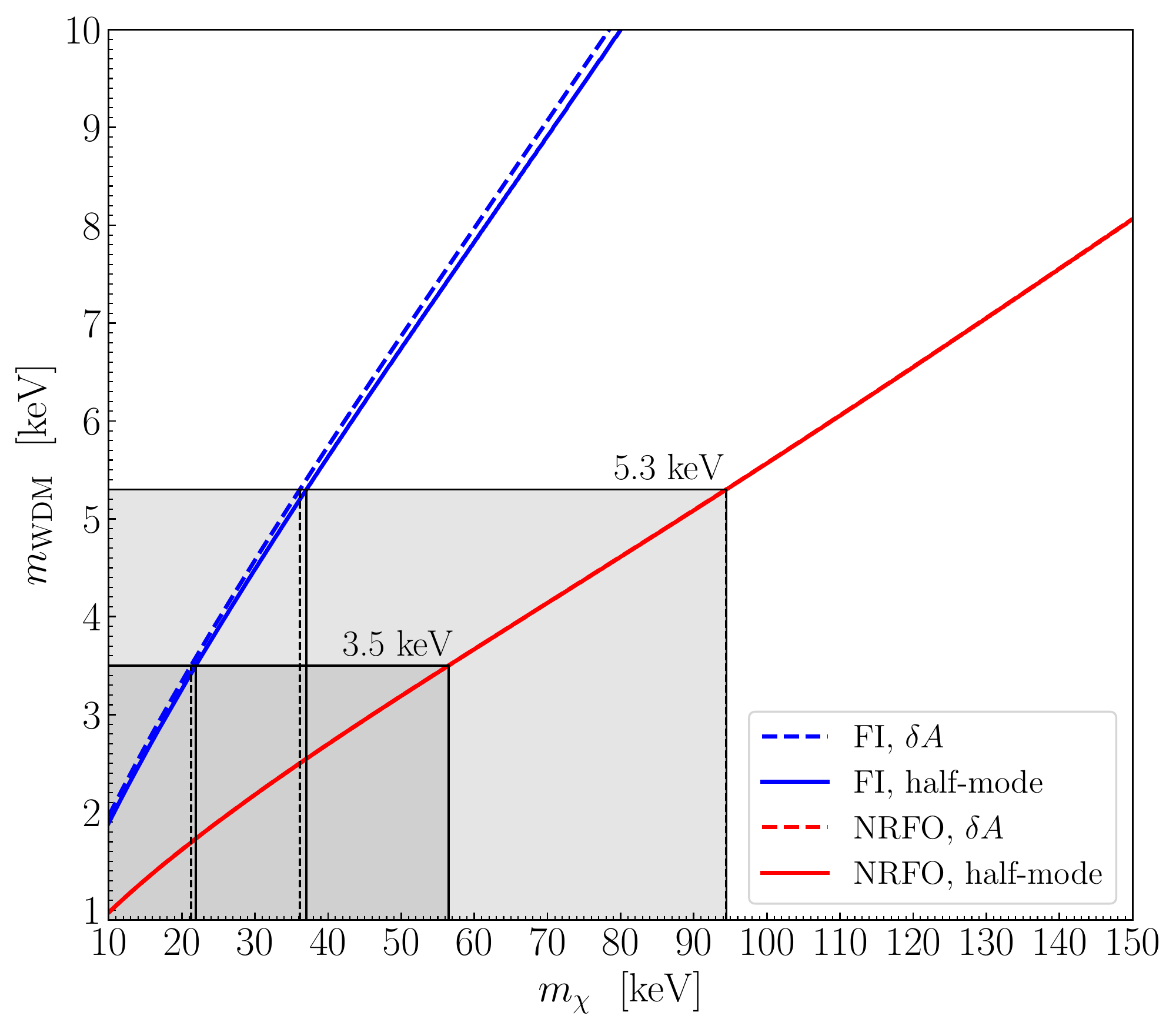}
\caption{Mappings between the constraint on $\mwdm$ and that on $m_\chi$ based on the half-mode analysis (solid curves) or the $\delta A$ analysis (dashed curves) in the freeze-in (blue) and the non-relativistic freeze-out (red) scenarios.
Notice that the two curves overlap completely in the latter scenario.
As indicated by the black lines which show the current bounds, a Lyman-$\alpha$ constraint on the WDM mass can be directly mapped to a constraint on $m_\chi$ in a particular scenario.
}\label{fg:mwdm_proj}
\end{figure}

Beyond the current bounds at 3.5 or 5.3 keV, the two methods can be more generally used to recast the Lyman-$\alpha$ constraint for any reference WDM mass, as is shown in Ref.~\cite{Dienes:2021cxp}.
Indeed, one can create a mapping between a WDM mass and $m_\chi$ in a specific scenario when their Lyman-$\alpha$ constraints are the same using one particular method.
In Fig.~\ref{fg:mwdm_proj}, we show such mappings between $m_\chi$ and $\mwdm$ in the freeze-in and the non-relativistic freeze-out scenarios obtained by allowing $\mwdm$ to vary in the two recasting procedures.
Thus, if the Lyman-$\alpha$ constraints on WDM is updated in the future, one can easily find the equivalent constraints for the freeze-in ore non-relativistic freeze-out scenarios.

\subsubsection{Halo mass function}

The distinction between the freeze-in scenario and the non-relativistic freeze-out scenario exists not only in the linear regime, but also in the non-linear regime.
One of the important physical quantity for assessing the non-linear evolution of the density perturbation is the halo mass function $dn/d\ln M$, defined as the comoving number density of dark-matter halos at a logarithmic interval of the halo mass $M$.
The halo mass function is in principle obtained from N-body or hydrodynamic simulations, which are computationally expensive.
Nevertheless, approaches have been developed to analytically calculate or parametrize the halo mass function.
For example,
the Press-Schechter formalism \cite{Press:1973iz}
enables one to assess the halo mass function using the linear matter power spectrum $P(k)$.
Recent study based on stochastic differential equations \cite{Lapi:2020juf},
and that based on the mass and energy cascade theory in dark matter flow \cite{Xu:2022vbs},
as well as the work towards a general parametrization \cite{Lovell:2019sfs,Lovell:2020bcy}
have shown good agreement with results from simulations.
In this paper, we follow the extended Press-Schechter framework \cite{Press:1973iz,Bond:1990iw,Lacey:1993iv,Sheth:1999mn,Sheth:1999su,Musso:2012qk} to derive the halo mass function.
In particular, we use a sharp-$k$ window function which has the advantage that the result is sensitive to the shape of the linear matter power spectrum rather than to the shape of the window function \cite{Schneider:2013ria}.
The halo mass function at the present-day thus takes the form
\begin{equation}
\frac{d n}{d \ln M}=\frac{\bar{\rho}}{12 \pi^2 M} \nu(M)\eta(M) \frac{P(1/R(M))}{\delta_c^2 R^3(M)}\,,
\end{equation}
where $\bar\rho=\Omega_m \rho_{\rm crit}$ is the average energy density of matter, $\delta_c \simeq 1.686$ is the critical overdensity,
the mapping between $R$ and $M$ is given by 
\beq
M=\frac{4\pi}{3}\bar\rho(c_W R)^3\,,\label{eq:M_R}
\eeq
with $c_W\approx 2.5$ \cite{Schneider:2014rda},
and $\nu(M) \equiv \delta_c^2/\sigma^2(t_0, M)$, 
with $\sigma^2(t_0, M)$ being the spatial average of the variance of $\delta(\vec x,t)$:
\begin{equation}
\sigma^2(t, R) \equiv \int_{-\infty}^{\infty} d \log k~  W^2(k,R) \frac{k^3 P(k, t)}{2 \pi^2}\,,
\end{equation}
in which the sharp-$k$ window function $W(k,R) = \Theta(1-kR)$.
The function $\eta(M)$ only depends on $M$ through the function $\nu(M)$:
\beq
\eta(M)=\sqrt{\frac{2 \nu(M)}{\pi}} A\left[1+\nu^{-\alpha}(M)\right] e^{-\nu(M) / 2}\,,
\eeq
where $A\simeq 0.3222$ and $\alpha = 0.3$.
A detailed derivation can be found in Refs.~\cite{Schneider:2011yu,Schneider:2013ria,Schneider:2014rda,Dienes:2021itb}.

\begin{figure}\centering
\includegraphics[width=0.49\textwidth]{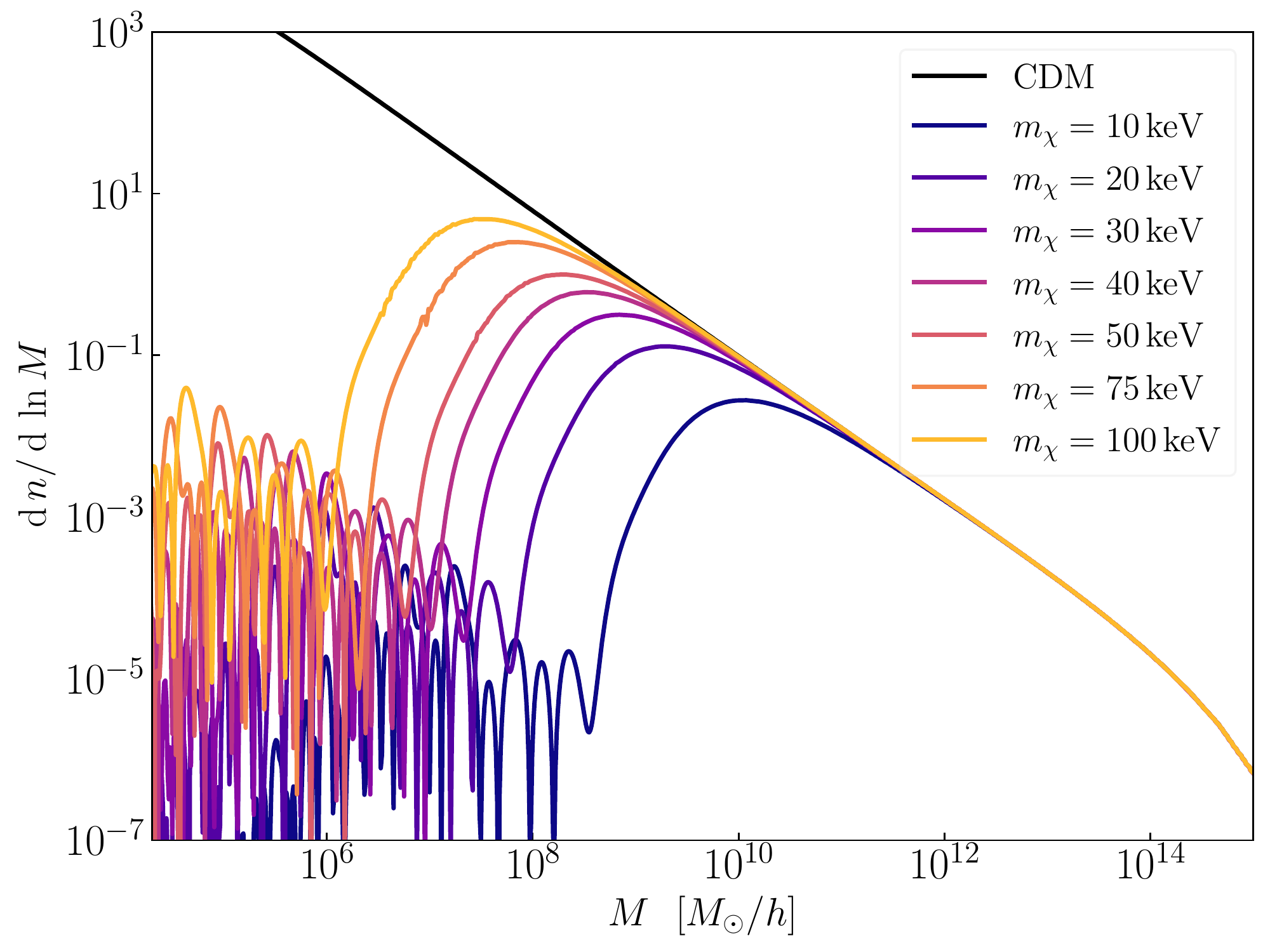}
\includegraphics[width=0.49\textwidth]{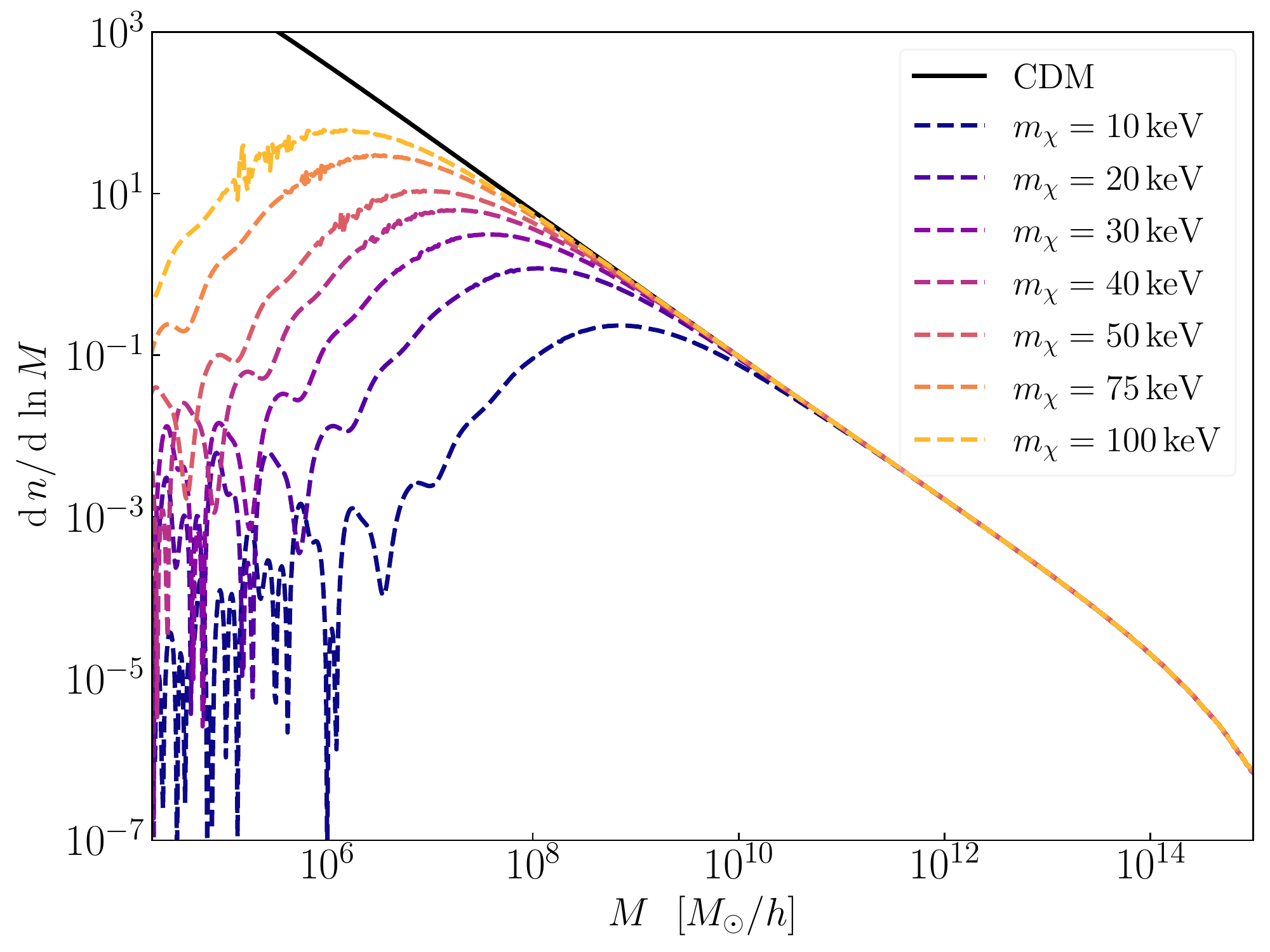}
\caption{The halo mass functions for non-relativistic freeze-out (left panel) and freeze-in (right panel) with dark-matter mass $m_\chi$ ranging from $10$ to $100~{\rm keV}$.}\label{fg:halo_mass}
\end{figure}

We present in Fig.~\ref{fg:halo_mass} the halo mass functions for a range of dark-matter masses from $10~{\rm keV} $ to $100~{\rm keV}$, where the results for the non-relativistic freeze-out and the freeze-in scenarios are shown in left and right panels, respectively. 
The halo mass function in the CDM scenario is shown in both panels by the black curve. 
Apparently, the halo mass function in the CDM scenario $(dn/d\ln M)_{\rm CDM}$ increases monotonically as the halo mass $M$ decreases. 
On the other hand, the number of dark-matter halos with smaller masses is clearly suppressed within the range of masses considered in both the freeze-in and the non-relativistic freeze-out scenarios.
For instance, if $m_\chi=10~\rm keV$, the number of halos starts to be suppressed below $M\lesssim 10^{11}~M_\odot/h$.
This suppression effect essentially originates from the suppression in the matter power spectrum $P(k)$ which enters the evaluation of $dn/d\ln M$ by itself and through the variance $\sigma^2(t,R)$.
Since the sharp-$k$ window function cuts off all contribution from modes whose wavenumber $k>1/R(M)$, it is not surprising that the halo mass function is only suppressed at a halo mass $M$ when the matter power spectrum is suppressed at the wavenumber $k<1/R(M)$.
Therefore, we find that the halo mass function
generally follows the trend in the matter power spectrum --- the halo mass functions in our scenarios coincide exactly with the $(dn/d\ln M)_{\rm CDM}$ at the larger-mass end, but gradually peel off from the CDM curve at smaller masses.
Similar to what we have seen in the squared transfer function, a smaller $m_\chi$ leads to stronger suppression at larger mass scales in each scenario, while for the same $m_\chi$ the suppression in $dn/d\ln M$ is always stronger in the non-relativistic freeze-out scenario.

It is also interesting to notice that for the same dark-matter mass, as the halo mass $M$ decreases, while the initial peel-off from $(dn/d\ln M)_{\rm CDM}$ appear at locations that are close in $M$ in both scenarios,
the decrease at further smaller $M$
is always more drastic in the non-relativistic freeze-out scenario than in the freeze-in scenario
before entering the regime of rapid acoustic oscillations where dark-matter particles are all free-streaming.
The reason behind this phenomenon can be glimpsed from the right panel of Fig.~\ref{fg:Nx_fx}, which shows that while 
the distribution function resulting from the freeze-in scenario is overall colder, 
it is also wider.
The tail of the freeze-in distribution function at larger momentum stretches close to the higher-momentum end of the distribution in the non-relativistic freeze-out scenario 
which gives rise to a peel-off location close to the one in the freeze-out case.
On the other hand, the lower-momentum tail of the freeze-in distribution function extends far into the low-momentum region 
which prevents a complete suppression in the halo formation on the corresponding scales.

Indeed, the shape of the halo mass function encodes the information of the phase-space distribution.
It is not difficult to conjecture that a correlation between the halo mass function and the phase-space distribution exists via the following mapping
\beq
M\rightarrow R(M) \rightarrow k=1/R(M)\rightarrow v(k)\,,
\eeq
where $v(k)$ stands for a dark-matter velocity whose corresponding free-streaming length $d_{\rm FSH}\sim 1/k$.
Recent developments on this correlation can be found in Refs.~\cite{Dienes:2020bmn,Dienes:2021itb}
where the information from structure formation can be exploited to reproduce the primordial phase-space distribution of dark matter.

\subsubsection{Milky-Way satellite counts}

\begin{figure}
\centering
\includegraphics[width=0.6\textwidth]{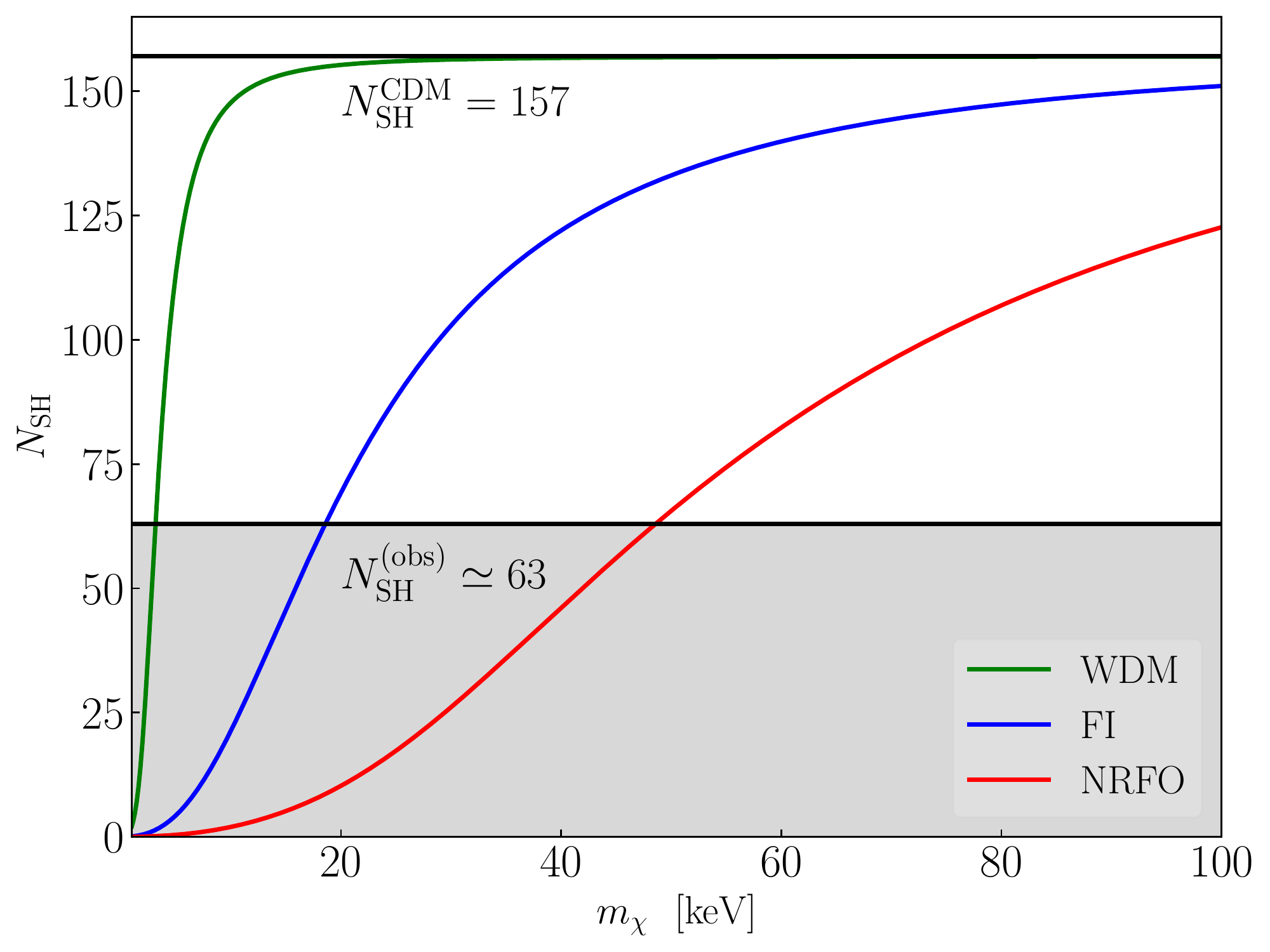}
\caption{The expected number of subhalos $N_{\rm SH}$ with masses $M> 10^{8} h^{-1} M_{\odot}$ 
within a MW-sized galaxy as a function of the dark-matter mass.
The results from the WDM, freeze-in, and non-relativistic freeze-out scenarios are colored in green, blue and red, respectively, whereas the two horizontal lines indicate the result from a pure CDM N-body simulation \cite{Lovell:2013ola}, 
and the estimate on the number of observed MW satellites.
 }\label{fg:subhalo}
\end{figure}

Another important physical observable relevant for the non-linear evolution of density perturbations is the number of satellite galaxies $N_{\rm SH}$ residing within a host galaxy of mass $M_0$.
Considering the observability, 
only the satellites that are more massive than a threshold $M_{\rm min}$ are taken into account. 
Thus, the expected number of detectable subhalos is given by
\beq
    N_{\rm SH} = \int_{M_{\rm min}}^{M_0} \frac{d N_{\mathrm{SH}}}{d M} dM \,,
\eeq
where ${d N_{\mathrm{SH}}}/{d M}$ is the subhalo mass function.
Following Refs.~\cite{Giocoli:2007gf,Schneider:2014rda,Dienes:2021itb},
for the shape-$k$ window function that we have been using, the subhalo mass function takes the form
\begin{equation}
\frac{d N_{\mathrm{SH}}}{d M}=  \frac{1}{6 \pi^2 \mathcal{N}_{\mathrm{SH}}}\left(\frac{M_0}{M^2}\right)  \frac{P(1 / R(M))}{R^3(M) \sqrt{2 \pi\left[\sigma^2(t_0,M)-\sigma^2(t_0, M_0)\right]}} \,,\label{eq:SHMF}
\end{equation}
where 
$\mathcal{N}_{\mathrm{SH}}$ is a normalization factor.

It is of particular interest to apply the above formula in the case of a MW-like galaxy of mass $M_0=1.77\times 10^{12}~M_\odot/h$,
for which the results from the N-body simulation in the Aquarius project \cite{Lovell:2013ola} predicted that $N_{\rm SH}=157$ for the number of subhalos above $M_{\rm min}=10^8~M_\odot/h$ in a pure CDM scenario.\footnote{Note that this approach does not include
effects from reionization, tidal stripping, the Galactic disk, and the Large Magellanic Cloud.
}
Based on this number, 
we obtain numerically that 
$\mathcal{N}_{\mathrm{SH}}\approx 46.9$.

On the observational side, we follow Refs.~\cite{Polisensky:2010rw,Schneider:2014rda,Merle:2015vzu,Schneider:2016uqi,Dienes:2021itb}
to estimate the observed number of MW satellites $N_{\rm SH}^{\rm (obs)}$ above $M_{\rm min}$ by multiplying the 15 ultra-faint satellites discovered by the Sloan Digital Sky Survey (SDSS) by a factor of 3.5 to account for the limited sky coverage and then add to it 11 classical satellites which were discovered pre-SDSS.
This method yields $N_{\rm SH}^{\rm (obs)}\approx 63$,
and thus for a viable dark-matter candidate, we expect that number of theoretically predicted MW subhalos $N_{\rm SH}>N_{\rm SH}^{\rm (obs)}$.

In Fig.~\ref{fg:subhalo}, we show the expected number of MW satellites as a function of the dark-matter mass $m_\chi$ in the freeze-in and non-relativistic freeze-out scenarios, 
where we have used the $\{\alpha,\beta,\gamma\}$ fitting formula in Eq.~\eqref{eq:T2_abc} together with the decomposition in Eq.~\eqref{eq:alpha12}
to obtain the matter power spectra for arbitrary $m_\chi$.
As we see, while the results in the freeze-in and non-relativistic freeze-out scenario both start from zero as $m_\chi\to 0$ and asymptotes to 
$N_{\rm SH}^{\rm CDM}=157$ at larger enough $m_\chi$,
their difference in the intermediate range, $m_\chi\sim \mathcal{O}(10)$ - $\mathcal{O}(100)~\rm keV$, is clearly distinct.
Similar to the discussion on the halo mass function, this difference in $N_{\rm SH}$ originates from the fact that for the same mass $m_\chi$ the distribution resulting from the non-relativistic freeze-out scenario is always overall hotter than that from the freeze-in scenario.
Therefore, the observed number of the MW satellites imposes a stronger constraint on the non-relativistic freeze-out scenario.
With $N_{\mathrm{SH}}^{\rm (obs)}\approx 63$, we find that
\beq
m_\chi \gtrsim
\begin{cases}
3.15~{\rm keV}& (\textbf{WDM}) \\
18.60~{\rm keV}& (\textbf{FI}) \\
48.60~{\rm keV}& (\textbf{NRFO})
\end{cases}\,.
\eeq
Comparing to the Lyman-$\alpha$ constraints in Eqs.~\eqref{eq:bounds_FI} and \eqref{eq:bounds_NRFO},
it is noticeable that the constraint from the MW satellite count is slightly weaker for the WDM scenario, but significantly weaker in either the freeze-in scenario or the non-relativistic freeze-out scenario.

\subsection{Improved constraints on decoupling temperatures}
\label{sec:improve}

Thus far in this section, we have assumed $\eta_{\rm dec}=1$ when deriving the constrains on the dark matter mass.
To extend our results to the regime with $\eta_{\rm dec}<1$ one in principle needs to solve Eq.~\eqref{eq:Boltzmann_fx} with a $T_\chi$ that is distinct from the temperature of the standard-model thermal bath.
However, we notice that the shape of the distribution in the non-relativistic freeze-out or the freeze-in scenario is very well described by the Maxwell-Boltzmann distribution or the distribution function in Eq.~\eqref{eq:f_fi}.
Therefore, instead of doing a computationally expensive scan with $\eta_{\rm dec}<1$,
it is possible to identify the result from one combination $\{m_\chi^{(1)}, \eta_{\rm dec}^{(1)},x^{(1)}_{\rm dec}\}$
to that from anther combination $\{m_\chi^{(2)}, \eta_{\rm dec}^{(2)},x^{(2)}_{\rm dec}\}$ provided that these two combinations produce approximately the same present-day velocity distribution.
Indeed, this identification amounts to requiring $m_\chi^{(1)}/T_{\chi}^{(1)}(t_0)=m_\chi^{(2)}/T_{\chi}^{(2)}(t_0)$
which leads to
\beq
\displaystyle\frac{m_\chi^{(1)}}{m_\chi^{(2)}}\left[\frac{g_{\star,s}(T^{(1)}_{\rm dec})}{g_{\star,s}(T^{(2)}_{\rm dec})}\right]^{1/3}\frac{\eta_{\rm dec}^{(2)}}{\eta_{\rm dec}^{(1)}}\sqrt{\frac{x_{\rm dec}^{(2)}}{x_{\rm dec}^{(1)}}}=1 
\eeq
in the non-relativistic freeze-out scenario, or
\beq
\displaystyle\frac{m_\chi^{(1)}}{m_\chi^{(2)}}\left[\frac{g_{\star,s}(T^{(1)}_{\rm dec})}{g_{\star,s}(T^{(2)}_{\rm dec})}\right]^{1/3}\frac{\eta_{\rm dec}^{(2)}}{\eta_{\rm dec}^{(1)}}=1
\eeq
in the freeze-in scenario after applying entropy conservation.

However, one potential issue arises as the decoupling process is not instantaneous in the non-relativistic freeze-out scenario (see the left panel of Fig.~\ref{fg:Nx_fx}), 
which leads to an ambiguity in defining $T_{\rm dec}$. 
To make a fair comparison with our analysis in Fig.~\ref{fg:T_dec_eta}, where we have assumed instantaneous decoupling,
we shall define $T_{\rm dec}$ to be the temperature at which the $N^{\rm eq}_\chi(T_{\rm dec})=N_\chi(T_0)$ --- when the dotted curve intersects the asymptotic value of $m_\chi N_\chi$ in Fig.~\ref{fg:Nx_fx}.
In this way, contours of average velocity and dark-matter mass remain unchanged.

\begin{figure}
\centering
\includegraphics[width=0.5\textwidth]{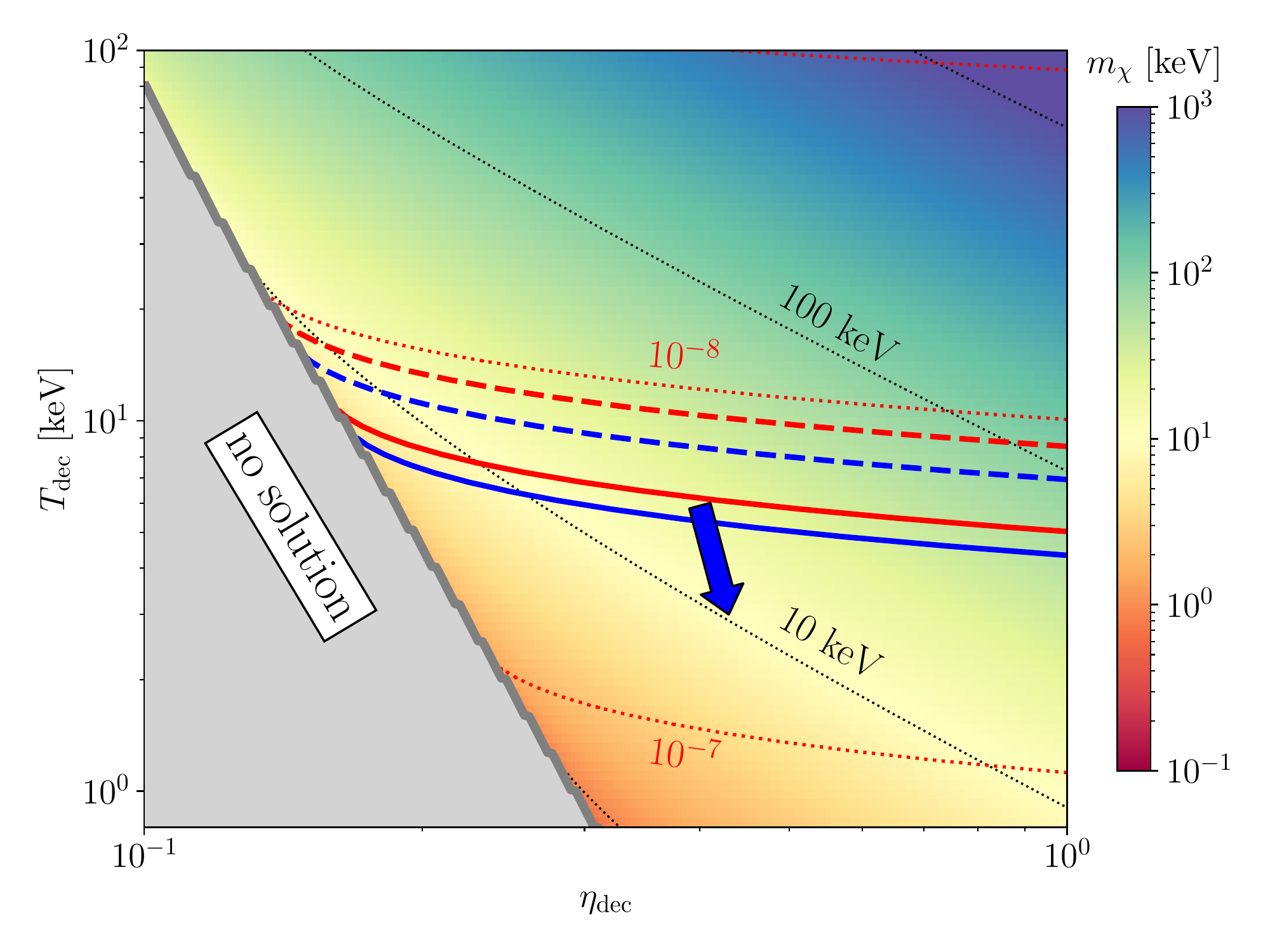}\includegraphics[width=0.5\textwidth]{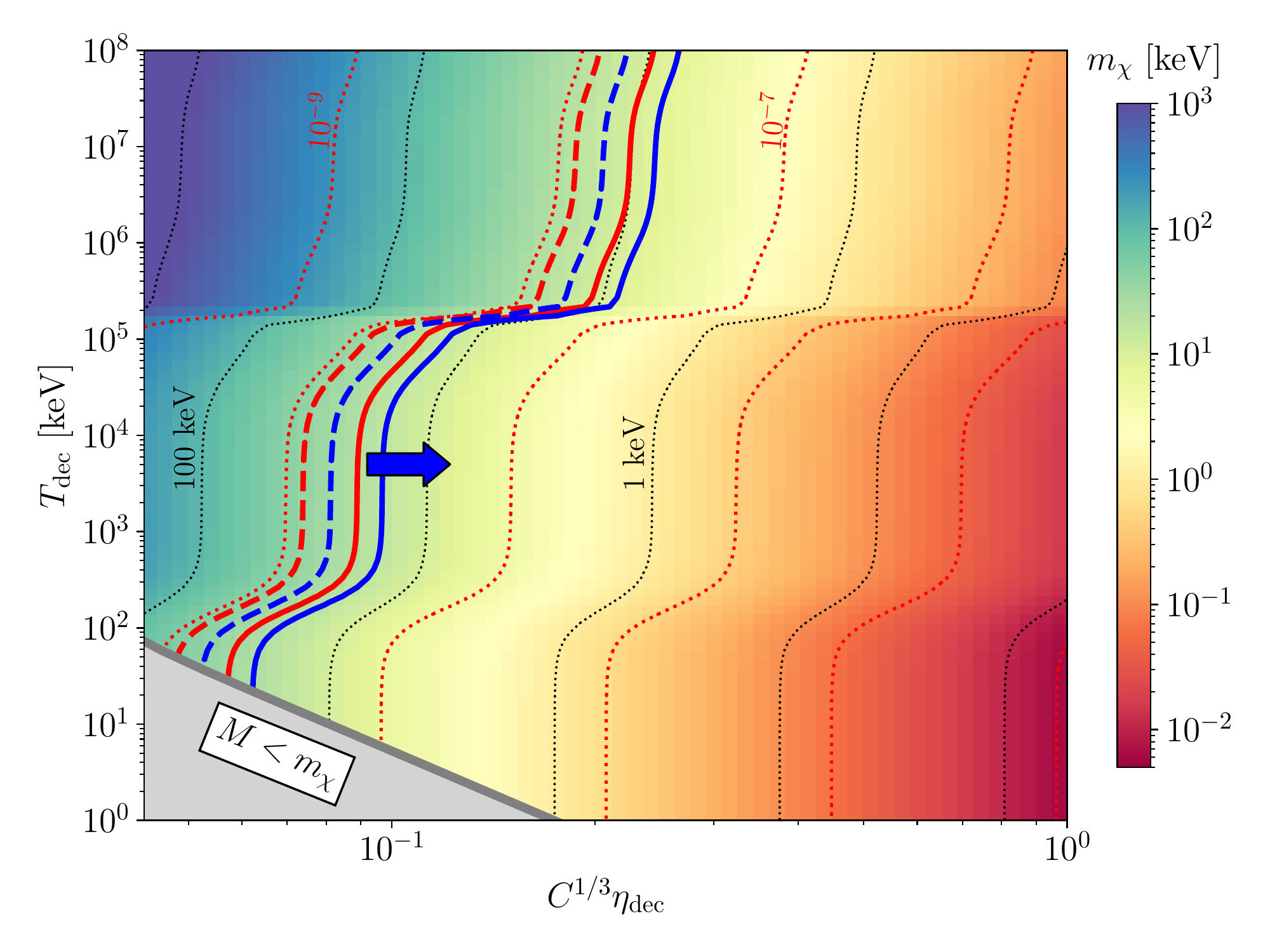}
\caption{Improved bounds on decoupling temperatures and dark-matter mass for the non-relativistic freeze-out (left panel) and the freeze-in (right panel) scenarios.
Since the constraints obtained from the $\delta A$ and the half-mode analysis are extremely similar, we only show the result from the latter method which are indicated by the blue contours.
All other contours are the same as those in Fig.~\ref{fg:T_dec_eta}.
}
\label{fg:improved_bounds1}
\end{figure}

With this mapping, the improved Lyman-$\alpha$ bounds on dark-matter mass and decoupling temperatures are presented in Fig.~\ref{fg:improved_bounds1} (solid and dashed blue contours) for the non-relativistic freeze-out and the freeze-in scenarios.
As expected, such constraints become weaker when one lowers $\eta_{\rm dec}$ below unity.
Comparing to the bounds derived from the average velocity (solid and dashed red thick contours, same as Fig.~\ref{fg:T_dec_eta}), it is clear that while the overall trend is similar in both scenarios,
the improved bounds are also noticeably weaker.

In particular, the 3.5 (5.3) keV WDM constraint is now projected roughly to a range from 5 to 56.5 keV (or from 10 to 94.5 keV) in the non-relativistic freeze-out scenario.
In the freeze-in scenario, assuming $\eta_{\rm dec}=1$, the WDM constraint is mapped to a range from 7.3 to 21.9 keV (or from 12.4 to 37.0 keV), which is shown by the solid (dashed) blue contour.
If we instead assume $C=1$, the contours of dark-matter mass can then be turned into contours of average velocity via the relation in Eq.~\eqref{eq:v0_rel_nrel_1}.
The half-mode analysis therefore put an upper limit on the present-day average velocity of a freeze-in distribution, $\expt{v}_0\lesssim 2.7 \times 10^{-8}$ (or $\expt{v}_0\lesssim 1.6 \times 10^{-8}$), which corresponds to $m_\chi\gtrsim 2.7~\rm keV$ (or $m_\chi\gtrsim 4.0~\rm keV$).

\section{Distinguishing freeze-in and freeze-out from observations}\label{sec:distinguish}

Thus far, we have seen that the constraints from structure formation give rise to significantly different bounds for the freeze-in and non-relativistic freeze-out scenarios which can be realized by the same $2\to 2$ annihilation with different interaction strength. 
While these bounds may help discriminate different production scenarios if additional information on the mass of dark matter can be obtained form any other source,
there is still a huge overlap in the allowed mass range in which it is essentially impossible to uniquely identify the thermal history of dark matter.
Therefore, in this section, we explore the extent to which one can unambiguously distinguish these two scenarios from future observations on the cosmic structure.
We emphasize that, while we use the comparison between the freeze-in and the non-relativistic freeze-out scenarios as an example, an analysis of this kind can be more generally applied to comparisons between many other models or scenarios.

Since the form of future data is somewhat arbitrary,
for generality, let us assume that future measurements on a physical quantity $X(y)$ at the location $y$ in the parameter space can be expressed as 
\beq
X^{+}(y)>X(y)>X^{-}(y)~~~~\text{with}~~
X^{\pm}(y)=X_{\rm ref}[1\pm \sigma_X^{\pm}(y)]\,,
\eeq
where $X^{\pm}(y)$ is the upper/lower limits of $X(y)$,
$X_{\rm ref}(y)$ is the best-fit reference value obtained from the data,
and $\sigma_X^{\pm}(y)$ is the relative uncertainty for the upper/lower bound.
In what follows, 
we shall explore two types of assumptions on future data,
with the first type 
({\bf Type I}) assumes a upper and lower limit on the mass of dark matter from a particular reference model;
and the second type ({\bf Type II}) 
assumes a series of error bars at a collection of wavenumbers $\{k_i\}$ on a reference matter power spectrum $P(k)$ (for example, see Refs.~\cite{Tegmark:2002cy,Tegmark:2008au,Hlozek:2011pc,Chabanier:2019eai,Munoz:2019hjh}).
For convenience, we shall take the WDM model as the reference model in both cases.
However, we emphasize that the WDM mass associated with the WDM model should simply be treated as a parameter that characterizes the deviation from CDM,
and, in principle, any other dark-matter model can be used a reference.

\subsection{Type I: analysis with a WDM mass range}
\begin{figure}\centering
\includegraphics[width=0.6\textwidth]{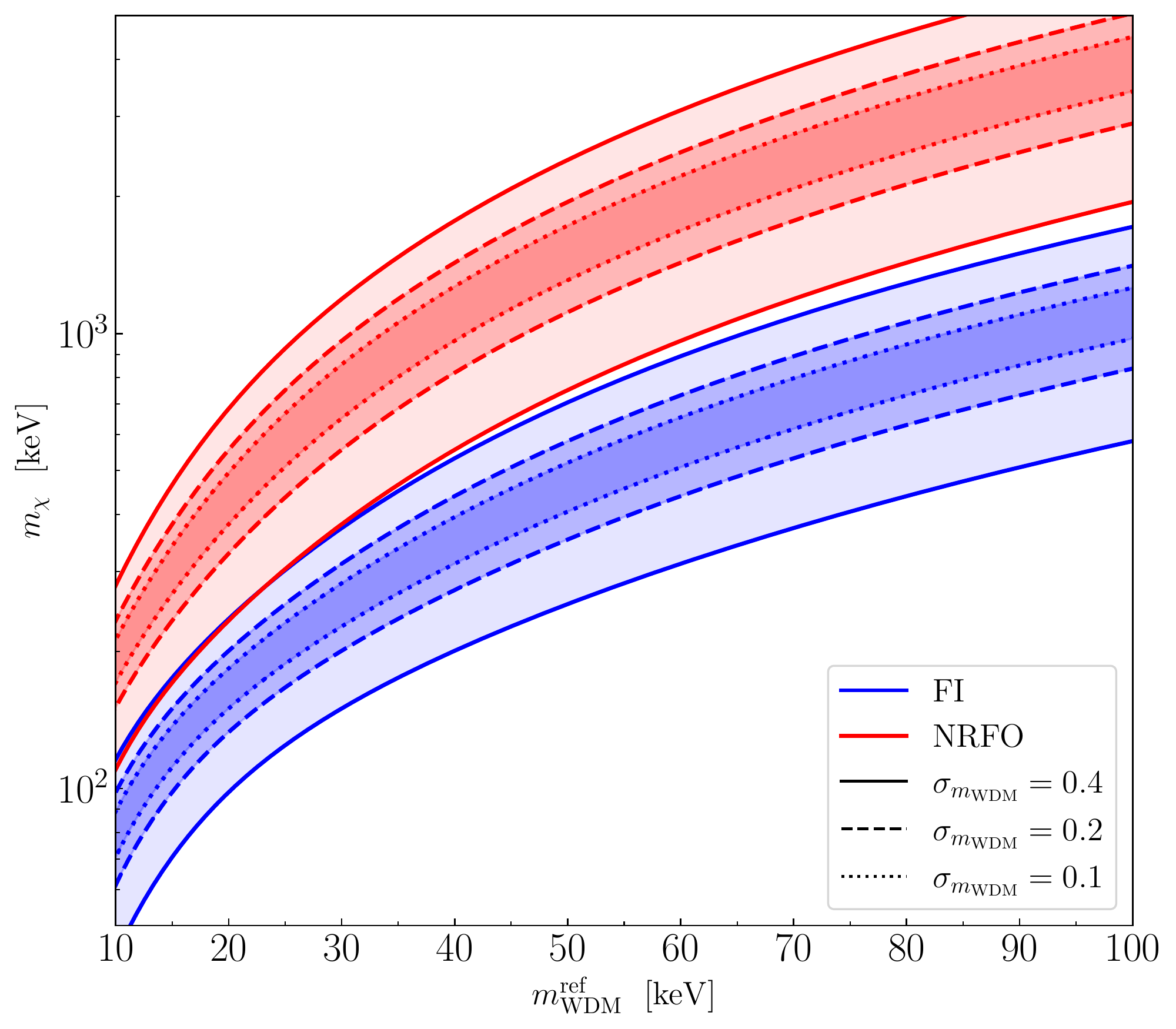}
\caption{
Mapping between the resulting allowed range of $m_{\rm WDM}$ to the allowed range of $m_\chi$ in the freeze-in (shaded in red) and the non-relativistic freeze-out (shaded in blue) scenarios based on the $\delta A$ analysis. 
The solid, dashed and dotted contours correspond to the choices $\sigma_{\rm WDM}=0.4,~0.2$ and $0.1$, respectively.}
\label{fg:mx_mwdm_dA}
\end{figure}

Let us assume that 
a deviation from CDM is detected in future observations, and this deviation is best fitted by a reference WDM model with a WDM mass $m_{\rm WDM}$.
We shall also assume that the relative uncertainty of the upper and lower limits in $m_{\rm WDM}$ is the same such that
\beq
m_{\rm WDM}^{\rm ref}(1+\sigma_{m_{\rm WDM}})> m_{\rm WDM}> m_{\rm WDM}^{\rm ref}(1-\sigma_{m_{\rm WDM}})\,.
\eeq

It is tempting to apply both the $\delta A$ and half-model analyses that we have shown in Sec.~\ref{sec:T2} to find the upper and lower limits on $m_\chi$ in our freeze-in and non-relativistic freeze-out scenarios.
However, while both methods are valid for recasting the lower bound,
it is not clear how to treat the upper bound on an equal footing in the half-mode analysis.
If one requires the squared transfer function associated with the upper bound to have a stronger power at any $k<k_{1/2}$,
the half-mode analysis can be overly conservative when $\sigma_{m_{\rm WDM}}$ is small for cases in which the test transfer function has a shape that differs significantly from the WDM transfer functions.
Therefore, we shall only apply the $\delta A$ method which is less sensitive to the detailed shape of the transfer function to recast the WDM bounds,
\ie~the upper/lower limit on $m_{\rm WDM}$ is mapped to the upper/lower limit on $m_\chi$ in our scenarios by demanding them to produce the same $\delta A$. 

We show in Fig.~\ref{fg:mx_mwdm_dA} the mapping between the allowed mass range for the freeze-in and the non-relativistic freeze-out   scenarios as a function of $m_{\rm WDM}$ with different choices of $\sigma_{m_{\rm WDM}}$.
Clearly, as long as $\sigma_{\rm WDM}$ is not too large ($\sigma_{\rm WDM} \lesssim 0.4$)
the allowed regions for the non-relativistic freeze-out and the freeze-in scenarios can be well separated as long as 
the reference WDM mass is not too small.
For example, the two regions can be separated if $m^{\rm ref}_{\rm WDM}\gtrsim 30~{\rm keV}$ for $\sigma_{m_{\rm wdm}}=0.4$,
and for $\sigma_{m_{\rm wdm}}=0.2$ or $0.1$, they are completely separated for the entire range shown in the figure.

Given the separation of the allowed regions, if additional information on the allowed range of dark-matter mass can be obtained from other sources, such as direct and indirect detection experiments, 
the constraints given by structure formation can be combined to distinguish between the freeze-in scenario and the non-relativistic freeze-out scenario.
Such a comparison can in principle be done using different reference models
and can be made between other scenarios.

It is worth pointing out that the $\delta A$ analysis is designed to impose Lyman-$\alpha$ constraints. 
If the future constraints are derived primarily from other sources, such as the large-scale 21-cm data (see, \eg~Ref.~\cite{Munoz:2019hjh}),
one would need a different proxy other than $\delta A$ to reliably recast the constraints.
It therefore motivates us to explore the more complicated Type-II assumptions on future data
which deal directly with data points on the matter power spectrum.

\subsection{Type II: analysis with future constraints on \texorpdfstring{$P(k)$}{P(k)}}

Future observations can be represented as constraints on the matter power spectrum (or transfer function) at a collection of wavenumbers $\{k_i\}$ which are consistent with a best-fitted matter power spectrum from a WDM model with a reference mass $m^{\rm ref}_{\rm WDM}$.
Thus for any $k\in\{k_i\}$, we have
\beqn
P_{\rm WDM}(k)[1+ \sigma_P^{+}(k)]> &P(k)&>P_{\rm WDM}(k)[1- \sigma_P^{-}(k)]\,,\nn\\
T^2_{\rm WDM}(k)[1+ \sigma_{P}^{+}(k)]>&T^2(k)&>T^2_{\rm WDM}(k)[1- \sigma_{P}^{-}(k)]\,.\label{eq:type2}
\eeqn
Obviously, we notice that the relative errors in the squared transfer function are always the same with that in the matter power spectrum since $T^2(k)=P(k)/P_{\rm CDM}(k)$.
To specify the upper and lower limits, and the dependence on the scales $k$,
we shall test three different possibilities on $\sigma_P^{\pm}(k)$:\\

\noindent 1.~Constant symmetric relative errors on $P(k)$

In this case, we demand the relative errors of the upper and lower bounds are constant and symmetric with respect to $P_{\rm ref}(k)$ on a linear scale.
Thus, we can write $\sigma_P^{\pm}(k)=\sigma_P$ to reduce the relative errors at different $k$ to a constant.
This method has the advantage that the error bars on either $P(k)$ or $T^2(k)$ do not exceed below zero when the value of the WDM reference quantity is small.
However, as we can see from Eq.~\eqref{eq:type2}, it tends to give more constraining power to error bars at the places where the reference $P_{\rm WDM}(k)$ or $T^2_{\rm WDM}(k)$ is small and thus might be sensitive to largest wavenumbers at which measurements are made.\\

\noindent 2.~Constant symmetric absolute errors on $\log P(k)$

Since the matter power spectrum is often presented on a logarithmic scale in literature,
we also include the possibility that the error bars on $P(k)$, when plotted on logarithmic scale, appear to be symmetric and constant.
In this case, one can define a constant quantity
\beq
\Delta_{\log P} \equiv \abs{\log \left[ \frac{P^{\pm} (k)}{ P_{\rm WDM} (k)} \right]}
\eeq
with which it is easy to find that the relative errors on $\log P(k)$ and $P(k)$, 
\beq
\sigma_{\log P}(k)=\abs{\frac{\Delta_{\log P}}{\log P_{\rm WDM}(k)}}\,,~~~ \sigma_P^{\pm}=\pm e^{\pm\Delta_{\log P}}\mp 1\,.
\eeq
Notice that while $\sigma_{\log P}(k)$ is symmetric with respect to $\log P_{\rm WDM}(k)$ but $k$-dependent,
$\sigma_P^{\pm}$ is $k$-independent but asymmetric with respect to $P_{\rm WDM}(k)$.
This approach is convenient when $\sigma_P(k)$ can exceed $\mathcal{O}(1)$, especially when the upper/lower limit is a few times larger/smaller than the best-fit value.
Nevertheless, the upper and lower limits might have unbalanced constraining power.
It also tends to give more constraining power to the data points at which $P_{\rm WDM}(k)$ or $T^2_{\rm WDM}$ is small. 
\\

\noindent 3.~Constant symmetric absolute errors on $T^2(k)$

In this case, we demand that the error bars on the transfer function are constant and symmetric with respect to $T^2_{\rm WDM}(k)$.
Similar to the second case, we can define a constant quantity
\beq
\Delta_{T^2}\equiv \abs{{T^2}^{\pm}(k)-T^2_{\rm WDM}(k)}
\eeq
such that
\beq
\sigma_P(k)=\sigma_{T^2}(k)=\frac{\Delta_{T^2}}{T^2_{\rm WDM}(k)}\,.
\eeq
Notice that the relative errors of $P(k)$ are symmetric but $k$-dependent --- $\sigma_P(k)$ approaches a constant for small $k$, but grows drastically at large $k$.
The advantages of this approach are that 
the constraining power is not dominated by the hypothetical data points at the largest $k$,
and that it seems to treat the upper and lower bounds equally when comparing transfer functions as long as $T^2_{\rm WDM}(k)$ is not sufficiently small such that
the lower limit ${T^2}^{-}(k)$ becomes formally negative.
In the case where the lower limit does turn negative beyond certain $k$,
the lower bounds will lose their constraining power completely.\\

In addition to the form of the uncertainties in each case, we assume that the hypothetical future data consist of 
$N$ data points evenly distributed along the $\log k$-axis within the interval
$[k_{\rm min},k_{\rm max}]$.
We then set the criterion that a test power spectrum is allowed only if $P^{+}(k_i)>P_{\rm test}(k_i)>P^{-}(k_i)$ for any $k_i$ within the collection of data points at $\{k_i\}$.
For our work, the test power spectra are those obtained from the freeze-in and the non-relativistic freeze-out scenarios.
In particular, since any data on $P(k)$ can be turned into data on $T^2(k)$,
with the help of Eq.~\eqref{eq:type2}, we shall base our analysis on the comparison between the squared transfer functions directly, using the fitting formulae in Eq.~\eqref{eq:T2_wdm} and Eq.~\eqref{eq:T2_abc}.

\begin{figure}\centering
\includegraphics[width=0.46\textwidth]{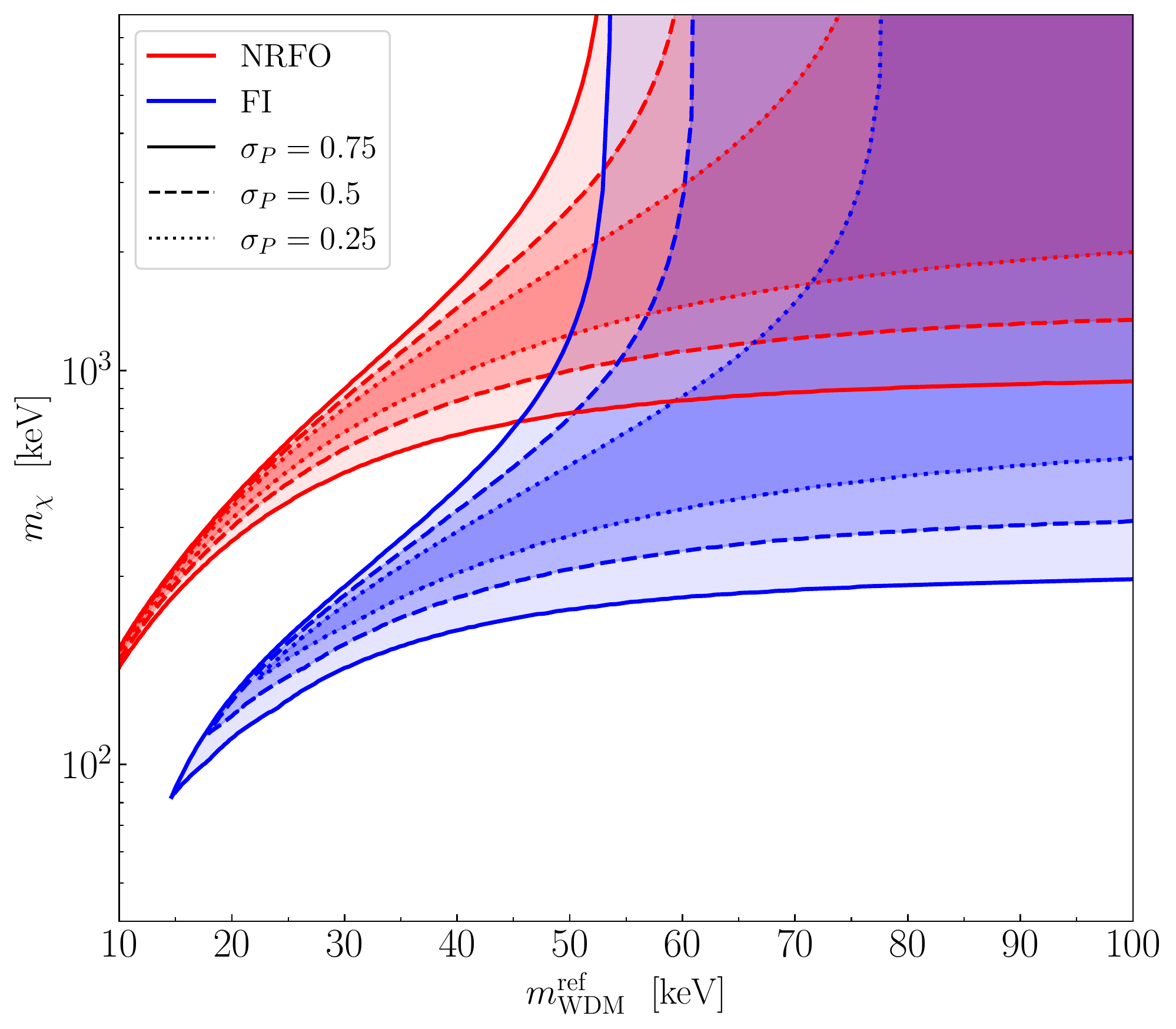}
\includegraphics[width=0.46\textwidth]{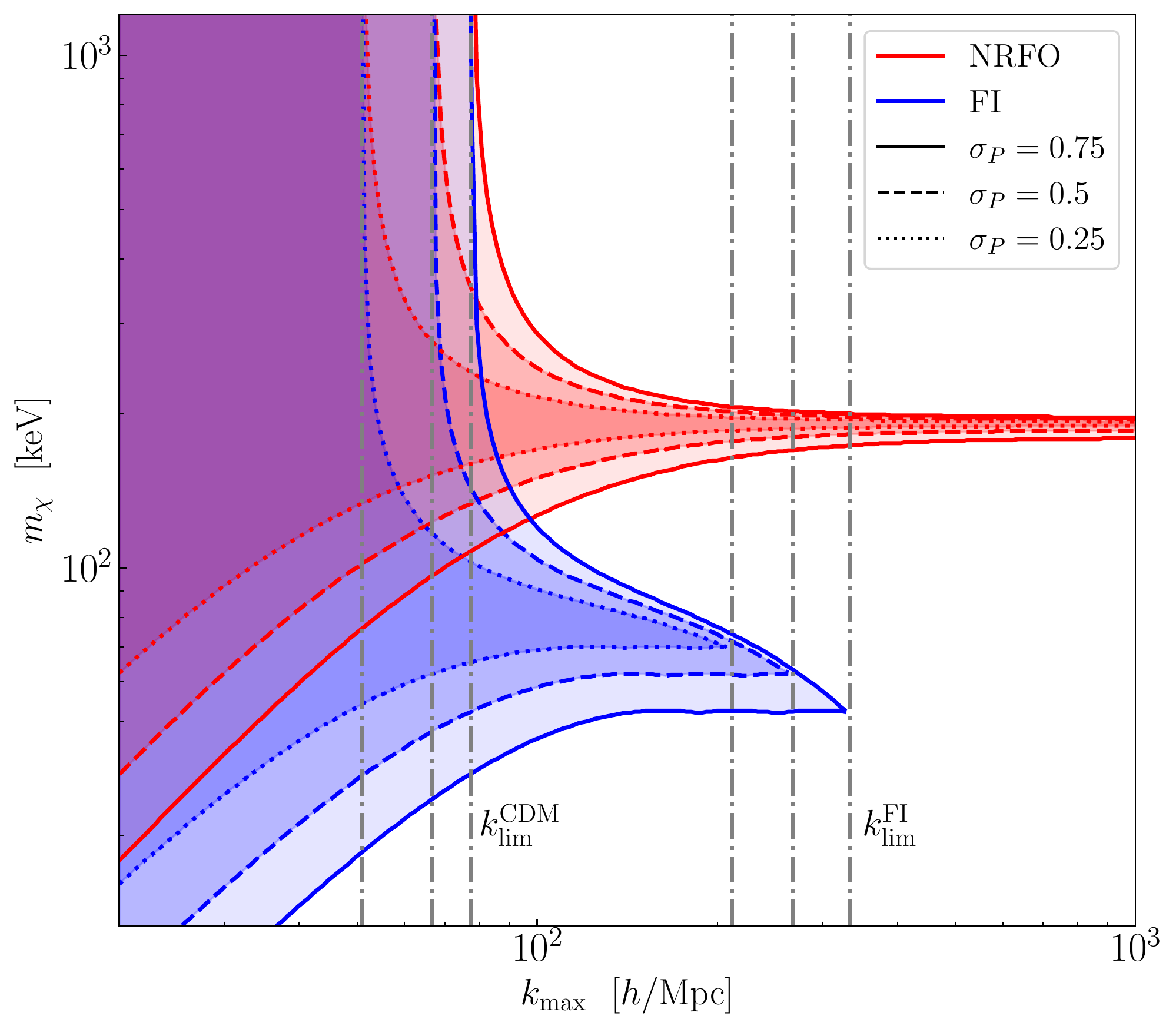}
\includegraphics[width=0.46\textwidth]{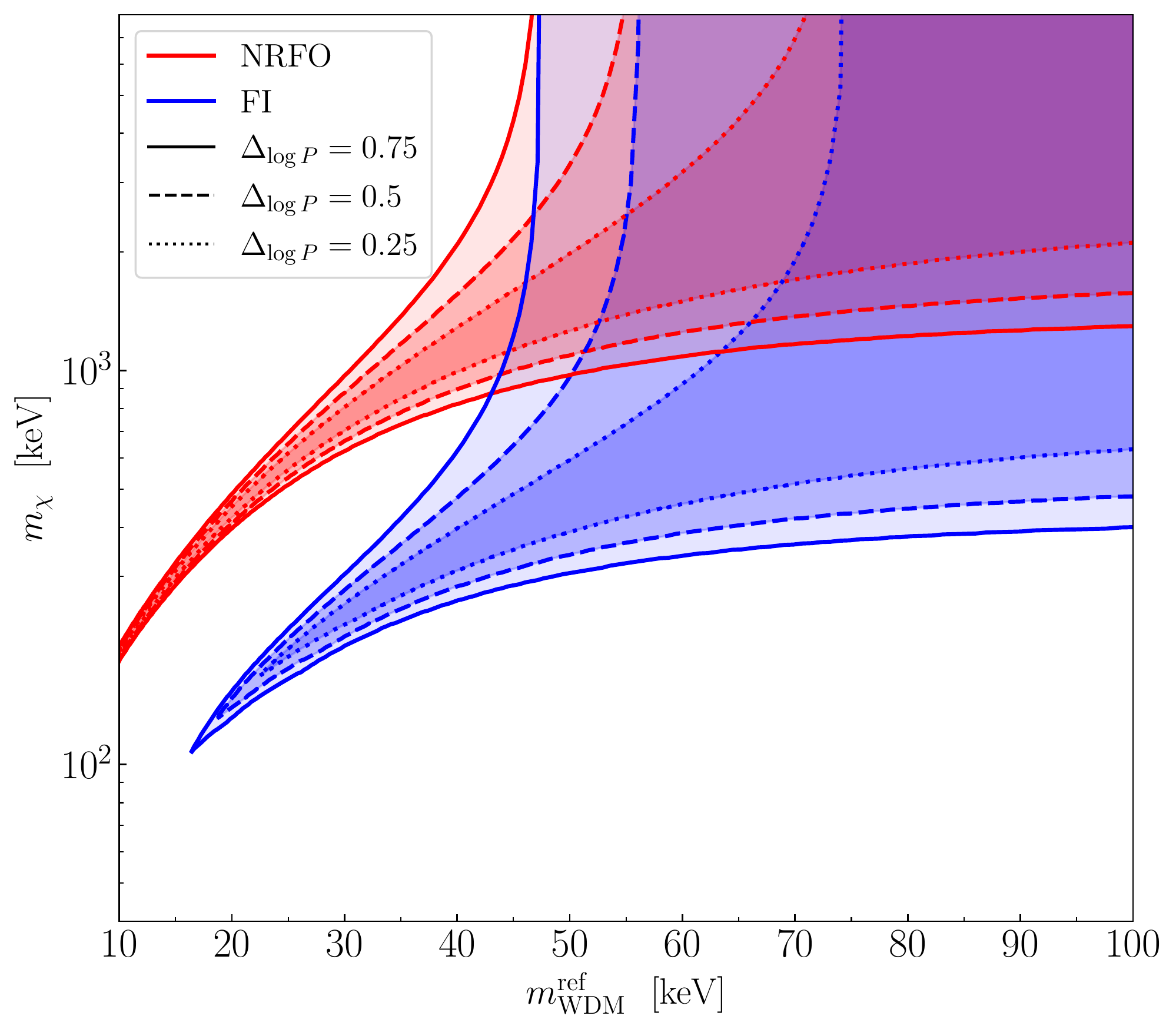}
\includegraphics[width=0.46\textwidth]{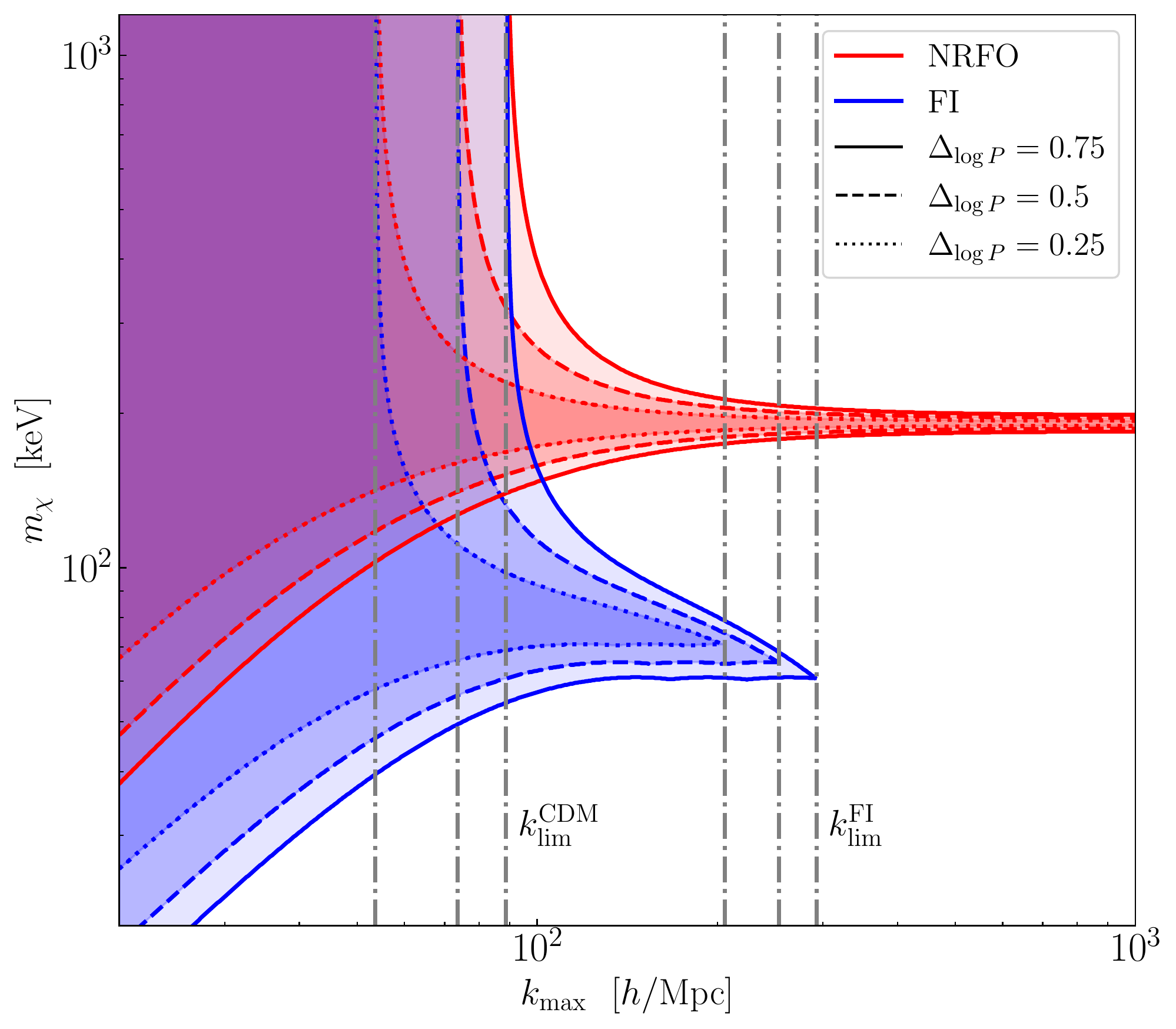}
\includegraphics[width=0.46\textwidth]{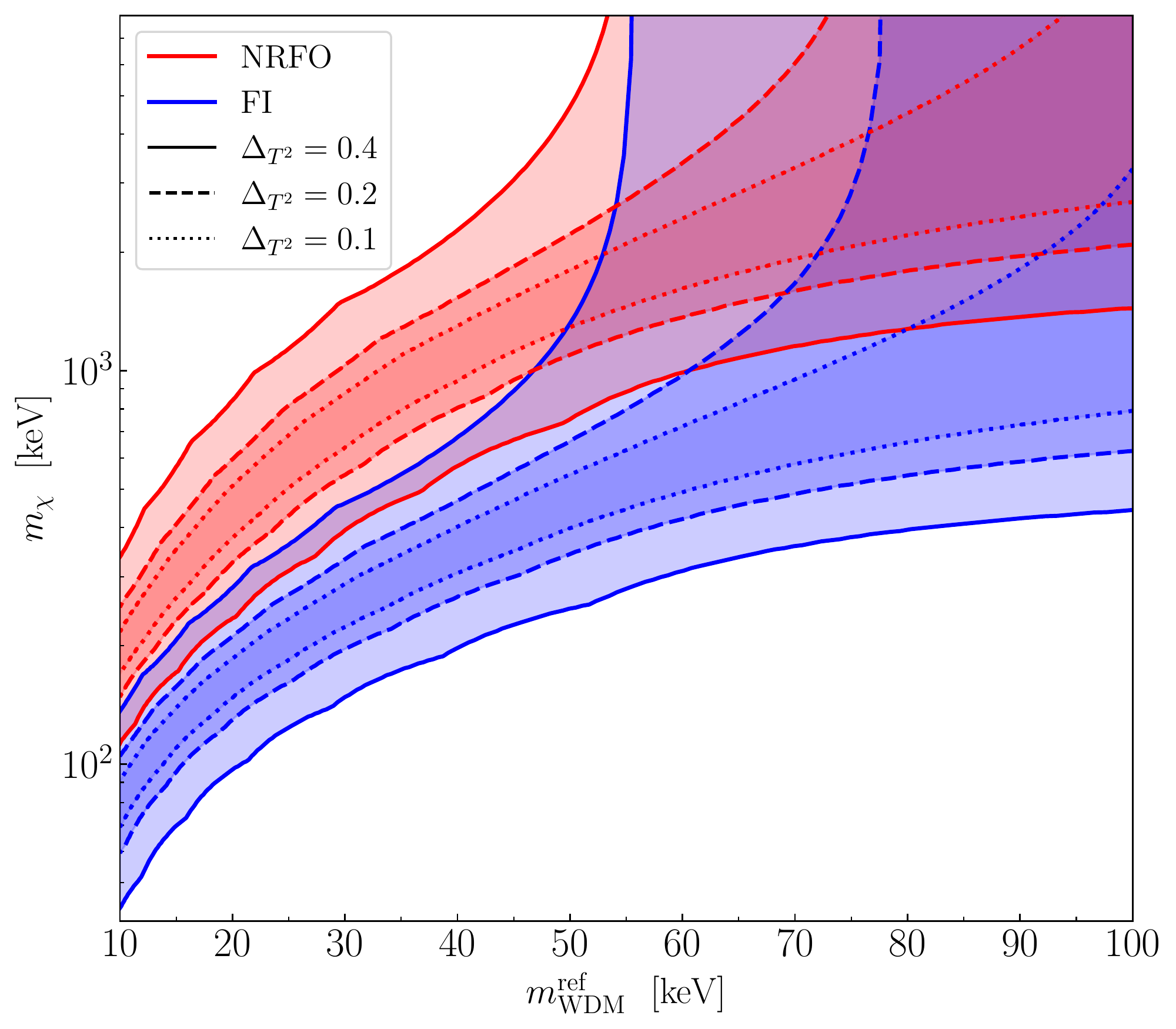}
\includegraphics[width=0.46\textwidth]{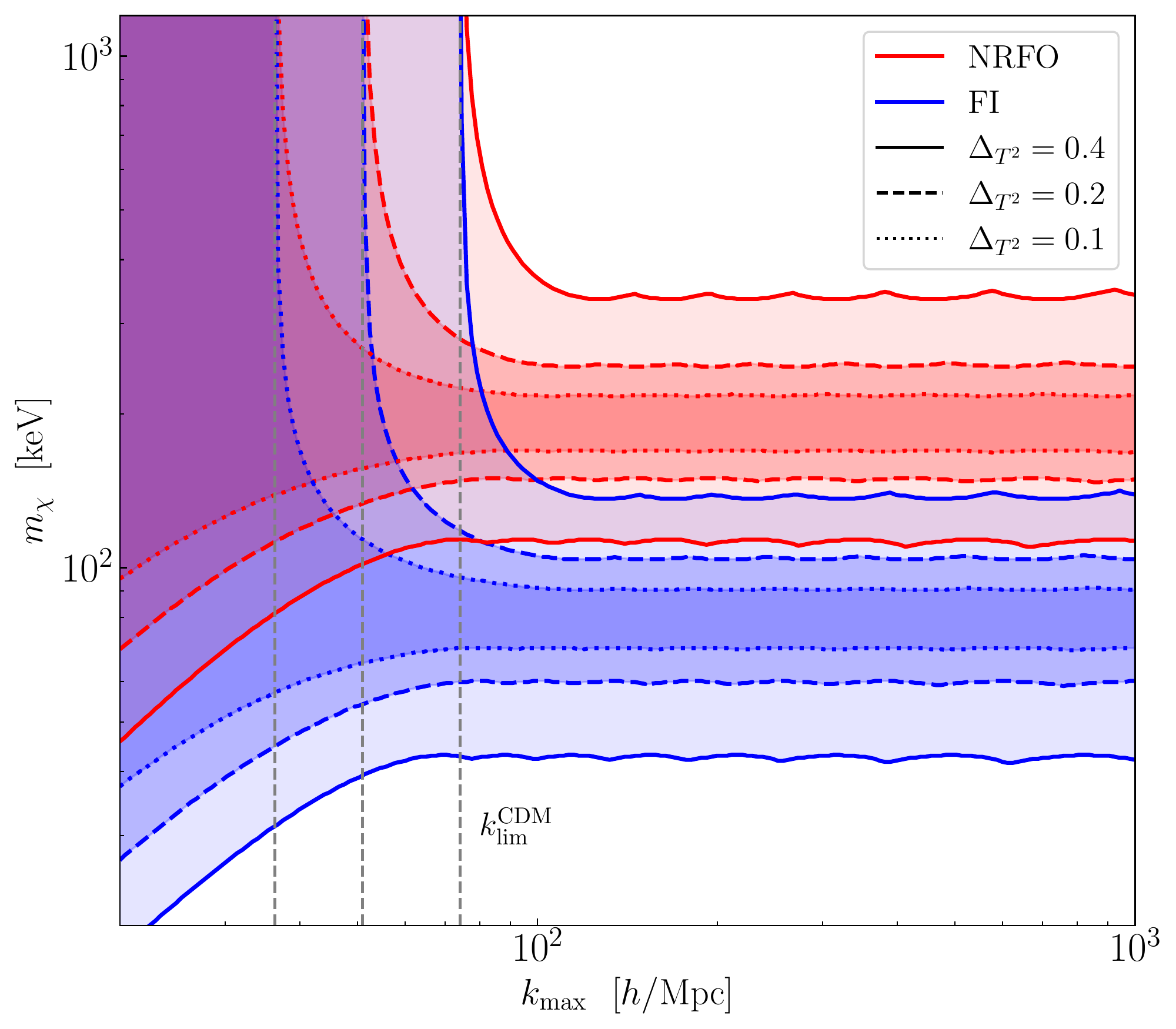}
\caption{
Mappings from the hypothetical constraints on the WDM matter power spectra to the allowed mass ranges for the freeze-in (blue shaded) and the non-relativistic freeze-out (red shaded) scenarios.
The results in the top, middle and bottom panels are obtained by assuming constant symmetric relative errors on $P(k)$,
constant symmetric absolute errors on $\log P(k)$ and constant symmetric absolute errors on $T^2(k)$, respectively. 
Contours with different line styles correspond to specific choices of $\sigma_P,~\Delta_{\log P}$ or $\Delta_{T^2}$ which characterize the errors on hypothetical data.
In the left panels, we fix $k_{\rm min}=1~h/\rm Mpc$ and $k_{\rm max}=500~h/\rm Mpc$ and assume $N=20$ hypothetical data points evenly distributed on log scale.
In the right panels, we fix $m_{\rm WDM}^{\rm ref}=10~\rm keV$ and vary $k_{\rm max}$.
}\label{fg:type2}
\end{figure}

We present in Fig.~\ref{fg:type2}
the allowed mass ranges for the freeze-in and the non-relativistic freeze-out scenarios for the three different assumptions on the uncertainties of the hypothetical future data.
We fix the number of data points $N=20$, and we have tested that further increasing $N$ do not bring appreciable change in the allowed regions. 
Similar to the previous analysis on Fig.~\ref{fg:mx_mwdm_dA}, we look for clear separation of the allowed mass ranges associate 
at the same reference WDM mass such that one is able to distinguish the two types of thermal history if additional information on dark-matter mass is available from other sources.

In the left panels, we fix $k_{\rm min}=1~h/\rm Mpc$ and $k_{\rm min}=500~h/\rm Mpc$.
We first notice in all cases that,
for sufficiently large $m_{\rm WMD}^{\rm ref}$, while one can obtain the lower limits for the dark-matter mass $m_\chi$ in corresponding scenarios,
it is not possible to obtain the upper limits as the contours become vertical.
This is not surprising since
the resulting bounds on the non-relativistic freeze-out and the freeze-in scenarios depend on both the observational precision and the scale at which $P_{\rm WDM}(k)$ deviates from $P_{\rm CDM}(k)$.
In the case where the upper error bars on $T^2(k)$ all exceed one, \ie~the data set is consistent with the CDM predictions within the range of $k$,
obtaining an upper limit for our scenarios is impossible since the free-streaming effects can only suppress the matter power spectrum.
In other words, an upper limit on the dark-matter mass can only be found if a deviation from CDM predictions is confirmed in this analysis.

As the reference WDM mass decreases,
the upper limits on $m_\chi$ do appear as the corresponding contours slowly turn in the horizontal direction
which indicates that the hypothetical data set starts to show deviation from CDM.
If we decrease $m_{\rm WDM}^{\rm ref}$ further, 
a distinct separation of the allowed mass ranges would eventually occur when both the uncertainty and are sufficiently small.

For the cases in the top and middle panels, this separation increases as the allowed mass range shrinks drastically when 
$m_{\rm WDM}^{\rm ref}$ decreases.
This is simply because for smaller $m_{\rm WDM}^{\rm ref}$, the part of the region where the transfer function is suppressed becomes larger within the fixed interval $[k_{\rm min},k_{\rm max}]$,
and thus the entire data set becomes more constraining.
In these two cases, we find that the freeze-in scenario can even be ruled out completely if $m_{\rm WDM}^{\rm ref}\lesssim 15~\rm keV$, leaving the non-relativistic freeze-out scenario as the only viable possibility between the two scenarios.
This happens when the uncertainties of the data set cannot accommodate any transfer function from the freeze-in scenario,
and it is worth pointing out that this constraining power is not possessed by the Type-I data (which essentially also include the type of analysis based on the present-day average velocity of dark matter).
However, it is more difficult to completely rule out the non-relativistic freeze-out scenario given that its corresponding transfer function has a shape that is extremely similar to that from a WDM scenario.
Nevertheless, an exclusion of the freeze-out scenario is in principle possible if the reference scenario is modified. 
On the other hand, in the bottom left panel, the shrinking of the allowed region is not very noticeable at small $m_{\rm WDM}^{\rm ref}$.
This is due to the fact that the absolute errors on $T^2_{\rm WDM}(k)$ is set to constant in this case, and thus the constraining power does not vary significantly when $m_{\rm WDM}^{\rm ref}$ decreases. 

While varying $\mwdmr$ would change the constraining power of the data set within a fixed range of $k$,
we are also interested in understanding the how the range of $k$ at which the data set is presented affects our analysis.
In the right panels of Fig.~\ref{fg:type2}, we show the mass ranges as a function of $k_{\rm max}$ while holding $\mwdmr = 10~\rm keV$ fixed. 

We notice in all cases that for a given uncertainty there exists a critical value of $\kmax$ at which the contours for the freeze-in and the non-relativistic freeze-out scenarios both become vertical, which means the upper limits for $m_\chi$ cannot be obtained if $\kmax$ is below this value.
Similar to the situation in the left panels, the appearance of this critical value simply suggests that the data set cannot exclude CDM.
We shall therefore call this critical value $\klim^{\rm CDM}$,
and it can be determined by solving the equation
\beq
T^2_{\rm WDM} ( k_{\rm lim}^{\rm CDM}) =\frac{1}{1+\sigma_P^+(\klim^{\rm CDM})}=
\begin{cases}
\displaystyle\frac{1}{1+ \sigma_P} & (\textbf{case 1}) \\
 e^{-\Delta_{\log P}} & (\textbf{case 2}) \\
1-\Delta_{T^2} & (\textbf{case 3})
\end{cases}\,.\label{eq:T2limCDM}
\eeq
Obviously, $\klim^{\rm CDM}$ depends on the reference WDM transfer function and thus on $\mwdmr$.
It is also straightforward to see that while one needs $\kmax>\klim^{\rm CDM}$ to allow a separation of the two mass ranges,
a smaller uncertainty on the future data set can decrease $\klim^{\rm CDM}$ and thus enable a scenario discrimination with a smaller data range.

Beyond $\klim^{\rm CDM}$, we also notice in the top and middle panels that for a given uncertainty there exist another critical value of $\kmax$, which we call $\klim^{\rm FI}$, such that the allowed mass region of the freeze-in scenario vanishes.
As we have explained for the similar behavior in left panels, 
the appearance of $\klim^{\rm FI}$ suggests that no transfer function from the freeze-in scenario can be fit within the error bars.
Apparently, a data set that covers a range beyond $\klim^{\rm FI}$ is able to rule out freeze-in completely,
and if the uncertainty of the data decreases, $\klim^{\rm FI}$ would also decrease making it easier to rule out the freeze-in scenario.
Similar to $\klim^{\rm CDM}$, $\klim^{\rm FI}$ also depends on $\mwdmr$ since a change in it would shift the entire reference WDM transfer function.
However, for both $\klim^{\rm CDM}$ and $\klim^{\rm FI}$, the dependence on $\mwdmr$ is somewhat degenerate with the dependence on the data uncertainty.

\begin{figure}\centering
\includegraphics[width=0.49\textwidth]{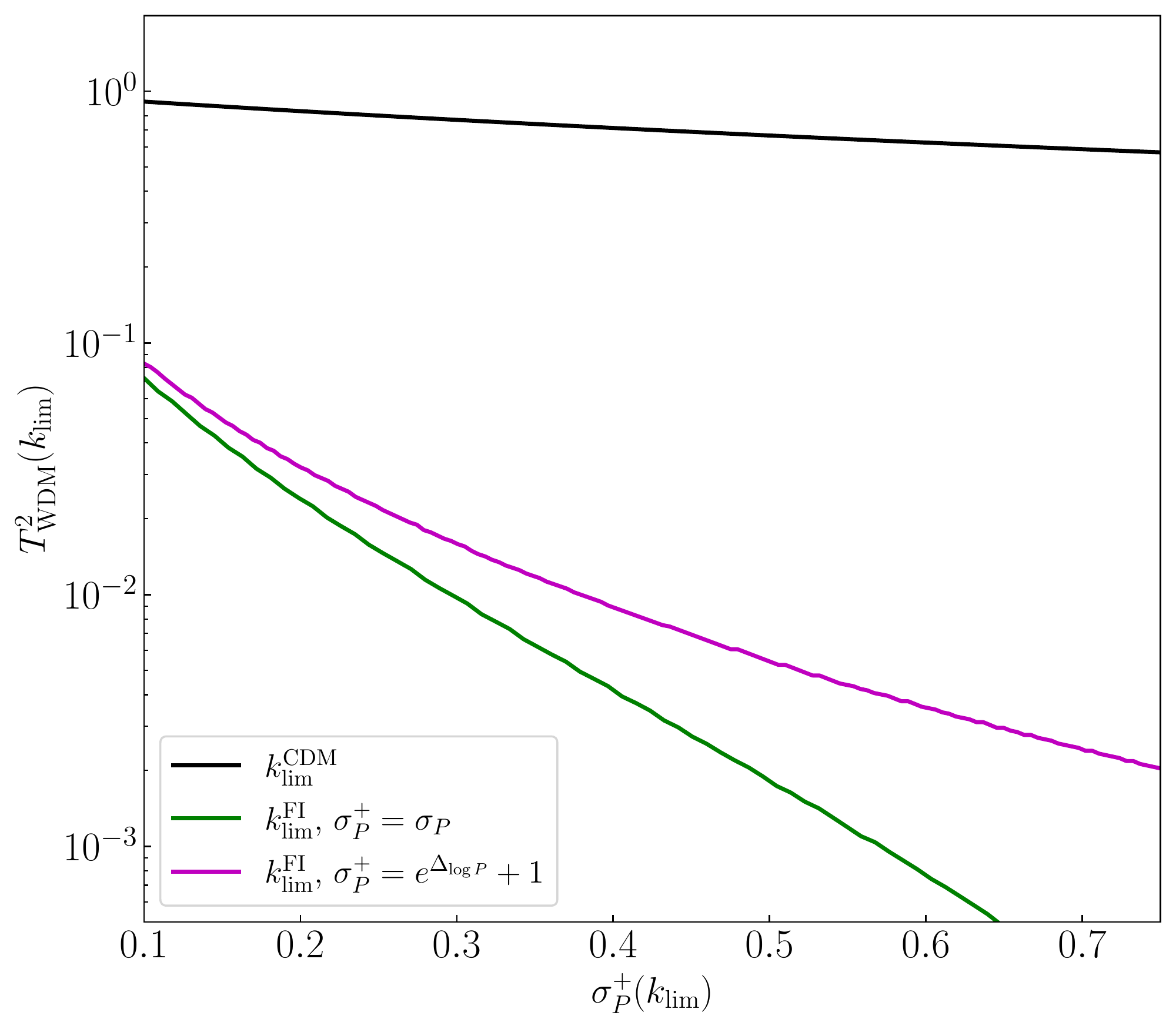}
\includegraphics[width=0.49\textwidth]{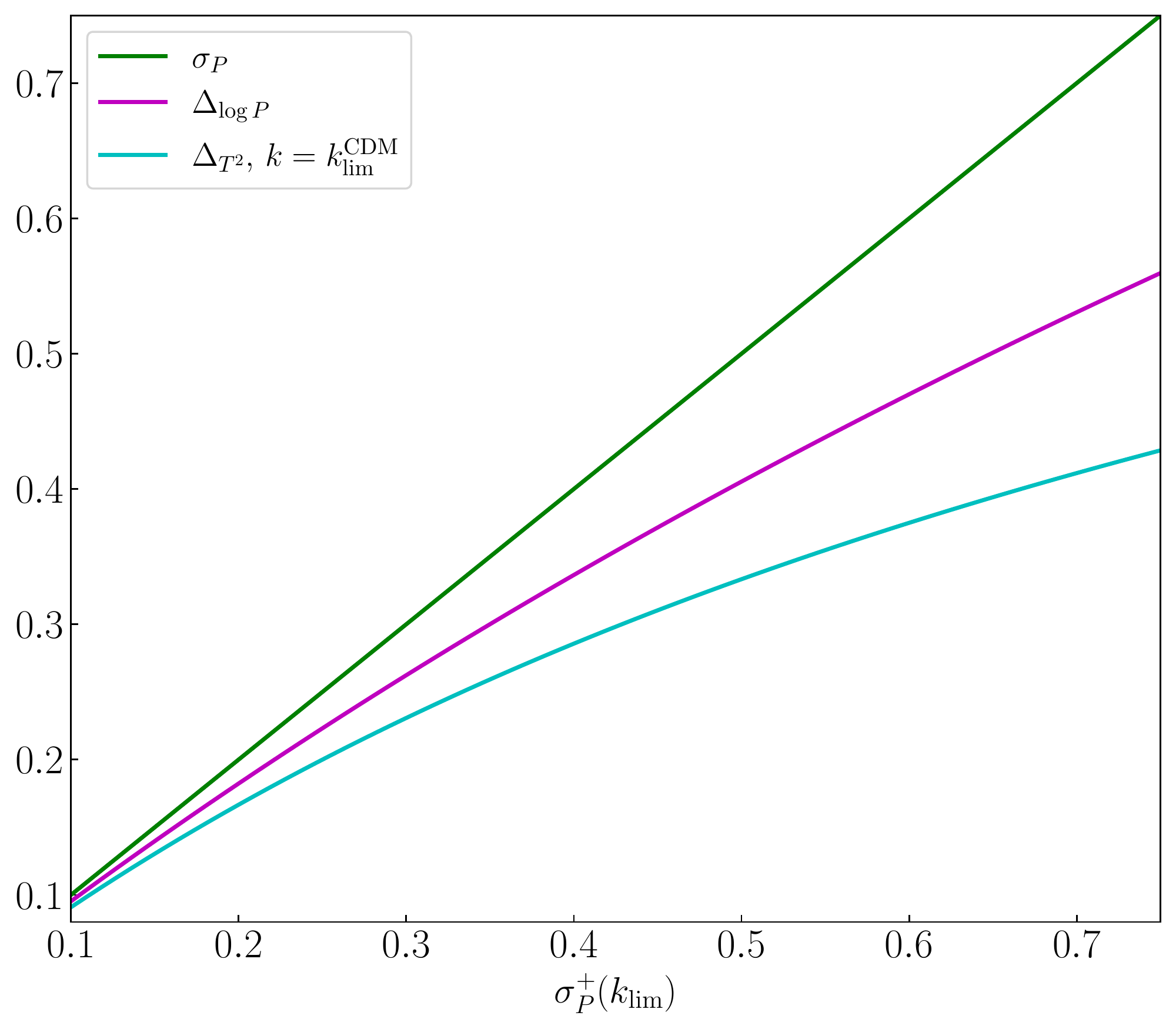}
\caption{The left panel shows the value of $T^2_{\rm WDM}(k)$ at critical wavenumbers $\klim^{\rm CDM}$ and $\klim^{\rm FI}$ as a function of the relative uncertainty $\sigma_P^+$ at these wavenumbers in different cases.
The right panels shows the relation between $\sigma_P^+(\klim)$ and the characteristic quantities $\sigma_P,~\Delta_{\log P}$ and $\Delta_{T^2}$.
Notice that the relations for $\sigma_P$ and $\Delta_{\log P}$ also hold true for any wavenumber $k$.}
\label{fg:T2_sigma}
\end{figure}

While the critical wavenumbers $\klim^{\rm CDM}$ and $\klim^{\rm FI}$ vary with the data uncertainty, the value of the transfer function at these critical points also depend on how precise these data are.
Since we have been using the fitting functions Eq.~\eqref{eq:T2_wdm} and Eq.~\eqref{eq:T2_abc} in our analysis,
we need to make sure that the uncertainties of the fitting functions do not surpass that of the data.
To examine this, we plot in the left panel of Fig.~\ref{fg:T2_sigma} the relation between $T^2_{\rm WDM} (k_{\rm lim})$ and $\sigma_P^+(\klim)$, where $\klim$ can be either the one for CDM or for freeze-in.

The relation at $\klim^{\rm CDM}$ can be obtained directly from Eq.~\eqref{eq:T2limCDM} which clearly shows that it does not depend on which kind of data set we use,
nor does it depend on the reference WDM mass.
It is also clear that $T_{\rm WDM}^2(\klim^{\rm CDM})\gtrsim 0.5$ throughout the entire region and does not decrease significantly as long as $\sigma_P^+(\klim^{\rm CDM})$ is not much bigger than $\mathcal{O}(1)$.
Therefore, the location of$\klim^{\rm CDM}$ can be reliably identified as the fitting function has a high accuracy in this regime with the relative error less than 5\%.

For the relation between $T_{\rm WDM}^2(\klim^{\rm FI})$ and $\sigma_P^+(\klim^{\rm FI})$,
we first find that it does not depend on the reference WDM mass
due to the translational symmetry in the transfer functions that we are considering (see Eq.~\eqref{eq:trans_sym}), and the fact that the most constraining data points for excluding the freeze-in scenario are always those close to $\kmax$.
However, the relation at $\klim^{\rm FI}$ is sensitive to the relative uncertainty $\sigma_P^+(\klim^{\rm CDM})$ as $T_{\rm WDM}^2(\klim^{\rm FI})$ quickly decreases when the uncertainty increases.
This quick drop also depends on the specific case in consideration. 
We notice that while the relative uncertainty at the lower end of the error bar $\sigma_P^-=\sigma_P^+$ for the data set with linearly symmetric errors, the data set with log-symmetric errors gives $\sigma_P^-=\sigma_P^+/(1+\sigma_P^+)<\sigma_P^+$.
This means that, for the same $\sigma_P^+$, the latter case is always more constraining than the former case, which explains why it allows a smaller deviation from CDM at $\klim^{\rm FI}$ to rule out the freeze-in scenario. 
Overall, we see that $T^2(k_{\rm lim}^{\rm FI})$ approaches $\sim 10^{-3}$ when $\sigma_P \gtrsim 0.6$.
On the other hand, we find numerically that the relative error of the fitting formulae in Eq.~\eqref{eq:T2_wdm} and Eq.~\eqref{eq:T2_abc} grows larger than $\mathcal{O}(0.1)$ when the corresponding transfer function $T^2(k)\sim 10^{-2}$.
To claim an exclusion of the freeze-in scenario with sufficient confidence, it is necessary to demand that the relative uncertainty in the fitting function is smaller than that of the hypothetical data.
This correspond to $0.1\lesssim \sigma_P^+(\klim^{FI}) \lesssim 0.3$ or $0.4$ depending on the data set. 

Finally, to allow a convenient comparison between Fig.~\ref{fg:type2} and the left panel of Fig.~\ref{fg:T2_sigma},
we plot on the right panel the relation between $\sigma_P^+(\klim)$ and the characteristic quantities $\sigma_P,~\Delta_{\log P}$ and $\Delta_{T^2}$.
Notice that while the relations for $\sigma_P$ and $\Delta_{\log P}$ do not rely on whether we choose $\klim^{\rm CDM}$ or $\klim^{\rm FI}$,
the relation for $\Delta_{T^2}$ is evaluated at $\klim=\klim^{\rm CDM}$ since there is not a critical wavenumber to rule out freeze-in in this case.
We can see that, for the values of the characteristic quantities shown in Fig.~\ref{fg:type2}, the values of $\sigma_P^+(\klim)$ need not be exceedingly small.
The discussion in the previous paragraph also suggests a choice of $0.1\lesssim \sigma_P \lesssim 0.3$ or $0.1\lesssim \Delta_{\log P} \lesssim 0.3$ to be able to reliably rule out freeze-in.

\section{Conclusion}\label{sec:conclusion}

Dark matter is largely believed to be in the form of certain particle,
but the particle-physics processes that it participates and the mechanism through which it is produced remain unknown.
With all the current pieces of its supporting evidence originate from its gravitational interaction only,
it is therefore pressing to understand what one can infer from the effects that only rely on the gravitational interaction of dark matter.

In this paper, we explore the connection between structure formation and the dark-matter thermal history using the phase-space distribution after dark matter is produced.
We first revisited the calculation to relate the present-day average velocity of dark matter to relevant physical quantities at the time when dark matter just decouples.
We find that the decoupling temperatures of the dark and the SM sectors, as well as the mass of dark matter are correlated with the present-day velocities.
This correlation can vary significantly across different types of dark-matter production mechanisms.
We therefore re-interpret the Lyman-$\alpha$ constraints in the well studied WDM scenario as constraints on the present-day average velocity of dark matter and then apply them to other production scenarios such as freeze-in and non-relativistic freeze-out.
We find that the constraints on the dark-matter mass can differ a lot among the two scenarios and the WDM scenario.
After mapping out the correlation between the physical quantities at the early and the late times,
we are able to project these preliminary bounds on dark-matter mass or average velocity to constrain the decoupling temperatures which concern the physics in the early universe.

While the present-day average velocity can provide a good estimate on the scale below which structure formation is suppressed,
it cannot capture all the effects from the entire phase-space distribution in general.
To provide more reliable constraints,
we numerically solve the Boltzmann equation to obtain the phase-space distribution $f_\chi$ after the production of dark matter.
By requiring the results to match the present-day energy density of dark matter, we find that, above certain threshold mass $m_\chi^{\rm min}$,
there are always two amplitudes for the same particle-physics processes that are able to generate the correct relic abundance.
These solutions are associated with either the freeze-in regime if the interaction strength is feeble, 
the freeze-out regime if the particle interactions are rapid, 
or the transition regime if the interaction strength is intermediate.
We then focus on pure freeze-in and the non-relativistic freeze-out scenarios,
and use the corresponding phase-space distributions as input for the \texttt{CLASS} code to obtain the matter power spectra and the transfer functions.
We find that the transfer functions in each scenario all possess the same shape and can be related to each other via a translational symmetry.
In addition, all of them can be well fitted using the $\{\alpha,\beta,\gamma\}$ fitting formula for which we have introduced the dependence on the dark-matter mass $m_\chi$ in the parameter $\alpha$.
The new fitting parameters then enable us to scan over a large range of the parameter space to constrain our scenarios.
In particular, we have used the well-known half-mode analysis and the $\delta A$ analysis to recast the current Lyman-$\alpha$ constraints on WDM as constraints in the freeze-in and the non-relativistic freeze-out scenarios.
These constraints are weaker than the preliminary ones from the average velocity but still show significant large differences when comparing the results in the WDM, the freeze-in, and the non-relativistic freeze-out scenarios.

With the matter power spectrum, we have also discussed the effects on the physical quantities in the non-linear regime such as the halo mass function and the MW satellite counts using an extended Press-Schechter approach.
The halos mass function is difficult to constrain directly from astronomical observations since dark-matter halos are not visible.
Efforts on constraining the halo mass function thus rely on 
the visible content that resides in the halos and the comparison with simulations.
For example, observables such as 
the galaxy cluster number counts, 
the galaxy cluster power spectrum, 
the lensing of Type Ia supernovae 
were exploited in Ref.~\cite{Castro:2016jmw},
the possibility of using the magnification bias effect on high-redshift submillimeter galaxies
were discussed in Ref.~\cite{Cueli:2021dai}.
Ref.~\cite{Shirasaki:2021orc} used the cumulative luminosity function of galaxies at redshift $z=6$ and placed a lower limit on WDM mass at 2.71 keV at $2\sigma$ level (or 2.27 keV for Sheth-Tormen \cite{Sheth:1999mn} and 1.96 keV for Press-Schechter mass function),
though this limit is sensitive on the minimum halo mass.
Ref.~\cite{Gu:2023jef} showed that the halo-mass-function parameters can be measured at a few percent level by weak lensing in future survey.
For the MW satellite count,
we found the constraints directly from it are weaker than that from the Lyman-$\alpha$ observations.
On the other hand, recent studies (see \eg~Refs.~\cite{DES:2020fxi,Newton:2020cog,Nadler:2021dft,Dekker:2021scf}) on MW satellite count are shown to be able to place more stringent constraint on WDM mass. 
In particular, effects such as the tidal stripping from the host halo, the disruption of subhalo due to the Galactic disk, and reionization that were included in these works are able to lower the subhalo abundance and thus tighten the constraint. 
Nevertheless, we stick to the simple approach with Eq.~\eqref{eq:SHMF} since it allows us to recast the WDM bounds in other scenarios analytically in a straightforward manner.
Future observations with better sensitivity and larger sky coverage may further improve this bound.

Since no deviation from the CDM predictions has been confirmed thus far from current astrophysical or cosmological data,
observational constraints can only provide lower bounds on the dark-matter mass in typical dark matter models.
Although such bounds can be quite different in different dark-matter scenarios which have already enabled a discrimination between different dark-matter thermal histories to some extent,
there is still a huge overlap in the allowed parameter space for these scenarios since no upper limit is present.
We therefore seek to explore the extent to which future data can better improve the ability to distinguish different thermal histories.

As the form of future data is unknown to us, we examine two general types of hypothetical future data
among which the Type-I data refer to upper and lower limits on the dark-matter mass of a reference model,
and the Type-II data refer to a series of data points with error bars superposed on a reference matter power spectrum or transfer function.
Obviously, the Type-I data should be seen as a fully processed result, whereas the Type-II data are closer to the raw data themselves though they are not the data on the actual physical observables.
Using WDM as a reference scenario, 
and with several different assumptions on the data uncertainties,
we find that in all cases one can significantly narrow down the allowed mass ranges for the freeze-in and the non-relativistic freeze-out scenarios ---
with reasonably small data uncertainties, 
the allowed regions for the two scenarios in consideration can be completely separated.
In the presence of Type-II data, we find that it is even possible to fully exclude one of the scenarios (in our analysis the freeze-in scenario) while maintaining the viability of the other scenario.
This ability, while requiring measurements with sufficiently small uncertainties to be made beyond certain critical wavenumber, is not possessed by the analysis with Type-I data which also includes the type of analysis based on the average velocity only.
Our results therefore suggest that there is a great opportunity in distinguishing different thermal histories of dark matter from future observational data.

\acknowledgments

We would like to thank B.\,Thomas for discussions.
FH wishes to acknowledge the hospitality of the Institute of Theoretical Physics, Chinese Academy of Science. This work is supported by the National Science Foundation of China under Grants No. 12022514, No. 11875003 and No. 12047503, and National Key Research and Development Program of China Grant No. 2020YFC2201501, No. 2021YFA0718304, and CAS Project for Young Scientists in Basic Research YSBR-006, the Key Research Program of the CAS Grant No. XDPB15.  

\appendix

\section{Boltzmann equation for the number density and the thermally averaged cross section}
\label{sec:formula}

The Boltzmann equation for the number density can be obtained from the Boltzmann equation for the phase-space distribution 
by integrating both side of Eq.~\eqref{eq:Boltzmann_fx} over the corresponding phase space. 
After that, the left-hand-side of the equation becomes
\beq
\int\frac{d^3 p}{(2\pi)^3} \left( \frac{\partial }{\partial t} - Hp\frac{\partial }{\partial p}\right)f_\chi  = \frac{dn_\chi}{dt}+3Hn_\chi\,.
\eeq
For the process $ 1+2\rightarrow 3+\chi $, the integration over the collision term reads
\beqn\label{numc1}
\int\frac{d^3 p_\chi}{(2\pi)^3}~\mathcal{C}[f] &=& \int d\pi_1 d\pi_2 d\pi_3 d\pi_\chi~ (2\pi)^4 \delta^{(4)}(p_1+p_2-p_3-p_\chi) \overline{\abs{\mathcal{M}_{1+2\rightarrow 3+\chi}}^2}\nn\\
&&\times \left( f_1 f_2 -f_3 f_\chi \right),
\eeqn
which cannot be integrated analytically since $ f_\chi $ is unknown a priori. 
Nevertheless, we can assume that all the particles are kept in the kinetic equilibrium with the thermal bath, such that $f_\chi\sim e^{-(E-\mu_\chi)/T} \sim f_\chi^{\rm eq} n_\chi/n_\chi^{\rm eq}$.
In this way, the above equation can be conveniently expressed as
\begin{equation}
\int\frac{d^3 p_\chi}{(2\pi)^3}~\mathcal{C} [f] = \left\langle \sigma v\right\rangle \left(n_3^{\rm eq} n_\chi^{\rm eq} \frac{n_1 n_2}{n_1^{\rm eq} n_2^{\rm eq}} - n_3 n_\chi \right) \,,
\end{equation}
where we have defined the thermally averaged cross section
\begin{equation}\label{key}
\left\langle \sigma v\right\rangle \equiv \frac{1}{n_3^{\rm eq} n_\chi^{\rm eq}}\int d\pi_1 d\pi_2 d\pi_3 d\pi_\chi ~(2\pi)^4 \delta^{(4)}(p_1+p_2-p_3-p_\chi) \overline{\abs{\mathcal{M}_{1+2\rightarrow 3+\chi}}^2}  f_1^{\rm eq} f_2 ^{\rm eq}\,.
\end{equation}
If the particle species 1 and 2 are in thermal equilibrium and the particle species 3 is also the dark matter $\chi$, 
the Boltzmann equation for the number density of dark matter can be simply expressed as
\begin{equation}
	\frac{dn_\chi}{dt}+3Hn_\chi =  \left\langle \sigma v\right\rangle \left[ \left( n_\chi^{\rm eq}\right) ^2  - n_\chi^2 \right] \,,
\end{equation}
which is often seen in the literature.

\section{A translational symmetry}\label{sec:shift}

When fitting the squared transfer function in Sec.~\ref{sec:T2},
we observed that the squared transfer function resulting from pure freeze-out (or freeze-in) possesses an approximate symmetry such that
varying the dark-matter mass $m_\chi$ amounts to rigidly translating the transfer function along the $\log k$-axis.
To be concrete, the translation operation can be expressed as
\beqn
m_\chi \to  m'_\chi\,,~~~
\alpha(m_\chi) \to \alpha'=\alpha(m'_\chi)\,, \nn\\
k \to  k'=\frac{\alpha}{\alpha'}k\,,~~~
T^2(k) \to {T'}^2(k')=T^2(k)\,.\label{eq:trans1}
\eeqn
In this appendix, we aim to show that the underlying reason for this symmetry is due to the fact that, the normalized momentum distributions $g_p(p)/(N_\chi)$ resulting from freeze-out (or freeze-in) are approximately the same for different $m_\chi$, as we have seen in Fig.~\ref{fg:Nx_fx}.\footnote{Since we shall discuss distribution functions in different spaces in this appendix, we reserve the subscript to indicate the corresponding phase space, such as momentum space, velocity space and wavenumber space.}

We first notice that the velocity distributions $g_v(v)\equiv g_p(p)\abs{d\log p/d\log v}\approx g_p(p)$ are identical to each other (after normalization by the area under the curve) up to some rigid translation along the $\log v$-axis when the distributions only have non-vanishing support for non-relativistic velocities.
In particular,
changing $m_\chi$ to $m'_\chi$ amounts to translating the velocity distribution as
\beqn
v\to v'=v\frac{m_\chi}{m'_\chi}\,,~~~
g_v(v)\to g'_v(v')=g_v(v)\,.
\eeqn
On the other hand, the velocity distribution can be mapped into a distribution $g_k(k)$ in the space of wavenumbers by mapping the velocity $v$ to the wavenumber $k$ using the freestreaming horizon $k_{\rm hor}$ \cite{Dienes:2020bmn,Dienes:2021itb}:
\beqn
k=k_{\rm hor}(v)=\xi \left[\int_{a_{\rm prod}}^1 \frac{da}{Ha^2}\frac{\gamma v}{\sqrt{\gamma^2v^2+a^2}} \right]^{-1}\,,
\eeqn
where $\xi$ is an $\mathcal{O}(1)$ constant factor.
One can then define a distribution function in the $\log k$ space
\beq
g_k(k)\equiv J(v)g_v(v)\,,
\eeq
in which $J(v)\equiv \abs{d\log v/d\log k}$ is the Jacobian.
With these, the translation in the $v$-space can be mapped to a translation in the $k$-space which goes like
\beqn
k=k_{\rm hor}(v) \to k'=k_{\rm hor}\left(v'\right)  \,,~~~
g_k(k) \to J(v')g'_v(v')=g_k'(k')\,.
\eeqn
In the limit $v/a_{\rm MRE}\ll 1 $, one can show numerically that  
since $\log (k'/k)$ is approximately a constant as $v$ changes,
the Jacobian
$J$ is also approximately a constant.
Therefore, 
$g'_k(k')\approx J(v)g_v(v)=g_k(k)$, 
\ie~the transformation in the $k$-space is also an approximately rigid translation.
Approximating $J$ as a true constant,
we can write
\beq
\log k'-\log k=(\log v -\log v') /J=\log \left(\frac{m'_\chi}{m_\chi}\right)^{1/J} \,,
\eeq
which means $k'=\zeta k$ with $\zeta=(m'_\chi/m_\chi)^{1/J}$.

The next step is to figure out how the squared transfer function transform.
In Ref.~\cite{Dienes:2020bmn}, it has been shown that the logarithmic derivative of the transfer function at a wavenumber $k$ is correlated with the fraction of dark-matter particles that are able to freestream a distance larger than $\sim 1/k$.
In particular, it is shown that the relation
\beq
\frac{d\log T^2}{d\log k}\approx F^2(k)+\frac{3}{2} F(k)\label{eq:T2F}
\eeq
holds to high accuracy, where
\begin{equation}
   F(k) ~\equiv~   \frac{\int_{\infty}^{\log k} d \log k'\, g_k(k')}
     {\int_{-\infty}^\infty d\log k'\, g_k(k')}
  \label{eq:hotfrac}
\end{equation}
is called the \emph{hot-fraction function}.
Integrating Eq.~\eqref{eq:T2F}, it is straightforward to show that
\beq
T^2(k)\approx \exp\left\{-\int_{-\infty}^{\log k} d\log k' \left[ F^2(k')+\frac{3}{2}F(k')\right]\right\}\,,\label{eq:T2F_int}
\eeq
where we have used the fact that
$\lim_{ k\to 0} \log T^2(k) = 0$.
Since $F(k)$ is the integral of $g_k(k)$, 
it is easy to verify that the $k$-space transformation implies
\beqn
k\to k'\,,~~~F(k)\to F'( k')=F(k)\,.
\eeqn
Therefore, the squared transfer function transforms as
\beqn
T^2(k) &\to& {T'}^2(k')\nn\\
&=&\exp\left\{-\int_{-\infty}^{\log k'} d\log k'' \left[ {F'}^2( k'')+\frac{3}{2}F'( k'')\right]\right\}\nn\\
&=&\exp\left\{-\int_{-\infty}^{\log k} d\log (\zeta^{-1} k'') \left[ F^2(\zeta^{-1} k'')+\frac{3}{2}F(\zeta^{-1} k'')\right]\right\}\nn\\
&\approx& T^2(k)\,,
\eeqn
where we have used $k=\zeta^{-1}k'$ in the third line.
The entire transformation then goes like
\beqn
&&m_\chi \to m'_\chi\,,~~~v\to v'=  \frac{m_\chi}{m'_\chi}v\,,~~~g_v(v)\to g'_v(v')=g_v(v)\,,\nn\\
&& k \to  k'= \zeta k \,,~~~
g_k(k)\to g_k'(k')=g_k(k)\,,~~~T^2(k) \to {T'}^2(k')=T^2(k) \,.
\eeqn
Noticing that 
$J\approx 1.1$ for $k\in [1,10^3]~h/\rm Mpc$ from numerical result,
it is therefore reasonable to identify
$\zeta=(m'_\chi/m_\chi)^{1/J}$ with $\alpha(m_\chi)/\alpha(m'_\chi)$.
Thus, we have approximately proved that the underlying reason for the translational symmetry 
is the invariance of $g_p(p)$
(which amounts to a rigid shift of $g_v(v)$ in the $\log v$ space) under the change of $m_\chi$ as we have seen in Fig.~\ref{fg:Nx_fx}.

\bibliographystyle{JHEP}
\bibliography{ref}

\end{document}